\RequirePackage{fix-cm}
\RequirePackage{fixltx2e}
\documentclass[a4paper,english,11pt]{article}
\usepackage[T1]{fontenc}
\usepackage[latin9]{inputenc}
\usepackage{geometry}
\usepackage{babel}
\usepackage{refstyle}
\usepackage{float}
\usepackage{calc}
\usepackage{mathrsfs}
\usepackage{amsmath}
\usepackage{amsthm}
\usepackage{amssymb}
\usepackage{dsfont}
\usepackage{upgreek}
\usepackage{slashed}
\usepackage{algorithm}
\usepackage{algpseudocode}
\usepackage{stmaryrd}
\usepackage{graphicx}

\usepackage[unicode=true,
 bookmarks=true,bookmarksnumbered=false,bookmarksopen=false,
 breaklinks=true,pdfborder={0 0 1},backref=false,colorlinks=false]
 {hyperref}
\hypersetup{pdftitle={{Measuring finite Quantum Geometries via Quasi-Coherent States}},
 pdfauthor={Lukas Schneiderbauer, Harold C. Steinacker}}

\makeatletter

\AtBeginDocument{\providecommand\secref[1]{\ref{sec:#1}}}
\AtBeginDocument{\providecommand\eqref[1]{\ref{eq:#1}}}
\AtBeginDocument{\providecommand\figref[1]{\ref{fig:#1}}}
\AtBeginDocument{\providecommand\chapref[1]{\ref{chap:#1}}}
\let\SF@@footnote\footnote
\def\footnote{\ifx\protect\@typeset@protect
    \expandafter\SF@@footnote
  \else
    \expandafter\SF@gobble@opt
  \fi
}
\expandafter\def\csname SF@gobble@opt \endcsname{\@ifnextchar[
  \SF@gobble@twobracket
  \@gobble
}
\edef\SF@gobble@opt{\noexpand\protect
  \expandafter\noexpand\csname SF@gobble@opt \endcsname}
\def\SF@gobble@twobracket[#1]#2{}
\floatstyle{ruled}
\newfloat{algorithm}{tbp}{loa}
\providecommand{\algorithmname}{Algorithm}
\floatname{algorithm}{\protect\algorithmname}
\RS@ifundefined{subref}
  {\def\RSsubtxt{section~}\newref{sub}{name = \RSsubtxt}}
  {}
\RS@ifundefined{thmref}
  {\def\RSthmtxt{theorem~}\newref{thm}{name = \RSthmtxt}}
  {}
\RS@ifundefined{lemref}
  {\def\RSlemtxt{lemma~}\newref{lem}{name = \RSlemtxt}}
  {}

\numberwithin{equation}{section}
\numberwithin{figure}{section}
\numberwithin{table}{section}
  \theoremstyle{definition}
  \ifx\thechapter\undefined
    \newtheorem{defn}{\protect\definitionname}
  \else
    \newtheorem{defn}{\protect\definitionname}[chapter]
  \fi
  \theoremstyle{plain}
  \newtheorem*{thm*}{\protect\theoremname}

\@ifundefined{showcaptionsetup}{}{%
 \PassOptionsToPackage{caption=false}{subfig}}
\usepackage{subfig}
\makeatother

  \providecommand{\definitionname}{Definition}
  \providecommand{\theoremname}{Theorem}
  
\def\R{{\realv}} \def\C{{\compv}} \def\N{{\naturaln}}
   \newcommand{\eq}[1]{(\ref{#1})}

\def\cH{{\cal H}}  

\global\long\def\com#1#2{[#1,#2]}
\global\long\def\hilbert{\mathscr{H}}
\global\long\def\bra#1{\left\langle #1\right|}
\global\long\def\ket#1{\left|#1\right\rangle }
\global\long\def\cont{\mathscr{C}}
\global\long\def\manifold{\mathscr{M}}
\global\long\def\realv{\mathds{R}}
\global\long\def\compv{\mathds{C}}
\global\long\def\integer{\mathds{Z}}
\global\long\def\naturaln{\mathds{N}}
\global\long\def\matunity{\mathds{1}}
\global\long\def\repiso{\doteq}
\global\long\def\iso{\cong}
\global\long\def\climit{\sim}
\global\long\def\acts{\triangleright}
\global\long\def\disp{\updelta}

\global\long\def\laplacianxx{\boxempty_{x}}
\global\long\def\diracxx{\slashed{D}_{x}}
\global\long\def\diracxxx#1{\slashed{D}_{#1}}

\begin{document}

\renewcommand{\title}[1]{\vspace{10mm}\noindent{\Large{\bf #1}}\vspace{8mm}} 
\newcommand{\authors}[1]{\noindent{\large #1}\vspace{5mm}} 
\newcommand{\address}[1]{{\itshape #1\vspace{2mm}}}

\begin{titlepage}
\begin{flushright}
 UWThPh-2015-36 
\end{flushright}
\begin{center}
\title{ {\Large Measuring finite Quantum Geometries via \\[1ex]  Quasi-Coherent States} }

\vskip 3mm

\authors{Lukas Schneiderbauer{\footnote{lukas.schneiderbauer@univie.ac.at}}, Harold C. Steinacker{\footnote{harold.steinacker@univie.ac.at}}
}
 
\vskip 3mm

 \address{ 

{\it Faculty of Physics, University of Vienna\\
Boltzmanngasse 5, A-1090 Vienna, Austria  }  
  }

\bigskip

\vskip 1.4cm

\textbf{Abstract}
\vskip 3mm

\begin{minipage}{14.6cm}%

We develop a systematic approach to determine and measure numerically the geometry
of generic quantum or ``fuzzy'' geometries realized by a set of 
finite-dimensional hermitian matrices.
The method is designed to recover the semi-classical limit of quantized symplectic spaces 
embedded in $\R^d$ including the well-known examples of fuzzy spaces, but it 
applies much more generally.
The central tool is provided by quasi-coherent states, which are defined as ground states of
Laplace- or Dirac operators corresponding to localized point branes in target space.
The displacement energy of these quasi-coherent states is used to extract the local 
dimension and tangent space of the semi-classical geometry, and provides a
measure for the quality and self-consistency of the semi-classical approximation.
The method is discussed and tested with various examples, and implemented in an open-source 
Mathematica package. 

\end{minipage}

\end{center}

\end{titlepage}

\tableofcontents{}

\section{Introduction}

It is generally expected that space-time should have some kind of quantum structure at very short distances. 
The nature of this quantum structure is not known, and there are many possibilities.
One interesting possibility is given by the ``matrix'' or ``quantum'' geometry realized by non-trivial solutions 
of Yang-Mills matrix models, such as the IKKT and BFSS model \cite{Ishibashi:1996xs,Banks:1996vh}. 
These configurations are defined  by a set of matrices  $X^a, \ a=1,...,d$, which 
are interpreted as quantized coordinates in a target space $\R^d$, equipped with the Euclidean 
metric $\delta_{ab}$ (or $\eta_{ab}$ in the Minkowski case).
Prominent examples are the fuzzy sphere $S^2_N$ \cite{hoppe1982QuaTheMasRelSurTwoBouStaPro,Madore:1991bw}, 
fuzzy tori $T^2_N$, and a wide range of similar spaces
which go under the name of ``fuzzy spaces''.
However this setting is not restricted to  group-theoretical spaces, and
the general concept is that of quantized 
symplectic spaces embedded in target space $\realv^{d}$  \cite{Steinacker:2010rh}.
If the underlying manifold is compact, then the $X^a$ are realized by finite-dimensional matrices
acting on an underlying Hilbert space $\cH \cong \C^N$.
These spaces  inherit an effective metric structure from their embedding in target 
space\footnote{From a string-theoretic point of view, the pull-back metric corresponds to the closed string metric, 
while the effective metric is in general different and corresponds (in a limit) to the open string metric 
 \cite{Steinacker:2008ri,Steinacker:2010rh}.}, and 
they are sufficiently general to realize rather generic 4-dimensional geometries, at least locally.

The main appeal of this matrix model approach  is that 
it provides a natural notion of a ``path'' 
integral over the space of geometries, given by the integral over the 
space of hermitian matrices,
with measure defined by the matrix model action. A discussion of some of these 
aspects can be found in \cite{Steinacker:2010rh}.

The aim of the present paper is to use
(suitably generalized) coherent states  as a  tool to understand the quantum geometry 
defined by such matrix configurations $X^a, \ a=1,...,d$. 
We propose a method which allows to distinguish ``non-geometric'' from ``geometric'' 
configurations, and to measure this geometry in some approximate sense. 
The idea is to look for ``optimally localized states'' with small dispersion;
then the $X^a$ can ``almost'' be simultaneously measured, and their expectation values 
provide the location of some variety $\manifold$ embedded in target space $\realv^d$. 
This allows to assign an approximate notion of geometry even to finite-dimensional
matrices, which is meaningful at low energy scales, in a distinctly Wilsonian sense.

Coherent states are well-known for many examples of fuzzy geometries. 
For  quantized coadjoint orbits of  compact Lie groups, the basic
concept goes back to Perelomov \cite{Perelomov:1986tf}, and was reinvented 
in the specific context of fuzzy spaces \cite{Grosse:1993uq}, cf.~\cite{Balachandran:2005ew}.
For generic matrix geometries however, the issue of coherent states is less clear.
In \cite{Ishiki:2015saa}, a concept of coherent states was proposed
for geometries realized as a series of matrix algebras, which in the limit recovers the classical space. 
However, we would like to deal with some fixed, given background geometry, {\em without} appealing to some limit.
From a physical point of view, this is very natural: the concepts of classical geometry need to emerge 
only in the low-energy limit, for distances much larger than the scale of noncommutativity
(i.e.\ the Planck scale, presumably). This is all we should expect in physics.

In the present paper, we propose a  notion of  
quasi-coherent states, which is applicable to generic, finite backgrounds. 
Adapting and generalizing ideas in  Berenstein \cite{Berenstein:2012ts}
and similar to \cite{Ishiki:2015saa}, we define quasi-coherent states 
as {\em ground states} of a matrix Laplacian or a matrix Dirac operator in the presence of a 
point-like test brane in target space.
Their eigenvalues are interpreted as displacement energy $E(\vec x)$,
or as energy of strings stretching between the test brane and the background brane.
This string-inspired concept turns out to be very powerful, and 
independent of more traditional 
notions in noncommutative geometry such as spectral geometry or differential calculi;
it can be seen as a special case of intersecting noncommutative branes \cite{Chatzistavrakidis:2011gs}.
However in contrast to previous work \cite{Berenstein:2012ts,Ishiki:2015saa}, we 
consider the generic case where this energy is non-vanishing and not constant on the brane.
This requires to select a subset of ``quasi-minimal''  states among all 
quasi-coherent states, which is 
achieved by considering the Hessian~$H$ of this energy function~$E(\vec x)$. 
We argue that for matrix backgrounds which define some approximate semi-classical ``brane'' geometry, 
the eigenvalues of $H$ must exhibit a clear {\em hierarchy} between small eigenvalues $\ll 1$ corresponding to 
tangential directions, and eigenvalues ${\cal O}(1)$ which correspond to directions 
transversal to the brane  in target space.
This hierarchy allows to scan the geometry in  a self-consistent way, and to measure the 
quality of the semi-classical approximation.
The resulting quasi-coherent states can then be used to 
measure the location of the brane in target space, and its geometrical properties similar as in 
\cite{Ishiki:2015saa,Steinacker:2010rh}.

We discuss two different realizations of this idea, one based on the Laplace operator $\laplacianxx$,
and one based on a Dirac operator $\diracxx$. 
Both arise naturally in the context of Yang-Mills matrix models.
The approach based on the Laplace operator is conceptually simpler, but appears less 
appealing at first because the corresponding ground state energy is strictly positive.
Nevertheless, it provides  useful information about the local 
dispersion and geometric uncertainty. The approach using the Dirac operator 
has the remarkable property that the location of the branes is often recovered 
from {\em exact} zero modes of $\diracxx$. This  property
was first pointed out by Berenstein for surfaces in $\realv^3$ \cite{Berenstein:2012ts}, 
but appears to hold much more generally.
We discuss this phenomenon in section~\ref{sec:exact-zero-modes}, and 
provide a heuristic argument for the existence of exact zero modes of 
$\diracxx$ in a generic setting. On the other hand, this does not provide information 
about the geometric uncertainty, and we consider both approaches as complementary and equally useful.

The general ideas in this paper are tested and elaborated in detail 
for the standard examples such as 
the fuzzy sphere $S^2_N$, fuzzy torus $T^2_N$, fuzzy $\C P^2_N$, and
squashed fuzzy $\C P^2_N$. The latter is a very interesting and non-trivial example which 
arises in ${\cal N}=4$ SYM with a cubic potential \cite{Steinacker:2014eua,Steinacker:2015mia}, 
and which does {\em not} correspond to a K\"ahler (nor an almost-K\"ahler) manifold.
In particular, we find that the (numerically obtained) quasi-coherent states 
have smaller dispersion than the Perelomov-type states, and we find  strong evidence 
that the semi-classical geometry is again recovered from exact zero modes of $\diracxx$.
We also find that $\diracxx$ and $\laplacianxx$ lead to slightly different 
but consistent results for the location of the brane. This can be viewed as 
a breaking of supersymmetry.

Finally, we provide an implementation  of  an algorithm in Mathematica, 
which very nicely reproduces the 
expected semi-classical geometry of  standard examples 
such as the fuzzy sphere, fuzzy tori, 
and even degenerate spaces such as squashed fuzzy $\compv P^2$.
This algorithm is basically a camera for fuzzy or quantum geometries.
It allows to numerically test any given matrix configuration for a possible 
approximate geometry,  to assess the quality of the geometric approximation, 
and to adjust its ``focus'' by various parameters. 
Some pictures taken by this algorithm are reproduced in the paper.
We hope that this provides a valuable tool and starting point to explore 
other unknown quantum geometries, such as those produced by 
Monte-Carlo simulations of the IIB matrix model \cite{Kim:2012mw,Ito:2015mxa,Kim:2011cr}.

\section{Non-commutative (matrix) geometry\label{sec:non-com-geom}}

Although the methods developed in this paper are more general, we 
will focus on quantum geometries realized as quantized symplectic manifolds
embedded in Euclidean  target space. This framework applies to
a large class of noncommutative spaces known as fuzzy spaces, 
which also arise in the matrix-model approach to string theory 
\cite{Ishibashi:1996xs,Banks:1996vh,Dijkgraaf:1997vv}. 
Their non-commutative  structure can be viewed as quantization 
of an underlying symplectic manifold, using the same mathematical structures as in quantum mechanics.
The metric structure of the geometry is induced by the metric in 
target space \cite{Steinacker:2010rh}.

\subsection{Quantization and semi-classical limit\label{sec:Geometric-Quantization}}

The quantization of a manifold $\manifold$ with Poisson or symplectic structure amounts to replacing 
the commutative algebra of
functions $\mathscr{C}(\manifold)$ on a manifold  with a non-commutative one, and a 
\emph{quantization map} $\mathcal{Q}$.  
For convenience we first recall the concept of a Poisson structure:
\begin{defn}
A bilinear map $\{.,.\}: \ \cont(\manifold)\times\cont(\manifold)\to\cont(\manifold)$
which satisfies
\begin{itemize}
\item \emph{antisymmetry} $\{f,g\}=-\{g,f\}$,
\item the \emph{Leibniz rule} $\{f\cdot g,h\}=f\cdot\{g,h\}+\{f,h\}\cdot g$,
\item and the \emph{Jacobi identity} $\{f,\{g,h\}\}+\{g,\{h,f\}\}+\{h,\{f,g\}\}=0$
\end{itemize}
for all $f,g,h \ \in\cont(\manifold)$
is called \emph{Poisson structure} or \emph{Poisson bracket} on $\manifold$.
A manifold $\manifold$ equipped with a Poisson structure 
is called a \emph{Poisson manifold $(\manifold,\{.,.\})$.}
\end{defn}
For fixed $f\in\cont(\manifold)$, the map $\{f,.\}$ is
a derivation on $\cont(\manifold)$, which defines
a (Hamiltonian) vector field on $\manifold$. 
We can therefore write the Poisson bracket  as bi-vector field in local coordinates,
\begin{equation}
\{f,g\}=\Theta^{\mu\nu}(\partial_{\mu}f)(\partial_{\nu}g),
\label{Poisson-tensor}
\end{equation}
with the usual index conventions. Then $\Theta^{\mu\nu}$ is
antisymmetric and obeys the relation 
\begin{equation}
\Theta^{\mu \sigma}\partial_{\sigma}\Theta^{\nu \eta}+\Theta^{\nu \sigma}\partial_{\sigma}\Theta^{\eta\mu}
+\Theta^{\eta\sigma}\partial_{\sigma}\Theta^{\mu \nu}=0
\end{equation}
 due to the Jacobi identity. If  $\Theta^{\mu\nu}$ is non-degenerate, its inverse defines a 
 symplectic structure, i.e.\ a closed 2-form $\omega\in\Omega^2(\manifold)$.

The quantization of such a Poisson or symplectic manifold 
should be defined in terms of a noncommutative 
operator algebra $End(\hilbert)$ with $\hilbert$ being a Hilbert space, and some 
way of associating observables to classical function such that the 
Poisson bracket on $\manifold$ is recovered as ``semi-classical limit'' of the 
commutators  in $End(\hilbert)$.
One way to define a quantization map $\mathcal{Q}$ could be 
as follows:\footnote{There are various variations 
of this theme, cf.~\cite{waldmann2007Poi}.}
\begin{defn}
\label{def:quant-map}Let $\manifold$ be a Poisson manifold. A linear map
\begin{equation}
\mathcal{Q}:\cont(\manifold)\to End(\hilbert)\label{eq:quant-map}
\end{equation}
 satisfying the axioms
\begin{enumerate}
\item $Q(\mathbf{1})=\mathds{1}$,
\item $Q(f)^{\dagger}=Q(f^{*})\quad\forall f\in\cont(\manifold)$,\label{enu:axioms_prop_3}
\item \emph{correspondence principle}: 
\begin{equation}
\lim_{\theta\to\text{0}}\frac{1}{\theta}\big(\com{\mathcal{Q}(f)}{\mathcal{Q}(g)}-i\mathcal{Q}(\{f,g\}\big)=0\quad\forall f,g\in\cont(\manifold),
\end{equation}
\begin{equation}
\lim_{\theta\to0}\big(\mathcal{Q}(fg)-\mathcal{Q}(f)\mathcal{Q}(g)\big)=0\quad\forall f,g\in\cont(\manifold),
\end{equation}
\label{enu:correspondence-principle}
\item \emph{irreducibility}: If $\{f_{i},g\}=0\;\forall i\in I$ implies
$g\propto\mathbf{1}$ then $\com{\mathcal{Q}(f_{i})}A=0\ \forall i\in I$
implies $A\propto\mathbf{1}$
\end{enumerate}
is called a \emph{quantization map}.
Here $\theta$ is a scale parameter of $\Theta^{\mu\nu} = \theta \bar\Theta^{\mu\nu}$.
\end{defn}
However, this is somewhat unsatisfactory, since 
there is in general no unique way to define a quantization map. 
Moreover, this formulation of the correspondence principle 
requires a family of quantizations for each~$\theta$. 
In physics, the quantization parameter $\theta$ (or $\hbar$) should be a fixed number, and 
one is actually faced with the opposite problem of 
extracting the {\em semi-classical} Poisson limit from a given quantum system.
If we had a quantization map $\mathcal{Q}$,
the semi-classical limit would be obtained by
$\mathcal{Q}^{-1}$ in the limit $\theta\to0$; this will be denoted by
the symbol $\climit$. However if $\theta$ is fixed,
the answer to this problem can only be approximate, 
and apply in some {\em scale regime} where higher-order contributions of $\hbar$ can be neglected.
The required scale in quantum mechanics is set by the Hamiltonian, 
such as $H= P^2 + X^2$.

Here we want to address the analogous problem:
Given some ``quantum'' or matrix geometry in terms of a set of observables $X^a \in End(\hilbert)$
(with finite-dimensional $\hilbert$), how can we 
extract an underlying semi-classical Poisson-manifold? 
Again the answer can only be approximate, and 
the extra structure which sets the scale is provided 
by the metric $\delta_{ab}$ on the target space $\realv^d$. This will  allow
to extract an effective semi-classical (Riemann-Poisson) geometry from suitable background
matrices~$X^a$.

\subsection{Embedded non-commutative (fuzzy) branes\label{sec:Embedded-Non-commutative-Brane}}

We are interested in the quantum geometry 
defined in terms of a set of finite-dimensional matrices $X^a, \ a=1,\ldots,d$. 
For example, consider a given symplectic manifold embedded in target space,
\begin{align}
 x^a: \ \ \manifold \hookrightarrow \realv^d, \qquad a=1,\ldots,d
\end{align}
and some quantization $\mathcal{Q}$ thereof along the lines
of definition~\ref{def:quant-map}. Then define  $d$  matrices or operators by
\begin{equation}
X^{a}:=\mathcal{Q}(x^{a}) \ \in \ End(\hilbert) \ .
\end{equation}
If $\manifold$ is compact, these will be finite-dimensional matrices. 
Our aim is to develop a systematic procedure to reverse this, and to recover 
approximately the underlying Poisson manifold and its embedding from the $d$ matrices~$X^{a}$.
Clearly, $\mathcal{Q}$ cannot be injective, but this is physically reasonable and corresponds
to an UV-cutoff (as well as an IR cutoff).
A formal expansions in $\Theta^{\mu\nu}$ obviously does not make sense, and 
the algebra $End(\hilbert)$ itself contains very little information about the underlying manifold.
However, it should still be possible to extract the semi-classical limit in some {\em low-energy regime},
corresponding to functions with sufficiently long wave-length.
We will see that the Euclidean metric $\delta_{ab}$ on target space $\realv^d$ allows to  
define a matrix Laplacian and 
a Dirac operator, which encode the crucial metric information\footnote{This is somewhat analogous to the 
Dirac operator in Connes' axiomatic approach to noncommutative geometry, and it may also be interpreted in terms 
of some differential calculus. However we choose a minimalistic approach here, trying to avoid any 
structural prejudice.}.
This will allow to obtain a  hierarchy of scales, and to extract the approximate semi-classical 
geometry in some regime.

Of course, not any set of  matrices $\{X^{a}|\,a=1,\ldots,d\}$ will admit such a geometric interpretation,
and we should be able to distinguish  ''geometric`` configurations from non-geometric ones.
We will offer at least a partial answer to this problem, by providing some measures for the 
quality of a semi-classical approximation.

\section{Examples}

Let us collect various examples of the above type of embedded quantum geometries, which will serve as testing grounds for 
our procedure to extract the semi-classical geometry.

\subsection{The fuzzy sphere $S^2_N$}
\label{sec:fuzzy-sphere}

One of the simplest examples is the so called \emph{fuzzy
sphere}~\cite{hoppe1982QuaTheMasRelSurTwoBouStaPro,Madore:1991bw}.
Let us begin with the usual two-sphere 
\begin{equation}
\manifold=S^{2}=\{x\in\realv^{3}|\sum_{i=1}^{3}x_{i}^{2}=1\}.
\end{equation}
It has a $SO(3)$ symmetry 
\begin{eqnarray}
SO(3)\times S^{2} & \to & S^{2}\\
(g,\vec{x}) & \mapsto & g\cdot\vec{x} ,   \nonumber 
\end{eqnarray}
which induces an action on its algebra of functions $\cont(S^{2})$
\begin{eqnarray}
SO(3)\times\cont(S^{2}) & \to & \cont(S^{2})\\
(g,f(\vec x)) & \mapsto &  (g\triangleright f)(\vec{x}):=f(g^{-1}\cdot\vec{x}) .\nonumber 
\end{eqnarray}
We can then decompose the algebra $\cont(S^{2})$ into irreducible
representations of $SO(3)$, leading to the well known spherical harmonics $Y_{m}^{l}$
\begin{equation}
\cont(S^{2})\repiso\bigoplus_{l=0}^{\infty}\langle\{Y_{m}^{l}|m=-l,\ldots,l\}\rangle .
\end{equation}

\paragraph{Quantization map.}

The quantization is defined such that $\mathcal{Q}: \cont(S^{2}) \to Mat_{n}(\compv)$ respects this symmetry, 
\begin{equation}
\mathcal{Q}(g\triangleright f)=g\triangleright\mathcal{Q}(f)
\qquad\forall g\in SO(3), \  f\in\cont(S^{2})
\end{equation}
where $Mat_{n}(\compv)$ is
equipped with the  $SU(2)$ action 
\begin{eqnarray}
SU(2)\times Mat_{n}(\compv) & \to & Mat_{n}(\compv)\\
(g,M) & \mapsto & \pi({g})\cdot M\cdot\pi({g})^{-1} .
\nonumber 
\end{eqnarray}
Here $\pi$ denotes the  
$n$-dimensional irreducible representation (irrep) of $SU(2)$. Then $Mat_{n}(\compv)$ is in general a
reducible representation. To decompose it, we recall that 
$Mat_{n}(\compv) \repiso End(\hilbert_{(n)}) \repiso\hilbert_{(n)}\otimes\hilbert_{(n)}^{*}$ as a vector
space and also as representation\footnote{We use the  symbol $\repiso$ to emphasize that the isomorphism
between spaces is also compatible with the action. For a usual isomorphism
the symbol $\iso$ is used.} with $\hilbert$ carrying the representation~$\pi$. This gives
\begin{equation}
Mat_{n}(\compv) \repiso\compv^{n}\otimes\compv^{n*}\repiso (1)\oplus(3)\oplus\ldots\oplus(2n-1)
\end{equation}
 with $(d)$ denoting the $d$-dimensional irrep of $SU(2)$.
We define the fuzzy spherical harmonics $\hat{Y}_{m}^{l}$ to be the
basis compatible with this decomposition, so that $(2l+1)=\langle\{\hat{Y}_{m}^{l}|m=-l,\ldots,l\}\rangle$.
Then we can define a quantization map $\mathcal{Q}$
which  preserves the $SO(3)$ symmetry:
\begin{eqnarray}
\mathcal{Q}:\cont(S^{2}) & \to & Mat_{n}(\compv)\\
Y_{m}^{l} & \mapsto & \begin{cases}
\hat{Y}_{m}^{l} & l<n-1 \\
0 & l\geq n-1
\end{cases}\nonumber 
\end{eqnarray}
It is clear that this
map is surjective, and there is  a natural built-in momentum cutoff at $l=n-1$.

\paragraph{Embedding functions.}

As mentioned before, we are especially
interested in the quantized embedding functions $X^{a}\climit x^{a}$.
To identify them we note that the embedding functions
$x^i: \ S^{2}\hookrightarrow\realv^{3}$ can be identified with the spin 1 spherical 
harmonics, $Y_{\pm1}^{1}=x^{1}\pm ix^{2}$ and
$Y_{0}^{1}=x^{3}$. Hence their quantization are given by  $\hat{Y}_{\pm1}^{1}=X^{1}\pm iX^{2}$
and $\hat{Y}_{0}^{1}=X^{3}$, or equivalently
\begin{align}
 X^a := \mathcal{Q}(x^a) = C\,\pi_{(n)}(T^a)   \qquad \in End(\hilbert_{(n)})
\end{align}
for some constant $C$,
where $T^a$ are the generators of $\mathfrak{su}(2)$.
 It follows that they are the generators of the $n$-dimensional
 irrep of $SU(2)$, and thus satisfy 
\begin{equation}
\com{X^{a}}{X^{b}}=i\,C\,\varepsilon_{abc}X^{c}\label{eq:fuzzysphere_emb}
\end{equation}
 here $\varepsilon_{abc}$ is the Levi-Civita
symbol. We fix the radius in the semi-classical limit to be $1$, 
\begin{equation}
\sum_{a=1}^{3}(X^{a})^{2}=\matunity,
\end{equation}
which implies $C=2/\sqrt{n^{2}-1}$. 
Comparing ~\eqref{fuzzysphere_emb} with the correspondence principle in definition~\ref{def:quant-map}
one can read off the Poisson structure 
\begin{equation}
\{x^{a},x^{b}\}_{S^{2}}=\frac{2}{n}\,\varepsilon_{abc}\,x^{c}
\end{equation}
which is of order $\theta \sim 1/n$. This corresponds to the unique\footnote{In general, the 
symplectic volume is quantized and determines the dimension of $\hilbert_{(n)}$. 
However this will not be needed here.} 
$SU(2)$-invariant
symplectic structure on $S^{2}$ with $\int \omega = 2\pi n$.

By considering inductive sequences of fuzzy spheres $(S^2_N)_{N\in\N}$ with appropriate 
embeddings, it can be shown that the quantization map axioms
defined in (\ref{def:quant-map}) are  satisfied. 
However we are interested here in a given, fixed space $S^2_N$. 
Then the relation with the classical case is only justified for low angular momenta, 
consistent with a Wilsonian point of view.
One should then only ask for estimates for the deviation from the classical case.

%
%
%
%

\subsection{Fuzzy $\protect\compv P^{2}_N$}

The sphere $S^{2}$ can  be seen as co-adjoint orbit of the Lie group $SU(2)$. 
This leads to a large class of generalizations given by quantized coadjoint orbits
of semi-simple Lie groups $G$, which can be realized in terms of a matrix algebra $End(\hilbert)$ as 
fuzzy branes embedded in $\realv^d = Lie(G)$.
Here we discuss in some detail the complex projective space $\compv P^{2}$
\cite{Alexanian:2001qj,Grosse:2004wm}, which is a 
coadjoint orbit of $SU(3)$; for  $\C P^n_N $ see e.g.~\cite{Balachandran:2001dd}.

\paragraph{Co-adjoint orbits.}

Let $G$ be a Lie group with Lie algebra $\mathfrak{g} = Lie(G)$. Then $G$ has a natural action
on $\mathfrak{g}^{*}$ called the co-adjoint action given by $g\acts\mu=\mu(g\cdot.\cdot g^{-1})$
for a $g\in G$ and $\mu\in\mathfrak{g}^{*}$.
The co-adjoint orbit $\mathcal{O}_{\mu}^{*}$ of the Lie group $G$
through $\mu\in\mathfrak{g}^{*}$ is then defined as 
\begin{equation}
\mathcal{O}_{\mu}^{*}:=\{\mu(g\cdot.\cdot g^{-1})\,|\,g\in G\}.
\end{equation}
By definition,  $\mathcal{O}_{\mu}^{*}$ is invariant
under the co-adjoint action. Every
orbit of $G$ goes through an element of the dual space of the Cartan
algebra $\mathfrak{g}_{0}^{*}$.
Co-adjoint orbits  carry a natural symplectic form (hence  a Poisson structure): 
The tangent space ~$T_{\mu}\mathcal{O}_{\mu}^*$
can be identified with $\mathfrak{g}/\mathfrak{K}_{\mu}$ where $\mathfrak{K}_{\mu}$
is the Lie algebra of the stabilizer group $K_{\mu}$ of $\mu$. Then the
 $G$-invariant symplectic form is
\begin{equation}
\omega_{\mu}(\hat X,\hat Y):=\mu(\com XY)
\end{equation}
where $\hat X$ is the vector field on $\mathfrak{g}^{*}$ generated by the action of $G$. 
This is an antisymmetric, non-degenerate and closed 2-form on $\mathcal{O}_{\mu}^{*}$.

Let us now consider the  case $G=SU(3)$. 
There are two different types of orbits, depending on the rank of the stabilizer of $\mu$.
For the 4-dimensional orbit,
we can choose $\mu=t_{8}^{*}$ where 
\[
t_{8}=\frac{1}{2\sqrt{3}}\begin{pmatrix}1 & 0 & 0\\
0 & 1 & 0\\
0 & 0 & -2
\end{pmatrix} .
\]
The stabilizer group amounts to $K_{t_{8}^{*}}=SU(2)\times U(1)$, so that the orbit $\mathcal{O}_{t_{8}^{*}}^{*}$ 
can be characterized as
\begin{equation}
\mathcal{O}_{t_{8}^{*}}^{*}\iso SU(3)/(SU(2)\times U(1)) .
\label{eq:orbitiso}
\end{equation}
To see the connection to $\protect\compv P^{2}$, 
consider $S^{5} \subset \compv^{3}$. This carries a natural  action
of $SU(3)$ by matrix multiplication, which is transitive on $S^5$.
Then  the point $(1,0,0)\in\compv^{3}$ is stabilized by $SU(2)$, so that
\begin{equation}
S^{5}\iso {SU(3)} /{SU(2)}.
\end{equation}
On the other hand the complex projective space $\compv P^{2}$ can be defined as 
\begin{equation}
S^{5}/U(1)\iso\compv P^{2}
\end{equation}
which is isomorphic to $\mathcal{O}_{t_{8}^{*}}^{*}$
due to \eq{eq:orbitiso}, and points in $\mathcal{O}_{t_{8}^{*}}^{*}$ 
can be related to the rank one projectors.

\paragraph{Embedding map.}

Let $T^a$ be an orthonormal basis of $\mathfrak{su}(3)$, which satisfy
\begin{align}
 [T^a,T^b] = i c^{abc} T^c
\end{align}
where $c^{abc}$ are the structure constants. 
We can consider them as (Cartesian) coordinate functions 
on $\realv^{8}\iso\mathfrak{su}(3)^* \ni T= x^{a}t_{a}^*$, which 
describe the embedding 
\begin{align}
 x^a: \ \mathcal{O} \cong \C P^2 \hookrightarrow \mathfrak{su}(3)^* \iso\realv^{8}
\end{align}
Since $SU(3)$ is a matrix group, the 
coadjoint orbit can be characterized through a characteristic equation,
\begin{equation}
T^{*}\in\mathcal{O}_{t_{8}^{*}}^{*} \ \iff \  T\cdot T+\frac{1}{2\sqrt{3}}T-\frac{1}{6}=0
\label{eq:characteristic_cp2}
\end{equation}
which follows easily from  the explicit form of $t_{8}^{*}$.
Expanding  $T=x^{a}t_{a}^*$, this gives 
\begin{eqnarray}
x^{a}x^{b}d^{abc} & = & -\frac{1}{\sqrt{3}}x^{c}\label{eq:charasteristic_su3_coords}\\
x^{a}x^{b}\delta^{ab} & = & 1\nonumber 
\end{eqnarray}
where $d^{abc}$ is the totally symmetric invariant tensor of $\mathfrak{su}(3)$\footnote{See appendix~\ref{sec:appendix_su3_def} for relevant conventions.}.
These equations define the embedding of $\compv P^{2}$ as $4$-dimensional
submanifold in $\realv^{8}$.
One can also characterize the decomposition of the algebra of functions $\cont(\compv P^{2})$
into irreps of $SU(3)$:
\begin{equation}
\cont(\compv P^{2})\repiso\bigoplus_{p=0}^{\infty}\hilbert_{(p,p)} .
\label{modes-CP2}
\end{equation}
Here $\hilbert_{(p,p)}$ denotes the irrep of
$SU(3)$ with highest weight $(p,p)$.

\paragraph{Quantization map.}

\begin{figure}
\begin{centering}
\subfloat[$(4,0)$ irrep of $SU(3)$. This triangle form is typical for $(n,0)$
representations. Its size increases with~$n$.\label{fig:rep_3_0}]{\begin{centering}
\includegraphics[width=0.45\textwidth]{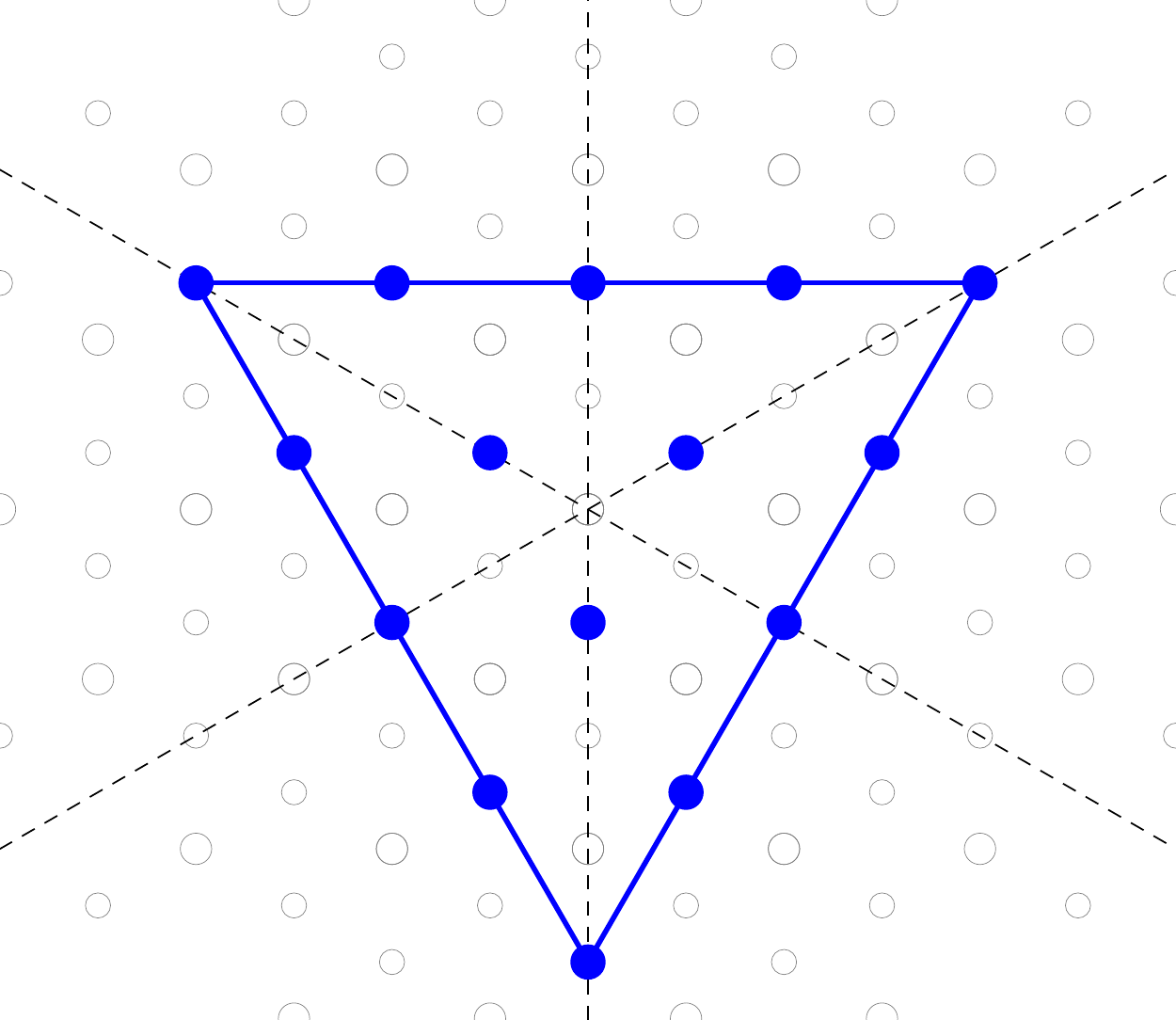}
\par\end{centering}

}\hfill{}\subfloat[$(2,2)$ irrep of $SU(3)$. All $(n,n)$ representations are of this
hexagon form. Its size increases with~$n$.
\label{fig:rep_2_2}]{\begin{centering}
\includegraphics[width=0.45\textwidth]{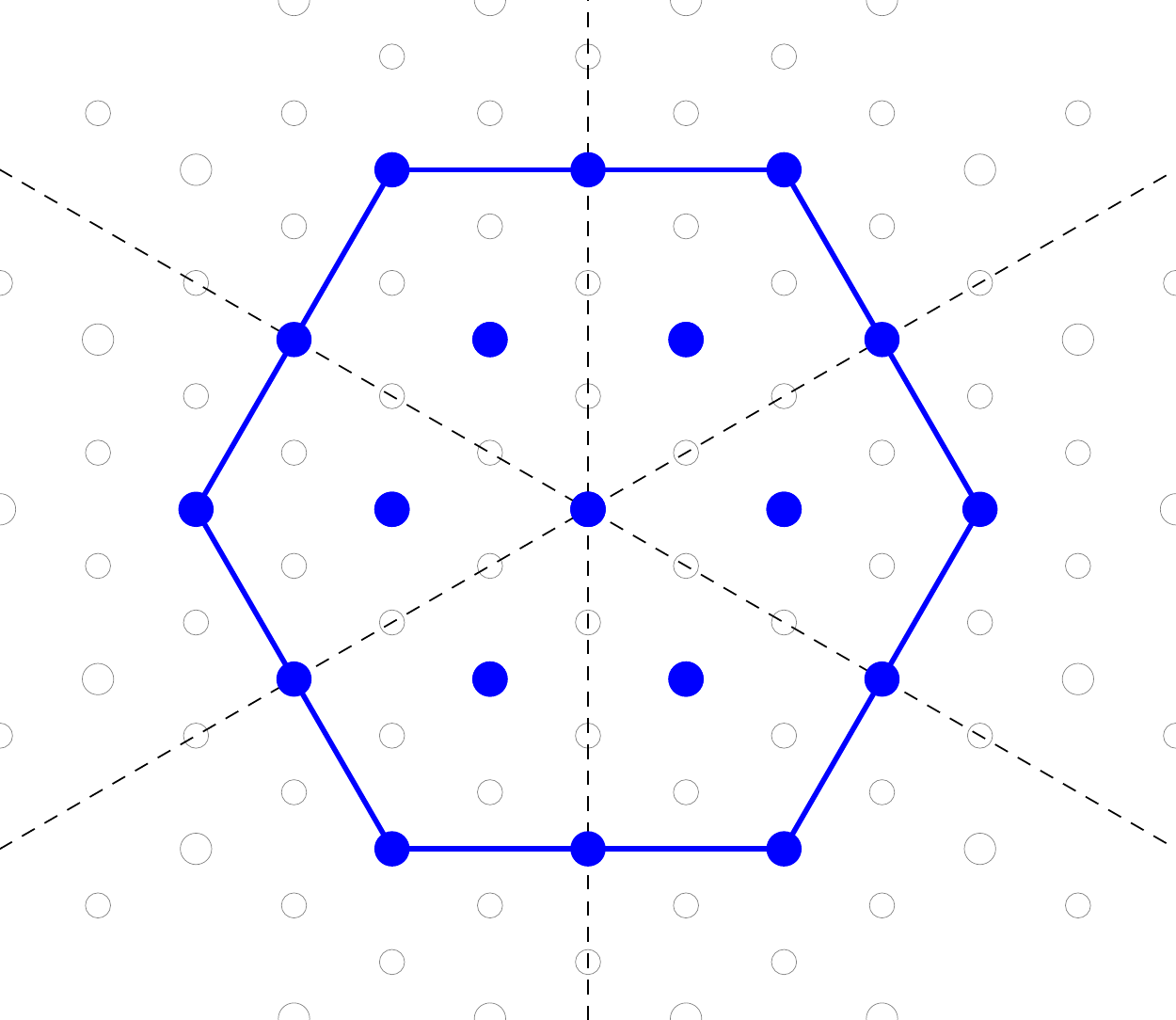}
\par\end{centering}

}
\par\end{centering}

\caption{Weight space diagrams of $(n,0)$ and $(n,n)$ -type representations
of $SU(3)$}
\end{figure}

As for all coadjoint orbits, the quantized algebra of functions is given by a finite matrix algebra 
$ Mat_{m}(\compv) \iso End(\hilbert)$, for an appropriate choice of $\hilbert$. The suitable $\hilbert$ are 
irreps with highest weight $\Lambda$ proportional to the $\mu$ which determines the 
coadjoint orbit. In the present case of $\C P^2$, this means that $\hilbert = \hilbert_{(0,n)}$.
Then the space of quantized functions
\begin{equation}
End(\hilbert_{(0,n)})\repiso\hilbert_{(0,n)}\otimes\hilbert_{(0,n)}^{*}\repiso\bigoplus_{p=0}^{n}\hilbert_{(p,p)}\label{eq:tensorproduct}
\end{equation}
is isomorphic\footnote{This works in general; for a proof see e.g.~\cite{Pawelczyk:2002kd} specialized to $q=1$.}
to \eq{modes-CP2} up to the cutoff $n$.
This justifies the above choice of $\hilbert$, and it provides us with a 
quantization map $\mathcal{Q}$ which is compatible with the $SU(3)$ symmetry,
\begin{eqnarray}
\mathcal{Q}: \ \cont(\compv P^{2}) & \to & Mat_{m}(\compv)\\
Y_{(p,p)}^{\Lambda} & \mapsto & \begin{cases}
\hat{Y}_{(p,p)}^{\Lambda} & p\leq n\\
0 & p>n \ .
\end{cases} 
\nonumber 
\end{eqnarray}
Here $Y_{(p,p)}^{\Lambda}$ respectively $\hat{Y}_{(p,p)}^{\Lambda}$
is an appropriate basis of the particular $\hilbert_{(p,p)}$, see~\figref{rep_2_2}.

\paragraph{Quantized embedding function.}

Since the coordinate functions $x^a$ transform in the adjoint, the 
same must hold for the quantized functions $X^{a}$.
This means that 
\begin{align}
 X^{a} =\mathcal{Q}(x^{a}) = C\,\pi_{(0,n)}(T^a)  \qquad \in End(\hilbert_{(0,n)})
\end{align}
are nothing but the 
$\mathfrak{su}(3)$ generators on $\hilbert_{(0,n)}$, which satisfy
\begin{equation}
\com{X^{a}}{X^{b}}=i\,C\,c_{abc}X^{c} .
\label{eq:com_rel}
\end{equation}
Again the  constant $C$  will determine the radius of  
$\compv P^{2}$ as embedded in $\R^8$.
Using the quadratic Casimir operator of $SU(3)$ one sees that setting
 $C=1/\sqrt{n(1+n/3)}$ yields a $\compv P^{2}$ with
radius $1$ in the semi-classical limit.
One can also derive a quantized version of \eq{eq:charasteristic_su3_coords} \cite{Alexanian:2001qj}.
This defines fuzzy $\compv P_{n}^{2}$ as an embedded fuzzy brane.
It is clear that the  commutation relations \eq{eq:com_rel}
should be viewed as quantization of the Poisson bracket
$\{x^{a},x^{b}\}=\,C\,c_{abc}x^{c}$ on $\C P^2$, which is precisely the 
canonical invariant symplectic form discussed above. 
This concludes  for now our  brief discussion of fuzzy $\compv P^{2}$.

The bottom line is that the 8 matrices $X^a$ contain enough information to reconstruct
the manifold $\C P^2 \subset \R^8$  in the semi-classical limit, provided
they are interpreted as quantized embedding functions $x^{a}$.

\subsection{Squashed $\protect\compv P^{2}_N$}
\label{sec:saqushed-cp2}

There is a simple  modification of $\compv P_{n}^{2}$, called \emph{squashed
}$\compv P_{n}^{2}$, which arises as a solution of  $\mathcal{N}=4$ supersymmetric Yang-Mills
theory with cubic flux term \cite{Steinacker:2014lma}.
Recall that fuzzy  $\compv P_{n}^{2}$ is defined by eight matrices $X^{a}= \pi_{(0,n)}(T^a)$
interpreted as quantized embedding functions $X^{a}\climit x^{a}$, where $T^a$ are the generators of 
$\mathfrak{su}(3)$. Then squashed $\compv P_{n}^{2}$ is defined 
by simply dropping two of these matrices corresponding to the Cartan generators, leaving the six
matrices $\{X^{a},\,a\in\mathcal{I}=\{1,2,4,5,6,7\}\}$ in the standard Gell-Mann basis. 
Geometrically, this amounts to a projection in target space
\begin{eqnarray}
\Pi: \ \realv^{8} & \to & \realv^{6}\\
(x^{a})_{a=1,\dots,8} & \mapsto & (x^{a}){}_{a\in\mathcal{I}}\nonumber 
\end{eqnarray}
resulting in a squashed 4-dimensional variety $\Pi(\compv P^{2})\hookrightarrow\realv^{6}$
embedded in $\realv^{6}$.
We will justify below that this interpretation is correct also in the 
fuzzy case, and we will recover the intricate self-intersecting geometry of
squashed $\compv P^{2}$ in the semiclassical limit both analytically and numerically.

It is worth noting that the Cartan generators $X^{3}$ and $X^{8}$ can be generated
by the $X^{a},\,a\in\mathcal{I}$. Therefore the
algebra of squashed $\compv P^{2}_N$ is the same
matrix algebra as in the non-squashed case.

\subsection{Fuzzy torus $T^2_N$}

An example for an embedded brane
which does not belong to the class of co-adjoint orbits
is the torus $T^{2}=S^{1}\times S^{1}$. It can be embedded in $\realv^{4}$
via the relations
\begin{gather}
x^{1}+ix^{2}=e^{i\varphi}\label{eq:def_torus}\\
x^{3}+ix^{4}=e^{i\psi}\nonumber 
\end{gather}
where $\varphi,\psi\in[0,2\pi)$, or more explicitly
\begin{align}
x:T^{2} & \hookrightarrow\realv^{4}\label{eq:torus_embedding}\\
(\varphi,\psi) & \mapsto\begin{pmatrix}(e^{i\varphi}+e^{-i\varphi})/2\\
-i\,(e^{i\varphi}-e^{-i\varphi})/2\\
(e^{i\psi}+e^{-i\psi})/2\\
-i\,(e^{i\psi}-e^{-i\psi})/2
\end{pmatrix}=\begin{pmatrix}\cos(\varphi)\\
\sin(\varphi)\\
\cos(\psi)\\
\sin(\psi)
\end{pmatrix}.\nonumber 
\end{align}
It is apparent by~\eqref{def_torus} that $T^{2}\subset S^{3}$.
Applying a generalized stereographic projection $S^{3}\to\bar{\realv}^{3}$ \eq{torus-projection}
yields a torus embedded in $\realv^{3}$ which resembles the usual
doughnut form.

It is again useful to identify the symmetry properties of $T^{2}$. The torus carries a $U(1)\times U(1)$ symmetry
which acts simply by a multiplication of a phase factor 
\begin{eqnarray}
(U(1)\times U(1))\times T^{2} & \to & T^{2}\\
((\Phi,\Psi),(\varphi,\psi)) & \mapsto & (\varphi+\Phi,\psi+\Psi)\nonumber 
\end{eqnarray}
which again induces an action on the algebra of functions on the torus
$\cont(T^{2})$. A basis of $\cont(T^{2})$ which respects the symmetry
is given by the functions $\phi_{l,k}$: 
\begin{equation}
\phi_{l,k}(\varphi,\psi):=e^{il\varphi}e^{ik\psi}.
\end{equation}
A function $f\in\cont(T^{2})$ expanded in this basis, i.e.\ $f(\varphi,\psi)=\sum_{l,k=-\infty}^{\infty}f_{l,k}\,\phi_{l,k}(\varphi,\psi)$,
is nothing but the Fourier series of $f$.

\paragraph{Construction of matrix algebra.}

To construct the appropriate quantization in terms of a matrix algebra $Mat_{n}(\compv)$,
we introduce the \emph{shift matrix} $U$ and the \emph{clock matrix}
$V$ 
\begin{equation}
U=\begin{pmatrix}0 & 1\\
 & 0 & 1\\
 &  & \ddots & \ddots\\
 &  &  & 0 & 1\\
1 &  &  &  & 0
\end{pmatrix},\qquad V=\begin{pmatrix}1\\
 & q\\
 &  & q^{2}\\
 &  &  & \ddots\\
 &  &  &  & q^{n-1}
\end{pmatrix}
\end{equation}
 with $q:=e^{2\pi i/n}$. These are unitary matrices with the property
$U^{n}=V^{n}=\matunity$ obeying the commutation relations 
\begin{equation}
\com UV=(q-1)\,V U.\label{eq:torus_com}
\end{equation}
It is clear that the set of matrices $\{\matunity,U,U^{2},\dots,U^{n}\}$
respectively $\{\matunity,V,V^{2},\dots,V^{n}\}$ form a representation
of the cyclic group $\integer_{n}$.
Now  define  matrices $\Phi_{l,k}$ as
\begin{equation}
\Phi_{l,k}:=U^{l} V^{k}
\end{equation}
 for $l,k\in\{-\frac{n-1}{2},\dots0,\dots,\frac{n-1}{2}\}$\footnote{We assume here and in the 
 following text that $n$ is odd. The required modifications for even $n$ are obvious.}. 
 These are $n^{2}$ linear independent matrices which form a
basis of $Mat_{n}(\compv)$. Let us consider now the adjoint action
of $\integer_{n}\times\integer_{n}\subset U(1)\times U(1)$ on $Mat_{n}(\compv)$
given by
\begin{align}
(\integer_{n}\times\integer_{n})\times Mat_{n}(\compv) & \to Mat_{n}(\compv)\\
((w^{r},w^{s}),M) & \mapsto V^{s}\cdot(U^{r}\cdot M\cdot U^{-r})\cdot V^{-s}.\nonumber 
\end{align}
 We can calculate the action on the basis $\Phi_{l,k}$ using~\eqref{torus_com}
and get 
\begin{equation}
(w^{r},w^{s})\acts\Phi_{l,k}=q^{sl+kr}\,\Phi_{l,k}.
\end{equation}
Therefore the one-dimensional subspaces spanned by the basis vectors
are invariant under the $\integer_{n}\times\integer_{n}$ action.

\paragraph{Quantization map.}

Now it is easy to construct a quantization map $\mathcal{Q}$ which
respects the $\integer_{n}\times\integer_{n}\subset U(1)\times U(1)$
symmetry on our algebras.
\begin{align}
\mathcal{Q}: \ \cont(T^{2}) & \to Mat_{n}(\compv)\\
\phi_{l,k} & \mapsto\begin{cases}
q^{-lk/2}\,\Phi_{l,k} & |l|,|k|\leq(n-1)/2\\
0 & \mathrm{otherwise}
\end{cases}.\nonumber 
\end{align}
 The factor $q^{-lk/2}$ is chosen such that
 $\mathcal{Q}(f^{*})=\mathcal{Q}(f)^{\dagger}$ holds.
The appropriate Poisson structure
on $\cont(T^{2})$ can be guessed from \eqref{torus_com}: Since
$\com{\mathcal{Q}(e^{i\varphi})}{\mathcal{Q}(e^{i\psi})}=\com UV=(q-1)\,U V=(q-1)\,\mathcal{Q}(e^{i\psi})\mathcal{Q}(e^{i\varphi})$
and $q=e^{2\pi i/n}=1+\frac{2\pi i}{n}+\mathcal{O}((\frac{1}{n})^{2})$,
the correspondence principle requires that 
\begin{equation}
2\pi i\,\mathcal{Q}(e^{i\psi})\mathcal{Q}(e^{i\varphi})=i\lim_{n\to\infty}n\,\mathcal{Q}(\{e^{i\varphi},e^{i\psi}\}),
\end{equation}
hence $\{e^{i\varphi},e^{i\psi}\}=\frac{2\pi}{n}e^{i\varphi}e^{i\psi}$,
and $\mathcal{Q}(f g)\to\mathcal{Q}(f)\mathcal{Q}(g)$ holds
as $n\to\infty$.
This allows us to conclude  
\begin{equation}
\{\varphi,\psi\}_{T^{2}}=-\frac{2\pi}{n}
\end{equation}
 which is obviously  $U(1)\times U(1)$ invariant.
It is then easy to see that the correspondence principle holds 
for all quantized functions in the limit $n\to\infty$.

\paragraph{Quantized embedding functions.}

We can directly read off the quantized
embedding functions $X^{a}=\mathcal{Q}(x^{a})$ from~\eqref{torus_embedding}:
\begin{align}
X^{1} & =(U+U^{\dagger})/2\\
X^{2} & =-i\,(U-U^{\dagger})/2\nonumber \\
X^{3} & =(V+V^{\dagger})/2\nonumber \\
X^{4} & =-i\,(V-V^{\dagger})/2 . \nonumber  
\end{align}

\subsection{Other types of matrix geometries}

The above examples are quantized symplectic spaces with a 
more-or-less regular immersion in target space.
However, there are many interesting  examples with degenerate embedding.
For example, it turns out that 
squashed $\C P^2$  has a triple self-intersection at the origin
in target space, 
which  leads to interesting physics \cite{Steinacker:2014eua,Steinacker:2015mia}. 
A more drastic example is the fuzzy four-sphere $S^4_N$ 
\cite{Castelino:1997rv}, which 
can be interpreted as a twisted $N$-fold degenerate embedding of 
fuzzy $\C P^3_N$ in $\realv^5$ \cite{Medina:2002pc,Steinacker:2015dra}.
The considerations in this paper are general enough to capture also such examples, possibly
with minor modifications\footnote{E.g.\ for $S^4_N$, there are $N$ degenerate 
coherent states at each point on $S^4$, which is interpreted in terms of an $N$-fold cover of $S^4$.
We will briefly address the issue of such degeneracies in sections 6 and 7.}.

\section{Perelomov coherent states\label{sec:Coherent-States}}

So far, we have discussed  examples of quantized spaces,
viewed as quantizations of symplectic manifolds embedded in $\realv^d$. 
Now we want to address the opposite problem of ``de-quantization'': Given some 
set of hermitian matrices $\{X^i, \ i=1,2,...,d\}$ as above (called {\em matrix background} henceforth), 
we want to extract their semi-classical geometry.
The key tool towards this goal is provided by  \emph{coherent states.}

Loosely speaking, coherent states are optimal localized states which are 
closest to the corresponding classical states, i.e.\ points in classical space. 
``Optimal localized''
typically means having minimum uncertainty.
Extensive treatments of coherent states from various points of view 
can be found e.g.\ in~\cite{Perelomov:1986tf,weyl1928GruundQua,
radcliffe1971Somprocohspista,Barut:1970qf,
PAM_doi:10.1007/BF00745155,Gazeau:2009zz}.
We will first describe the group theoretical approach of Perelomov~\cite{Perelomov:1986tf},
which applies to the basic examples of fuzzy geometries based on 
group theory. In this approach, coherent states are defined in terms of some algebraic condition, 
corresponding to highest weight states; cf.~\cite{Grosse:1993uq} for early work in the present context.

However for generic matrix geometries, 
we will need to relax this algebraic approach and consider a suitable generalization
of coherent states.
As opposed to previous work in this context such as \cite{Ishiki:2015saa}, 
we do not want to rely on some sort of limit $N\to\infty$; instead we assume some fixed 
configuration of $N\times N$~matrices.
This is essential to extract the geometry of 
some given matrix configurations, as obtained e.g.\ in 
non-perturbative numerical simulations \cite{Kim:2012mw} in the matrix-model approach 
to the theory of space-time and matter.

\subsection{Localization and Dispersion}

Assume we are given a set of $d$ matrices $\{X^{a},\ a=1,\dots,d\}$, which we want to interpret
as quantized embedding functions~$X^{a}\climit x^{a}$ of a classical
manifold embedded in $\realv^{d}$. The $X^a$ generate a matrix algebra
$\mathscr{A}\subset Mat_{N}(\compv)$ acting on a Hilbert space $\hilbert_{N}$.
We can associate to any normalized vector $\ket{\Psi}\in\hilbert_{N}$ 
the projector $\rho_\Psi :=\ket{\Psi}\bra{\Psi}\in\mathscr{A}$, and calculate the expectation value of $X^{a}$ 
in this state 
\begin{equation}
\vec{\bf x}(\Psi)_{a}:=\langle X^{a}\rangle_{\Psi}=\mathrm{tr}(X^{a}\rho_\Psi)=\bra{\Psi}X^{a}\ket{\Psi},
\end{equation} 
as in  quantum mechanics. Similarly we can calculate
the square of its standard deviation~$\left(\Delta_{\Psi}X^{a}\right)^{2}$
as 
\begin{equation}
\left(\Delta_{\Psi}X^{a}\right)^{2}:=\bra{\Psi}(X^{a})^{2}\ket{\Psi}-\bra{\Psi}X^{a}\ket{\Psi}^{2}.
\end{equation}
A good measure for the localization of $\ket{\Psi}$ is provided by the \emph{dispersion}~$\disp(\Psi)$
\begin{equation}
\disp(\Psi):=\sum_{a=1}^{d}\left(\Delta_{\Psi}X^{a}\right)^{2}\geq0\label{eq:dispersion}
\end{equation} 
which will be our guideline for the definition of 
coherent states. It is natural to require coherent states to have minimal dispersion.  
While this works perfectly well for fuzzy 
spaces with sufficient symmetry\footnote{It works in general for all  fuzzy branes obtained as co-adjoint orbits of
Lie groups.}, it will be useful in the following to slightly relax this condition.

\subsection{Coherent states on the fuzzy sphere $S_{N}^{2}$\label{sec:coh_states_fuzzy_sphere}}

Recall the fuzzy sphere $S_{N}^{2}$ defined in section~\ref{sec:fuzzy-sphere}.
In this case, we can simplify the dispersion~(\ref{eq:dispersion})
to
\begin{equation}
0 \leq\disp(\Psi)=\sum_{a=1}^{3}(\Delta_\Psi X^{a})^{2}
=\bra{\Psi}\sum_{i=1}^{3}(X^{a})^{2}\ket{\Psi}-\vec{\bf x}(\Psi)^{2}
=1-|\vec{\bf x}(\Psi)|^{2},
\label{eq:dispersion-fuzzysphere}
\end{equation}
recalling that $\sum_{i=1}^{3}(X^{a})^{2}=\matunity$. Since $\disp(\Psi) \geq 0$, this relation also
shows that $|\vec{\bf x}(\Psi)|\leq1$. 
We see explicitly that  the dispersion is minimized
for states whose expectation value $\vec{\bf x}(\Psi)$ is closest to the unit sphere $S^{2}$.

Since $(X^{a})_{a=1,2,3}$ is a vector operator, $\vec{\bf x}$ also transforms
as a vector, i.e.\ for $g\in SU(2)$ and $\pi_{N}(g)$ its $N$-dimensional
representation we have 
\begin{equation}
{\bf x}^{a}(\pi_{N}(g)\cdot\Psi)=\pi_{3}(g)_{\;b}^{a}\,{\bf x}^{b}(\Psi) \ .
\label{eq:vector_op}
\end{equation}
Clearly, $\disp(\Psi)$ is invariant under the $SU(2)$
action. Now let $\Psi_0$ be a state with minimal dispersion $\disp$. Acting with $SU(2)$, 
we obtain a class of states
\begin{equation}
\mathcal{O}_{\Psi_{0}}:=\{\pi_N(g)\cdot\Psi_{0}\,|\,g\in SU(2)\} 
\end{equation}
 which are all optimally localized, and the expectation values $\vec {\bf x}(\Psi_0)$
 of this class form a sphere $S^2$. Thus after some rotation, we can assume 
 that $\vec {\bf x}(\Psi_0) \sim (0,0,1)$ is at the ``north pole''. Then the dispersion is minimal if and only if 
 $\Psi_0$ is an eigenstate of $X^3$ with maximal absolute eigenvalue,
 due to \eq{eq:dispersion-fuzzysphere}. 
 Then the stabilizer group $K_{\Psi_{0}}\subset SU(2)$ of $\rho_{\Psi_0}$ is $U(1)$, and the space of such 
 optimally localized states $\rho_{\Psi_0}$ is given by the coadjoint orbit ${SU(2)}/{U(1)} \iso S^2$. Therefore there is precisely one 
 such coherent state  for each point on $S^2$. 
 Moreover, they are extremal weight states which satisfy an annihilation equation, which for the
 state at the north pole reads
 \begin{align}
  X^+ \Psi_0 = 0 ,
 \end{align}
 where $X^+$ is the standard $SU(2)$ raising operator.
This extremal weight property characterizes the coherent states as defined by Perelomov 
\cite{Perelomov:1986tf}.
In a suitable scaling limit, one recovers the standard coherent states 
on two-dimensional phase space known from quantum mechanics.

Let us calculate the dispersion and the expectation
value for this orbit. Consider the highest weight
vector 
\[
\Psi_{0}=\ket{\frac{N-1}{2},\frac{N-1}{2}}\in\hilbert_{N},
\]
written in standard quantum mechanics notation. One can calculate the expected location
of this state and the dispersion using standard  $SU(2)$ representation theory, which gives 

\begin{equation}
\vec{\bf x}(\Psi_{0})=\begin{pmatrix}0\\
0\\
\sqrt{\frac{N-1}{N+1}}
\end{pmatrix}=\begin{pmatrix}0\\
0\\
1
\end{pmatrix}+\mathcal{O}(\frac{1}{N}),\label{eq:psi0_loc}
\end{equation}
\begin{equation}
\disp(\Psi_{0})=1-\frac{N-1}{N+1}=\frac{2}{N+1}=\mathcal{O}(\frac{1}{N}).
\end{equation}
Together with rotation invariance, 
we see that the expectation values $\vec {\bf x}(\Psi_0)$ of the coherent states form a sphere with 
radius 
\begin{align}
 r_N = \sqrt{\frac{N-1}{N+1}}=1+\mathcal{O}(\frac{1}{N}).
\end{align}
Furthermore, their dispersion~$\disp(\mathcal{O}_{\Psi_{0}})$
goes to zero as $N \to\infty$. Therefore  the coherent states
are localized at the unit sphere $S^{2}$ in the limit $N\to\infty$. 
Interpreting them as quantized functions, they  become
Dirac-$\delta$-functions localized at the unit sphere. Thus, clearly, the geometry of $S^{2}$ 
is recovered from fuzzy
$S_{N}^{2}$ in the limit $N\to\infty$. However, we have seen that the 
coherent states allow to extract the geometry of a sphere from $S^2_N$
{\em even for finite $N$}, up to the precision set by $\disp(\Psi_{0})$.
This is all we should expect on physical grounds, and this is what we would like  to extract
from generic matrix geometries.

\subsection{Coherent states on fuzzy $\protect\compv P_{N}^{2}$\label{sec:coherent_states_fuzzycp2}}

Fuzzy $\compv P_{N}^{2}$ was defined by 8 matrices $X^{a}$ which
obey the commutation relation 
\begin{equation}
\com{X^{a}}{X^{b}}=i\,\frac{1}{\sqrt{N(1+N/3)}}\,c_{abc}X^{c},\label{eq}
\end{equation}
 $c_{abc}$ being the antisymmetric structure constants of $SU(3)$.
They generate the matrix algebra  $End(\hilbert_{(0,N)})$.
Coherent states on $\compv P_{N}^{2}$ are constructed
in complete analogy with the fuzzy sphere $S_{N}^{2}$, and the method applies for
all quantized spaces which arise from co-adjoint orbits~\cite{Perelomov:1986tf}.

The recipe is as follows: Take the highest weight vector $\Psi_{\text{0}}\in\hilbert$
of the given representation. Consider the orbit $\mathcal{O}_{\Psi_{\text{0}}}$
generated by the group action. One easily recognizes that the orbit
$\mathcal{O}_{\Psi_{0}}$ is isomorphic to the co-adjoint orbit of
$G$ (this is how $\hilbert$ is chosen). Furthermore, $\Psi_{\text{0}}$ minimizes the dispersion~$\disp$
defined in~\eqref{dispersion}; for the present case $G=SU(3)$ this follows by the 
same argument as above. Since
$\disp$ is group invariant, the whole orbit $\mathcal{O}_{\Psi_{0}}$
minimizes the dispersion~$\disp$.

\paragraph{Explicit calculation for $\protect\compv P_{N}^{2}$.}

Let us evaluate the expectation value and the dispersion of the highest
weight vector $\Psi_{\text{0}}\in\hilbert_{(0,N)}$ explicitly for
$\compv P_{N}^{2}$. Clearly \eq{eq:dispersion-fuzzysphere} generalizes replacing the $SU(2)$
generators with $SU(3)$ ones:

\begin{equation}
0\leq\disp(\Psi)=\sum_{a=1}^{8}(\Delta_{\Psi}X^{a})^{2}
=1-|\vec{\bf x}(\Psi)|^{2}.
\label{eq:dispersion_cp2}
\end{equation}
For the expectation value we get 
\begin{equation}
\vec{\bf x}(\Psi_{0})=\begin{pmatrix}0\\
0\\
0\\
0\\
0\\
0\\
0\\
\sqrt{\frac{N}{3+N}}
\end{pmatrix}=\begin{pmatrix}0\\
0\\
0\\
0\\
0\\
0\\
0\\
1
\end{pmatrix}+\mathcal{O}(\frac{1}{N}),\label{eq:psi0_loc-cp2}
\end{equation}
 and therefore, 
\begin{align}
{\bf x}^{a}(\Psi_{0})\,{\bf x}^{b}(\Psi_{0})\,\delta^{ab} & =\frac{N}{3+N}=1+\mathcal{O}(\frac{1}{N}),\\
{\bf x}^{a}(\Psi_{0})\,{\bf x}^{b}(\Psi_{0})\,d^{abc} 
& = -\sqrt{\frac{N}{3(N+3)}}\, {\bf x}^{c}(\Psi_{0}) =- \frac 1{\sqrt{3}} {\bf x^{c}}(\Psi_{0})+\mathcal{O}(\frac{1}{N}),\nonumber 
\end{align}
which reproduces the characteristic equations of $\compv P^{2}$ \eqref{charasteristic_su3_coords}
in the limit $N\to\infty$.\footnote{Choosing $\hilbert_{(N,0)}$ instead of $\hilbert_{(0,N)}$
for fuzzy $\compv P^{2}_N$ would lead to the mirror image of the embedding of $\compv P^{2}$.} 
The dispersion~$\disp$ is then easily calculated: 
\begin{equation}
\disp(\Psi_{0})=1-\frac{N}{N+3}=\frac{3}{N+3}=\mathcal{O}(\frac{1}{N})\label{eq:disp_extremalsteight_cp2}
\end{equation}
 which goes to zero as $N\to\infty$  as expected.
 By  $SU(3)$ invariance,
this is valid for the entire orbit $\mathcal{O}_{\Psi_{0}}\iso\compv P^{2}$.

Note that $\hilbert_{(0,N)}$ has three extremal weight vectors (see~\figref{rep_3_0}),
and we could have used any of them since they lie on the
same orbit generated by the $SU(3)$-action.

\subsection{Perelomov states on squashed $\protect\compv P_{N}^{2}$\label{sec:Coherent-States-on_scp2}}

For the squashed co-adjoint orbit the situation is less obvious,
since there is no longer a $SU(3)$
symmetry. Nevertheless, it turns out that explicit calculations  are possible even
for squashed $\compv P^{2}$.

The dispersion $\disp$ still serves
as a measure for the localization and is defined as in the general case
(\ref{eq:dispersion}) 
\begin{align}
\disp(\Psi) &:=\sum_{a\in\mathcal{I}}(\Delta_{\Psi}X^{a})^{2} 
=\bra{\Psi}\underbrace{\sum_{a=1}^{8}(X^{a})^{2}}_{\matunity}\ket{\Psi}
-\sum_{i=3,8}\bra{\Psi}(X^{i})^{2}\ket{\Psi}-\left|\vec{\bf x}(\Psi)\right|^{2} .
\label{eq:disp_squashed_gen}
\end{align}
While this is no longer invariant under $SU(3)$, it is still
invariant under the adjoint action of the $U(1)\times U(1) \subset SU(3)$ subgroup
generated by $X_{8}$ and $X_3$, which are  the Cartan generators that have been dropped.
The norm $\left|\vec{\bf x}(\Psi)\right|$
is also invariant under this $U(1)\times U(1)$ transformations.
Evaluating the expectation value $\vec{\bf x}(\Psi)$ and the dispersion
$\disp(\Psi)$ explicitly for the highest weight state $\ket{\Psi_{0}}$ yields 
\begin{equation}
\vec{\bf x}(\Psi_{0})=0
\end{equation}
and 
\begin{equation}
\disp(\Psi_{0}) =\frac{3}{N+3}
=\mathcal{O}(\frac{1}{N}).\label{eq:disp_highestweight}
\end{equation}
We see that in the squashed case the extremal weight states are now
located at the origin in target space, while the dispersion is the same as in the non-squashed
case (cf.~\eqref{disp_extremalsteight_cp2}). Since  $\disp(\Psi_{0}) \to 0$ as
 $N \to\infty$, the highest weight state~$\ket{\Psi_{0}}$ can still be considered as a 
coherent state.  The same is true for the
other two extremal states due to the remaining $SU(3)$-Weyl symmetry, which is preserved 
on squashed $\compv P_{N}^{2}$.

\paragraph{Rotations of the highest weight state.}

Let us investigate how $SU(3)$ rotations of the highest weight state
affect the location and the dispersion. First of all, since $T_{3}^{\pm}=T^{1}\pm iT^{2}$
annihilate $\ket{\Psi_{0}}$ and $T^{3}$, $T^{8}$  only act
via a phase shift\footnote{This corresponds to the stabilizer $\mathscr{K} =SU(2)\times U(1)$ of 
$\rho_{\Psi_{0}}$.},
we are left with 4 non-trivial directions\footnote{The same  is true for the non-squashed $\compv P^{2}$
which is one way to see that it is a 4-dimensional manifold.} corresponding to the generators $T^{4},\,T^{5},\,T^{6},\,T^{7}$.
Calculating the expectation values $\vec{\bf x}$ of the rotated vectors
\begin{equation}
\ket{\boldsymbol{\varphi}}:=\exp\left(i\varphi_{1}T^{4}+i\varphi_{2}T^{5}+i\varphi_{3}T^{6}+i\varphi_{4}T^{7}\right)\ket{\Psi_{0}}
\end{equation}
 with $\boldsymbol{\varphi}=(\varphi_{1},\varphi_{2},\varphi_{3},\varphi_{4})$
yields (see appendix \chapref{scp2_calc} for details) 
\begin{align}
\vec{\bf x}(\boldsymbol{\varphi}) & =c_{N}\frac{N}{2}\frac{1}{|\boldsymbol{\varphi}|}\begin{pmatrix}\frac{(\varphi_{1}\varphi_{3}+\varphi_{2}\varphi_{4})}{|\boldsymbol{\varphi}|}(\cos|\boldsymbol{\varphi}|-1)\\
2\frac{(\varphi_{1}\varphi_{4}-\varphi_{2}\varphi_{3})}{|\boldsymbol{\varphi}|}\sin^{2}\frac{|\boldsymbol{\varphi}|}{2}\\
\varphi_{2}\,\sin|\boldsymbol{\varphi}|\\
-\varphi_{1}\,\sin|\boldsymbol{\varphi}|\\
\varphi_{4}\,\sin|\boldsymbol{\varphi}|\\
-\varphi_{3}\,\sin|\boldsymbol{\varphi}|
\end{pmatrix},\label{eq:location_scp2}
\end{align}
where $c_{N}\frac{N}{2}=\frac{1}{2}\sqrt{\frac{N}{1+N/3}}$. 
For small $\boldsymbol{\varphi}$, this clearly spans a 4-dimensional manifold 
whose tangent space at the origin is given by the  $4567$ plane\footnote{Remember: 
the numbering was chosen to be consistent with the non-squashed
scheme. Since $x_{3}$ and $x_{8}$ were ``projected away'', the
third component in $\vec{\bf x}$ corresponds to $x_{4}$ and so on.};
globally, it turns out to have a triple self-intersection at the origin.
In particular, the
norm $|\vec{\bf x}(\boldsymbol{\varphi})|$ is 
\begin{align}
|\vec{\bf x}(\boldsymbol{\varphi})|^{2} & =\underbrace{(c_{N}\frac{N}{2})^{2}}_{\frac{3}{4}+\mathcal{O}(\frac{1}{N})}
\left(\frac{(\varphi_{1}\varphi_{3}+\varphi_{2}\varphi_{4})^{2}}{|\boldsymbol{\varphi}|^{2}}(\cos|\boldsymbol{\varphi}|-1)^{2}+4\frac{(\varphi_{1}\varphi_{4}-\varphi_{2}\varphi_{3})^{2}}{|\boldsymbol{\varphi}|^{2}}\sin^{4}\frac{|\boldsymbol{\varphi}|}{2}+\sin^{2}|\boldsymbol{\varphi}|\right).
\label{eq:localization_norm}
\end{align}
To calculate the dispersion $\disp$ we have to take care of the second
term in (\ref{eq:disp_squashed_gen}). After some computations
(see appendix \chapref{scp2_calc} for more details) we find
\begin{align}
\disp(\boldsymbol{\varphi})= & \frac{3}{8\,(3+N)}\frac{1}{|\boldsymbol{\varphi}|^{4}}\left\{ 4(\varphi_{1}^{2}+\varphi_{2}^{2})(\varphi_{3}^{2}+\varphi_{4}^{2})\cos|\boldsymbol{\varphi}|\right.\nonumber \\
+ & \left.\left(\varphi_{1}^{4}+\varphi_{2}^{4}+\varphi_{1}^{2}(2\,\varphi_{2}^{2}+\varphi_{3}^{2}+\varphi_{4}^{2})+\varphi_{2}^{2}(\varphi_{3}^{2}+\varphi_{4}^{2})+(\varphi_{3}^{2}+\varphi_{4}^{2})^{2}\right)\cos2|\boldsymbol{\varphi}|\right.\nonumber \\
+ & \left.\left(7\,(\varphi_{1}^{4}+\varphi_{2}^{4})+7\,(\varphi_{3}^{2}+\varphi_{4}^{2})^{2}+11\,\varphi_{2}^{2}(\varphi_{3}^{2}+\varphi_{4}^{2})\right)\right\} .\label{eq:disp_scp2}
\end{align}
The important point is 
that the dispersion $\disp$ vanishes as $N\to\infty$. This means that even after 
squashing, the rotated highest weight (Perelomov) states
can be considered to be coherent, as they become completely localized
in the semi-classical limit. This justifies the claim that
the semi-classical geometry of squashed~$\compv P_{N}^{2}$ is indeed 
$\Pi(\compv P^{2})$.
Moreover, one can check that the dispersion (\ref{eq:disp_scp2})
satisfies the sharp inequality 
\begin{equation}
\frac{2}{3+N}\leq\disp(\boldsymbol{\varphi})\leq\frac{3}{3+N} .
\end{equation}
Somewhat surprisingly, this means that the highest weight vector~$\ket{\Psi_{0}}$
located at the origin
actually has the highest dispersion in this class of states (and not
the lowest as one might have guessed).

Let us discuss some limits to check the formulas
(\ref{eq:disp_scp2}) and (\ref{eq:location_scp2}).

\paragraph{Limit $|\boldsymbol{\varphi}|\to0$.}

In this limit,
the expectation value $\vec{\bf x}(\boldsymbol{\varphi})$ goes to zero
as expected. Furthermore,  
\begin{equation}
\lim_{|\boldsymbol{\varphi}|\to0}\disp(\boldsymbol{\varphi})=\frac{3}{3+N}
\end{equation}
which correctly reproduces the dispersion of the highest weight state
in formula (\ref{eq:disp_highestweight}).

\paragraph{Limit $\varphi_{1},\varphi_{3}\to0$.}

\begin{figure}
\begin{centering}
\includegraphics[width=0.5\textwidth]{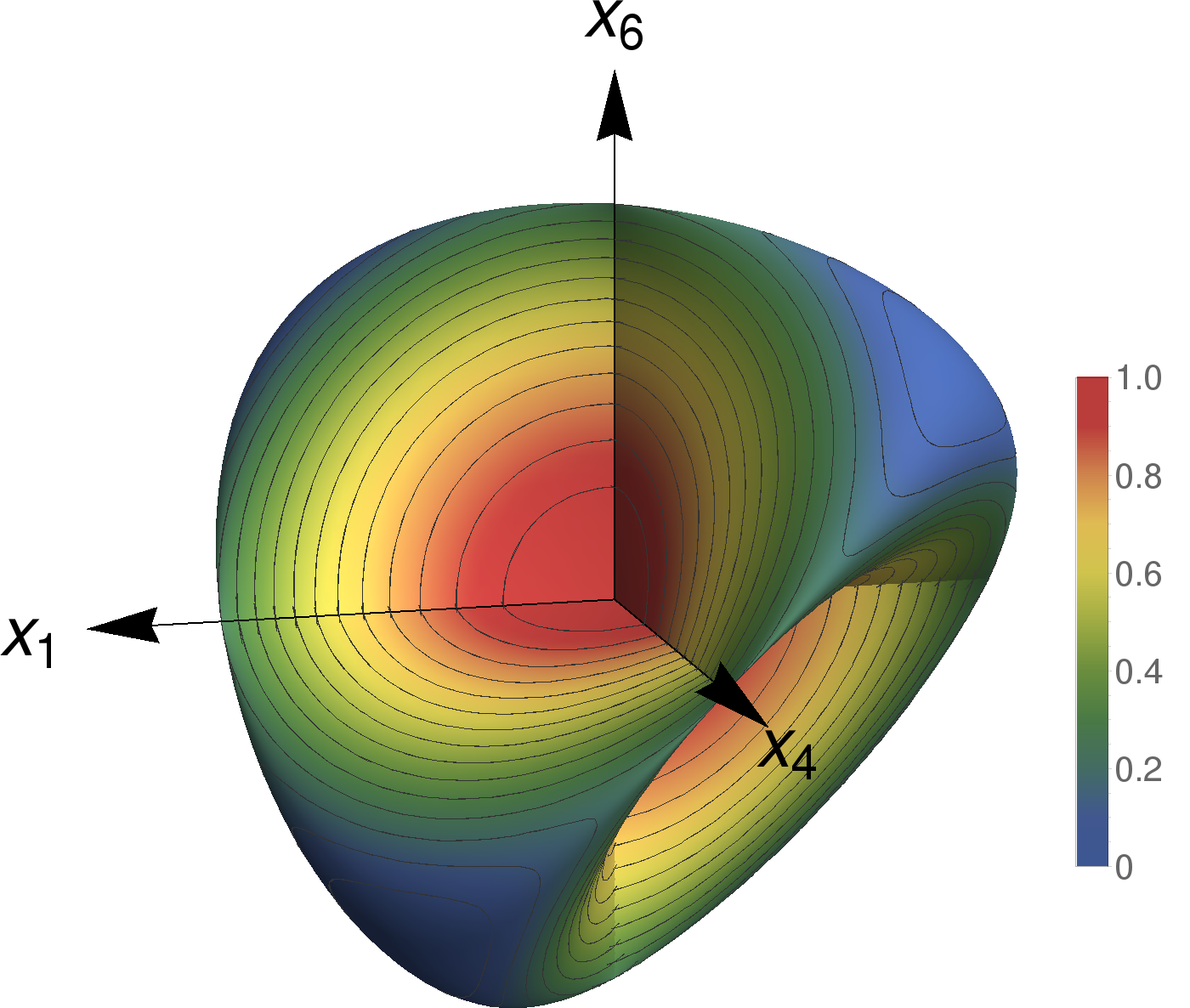}
\par\end{centering}
\caption{A 3-dimensional section of $\Pi(\protect\compv P^{2})$ through $x_{2}=x_{5}=x_{7}=0$
plane, first printed in~\cite{Steinacker:2014lma}.
The color indicates the corresponding scaled and shifted dispersion
$\bar{\disp}$, which has its minimum at $0$ ($\triangleq\frac{2}{3+N}$)
and maximum at $1$ ($\triangleq\frac{3}{3+N}$). The lines represent
contour lines of $\bar{\disp}$.\label{fig:scp2_3section}}
\end{figure}

In the limit $\varphi_{1}\to0$, $\varphi_{3}\to0$ the expectation
values read 
\begin{align}
\lim_{\varphi_{1},\varphi_{3}\to0}\vec{\bf x}(\boldsymbol{\varphi}) & =c_{N}\frac{N}{2}\frac{1}{|\boldsymbol{\varphi}|}
\begin{pmatrix}\frac{\varphi_{2}\varphi_{4}}{|\boldsymbol{\varphi}|}(\cos|\boldsymbol{\varphi}|-1)\\
0\\
\varphi_{2}\,\sin|\boldsymbol{\varphi}|\\
0\\
\varphi_{4}\,\sin|\boldsymbol{\varphi}|\\
0
\end{pmatrix}\label{eq:location_scp2_limit} .
\end{align}
This corresponds to a section of squashed $\compv P^{2}$ through
the $x_{2}=x_{5}=x_{7}=0$ hyperplane. Plotting this 2-dimensional manifold reproduces \figref{scp2_3section}
first published in~\cite{Steinacker:2014lma}.
The dispersion (\ref{eq:disp_scp2}) reduces in the limit to
\begin{align}
\lim_{\varphi_{1},\varphi_{3}\to0}\disp(\boldsymbol{\varphi})= & \frac{3}{8\,(3+N)}\frac{1}{|\boldsymbol{\varphi}|^{4}}\left\{ 4\,\varphi_{2}^{2}\,\varphi_{4}^{2}\,\cos|\boldsymbol{\varphi}|\right.\nonumber \\
+ & \left.\left(\varphi_{2}^{4}+\varphi_{2}^{2}\,\varphi_{4}^{2}+\varphi_{4}^{4}\right)\cos2|\boldsymbol{\varphi}|\right.
+  \left.7\,(\varphi_{2}^{4}+\varphi_{4}^{4})+11\,\varphi_{2}^{2}\,\varphi_{4}^{2}\right\} \label{eq:disp_scp2-limit}
\end{align}
whose global minima lie at
\begin{equation}
|\varphi_{2}|=|\varphi_{4}|=\sqrt{2}\arctan(\sqrt{2}).
\end{equation}
 This corresponds to 4 points on $\Pi(\compv P^{2})$, given by
\[
c_{N}\frac{N}{3}\begin{pmatrix}-1\\
0\\
1\\
0\\
1\\
0
\end{pmatrix},\quad c_{N}\frac{N}{3}\begin{pmatrix}1\\
0\\
1\\
0\\
-1\\
0
\end{pmatrix},\quad c_{N}\frac{N}{3}\begin{pmatrix}-1\\
0\\
-1\\
0\\
-1\\
0
\end{pmatrix},\quad c_{N}\frac{N}{3}\begin{pmatrix}1\\
0\\
-1\\
0\\
1\\
0
\end{pmatrix},
\]
 which can be seen in \figref{scp2_3section} as centers of the blue
zones.

\paragraph{Limit $\varphi_{3},\varphi_{4}\to0$.}

\begin{figure}
\begin{centering}
\includegraphics[width=0.5\textwidth]{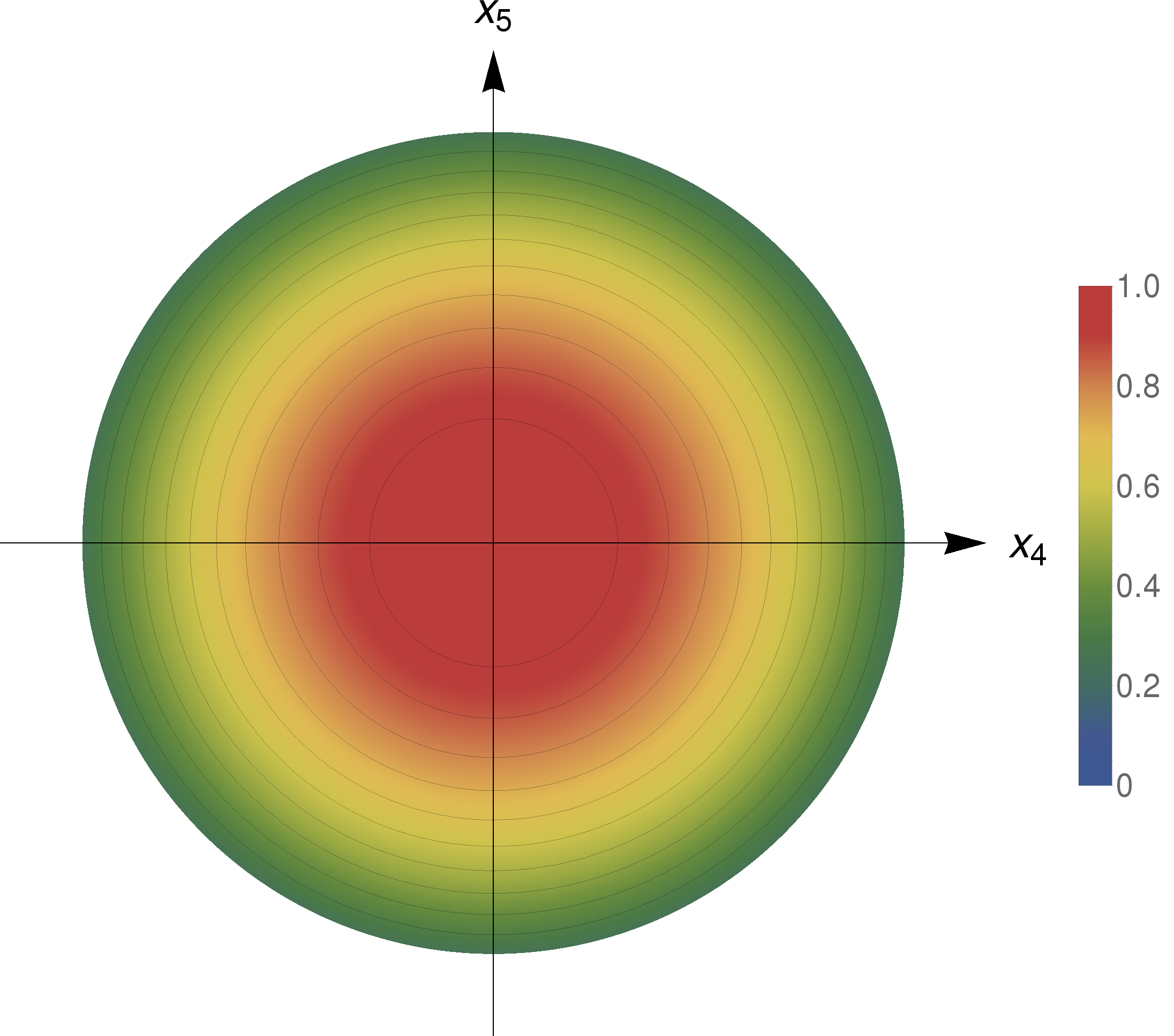}
\par\end{centering}

\caption{Semi-classical limit of the squashed fuzzy sphere. The color indicates
the corresponding scaled dispersion $\bar{\disp}$ which has its
minimum at $0$ ($\triangleq\frac{2}{3+N}$) and maximum at $1$ ($\triangleq\frac{3}{3+N}$).
The lines represent contour lines of $\bar{\disp}$.\label{fig:ss2}}
\end{figure}

Another interesting case is the limit $\varphi_{3},\varphi_{4}\to0$, corresponding to 
rotations by $T^{4}$ and $T^{5}$. Since
$\{T^{4},T^{5},T^{8}\}$ form a $\mathfrak{su}(2)$ subalgebra of $\mathfrak{su}(3)$,
this  essentially reduces to the squashed fuzzy 
sphere\footnote{The squashed fuzzy sphere \cite{Andronache:2015sxa} is defined in analogy to squashed $\compv P^{2}_N$
by omitting the Cartan generator $X^{3}$ from  fuzzy $S^2_N$.}. Here 
\begin{align}
\lim_{\varphi_{3},\varphi_{4}\to0}\vec{\bf x}(\boldsymbol{\varphi}) & =c_{N}\frac{N}{2}\frac{1}{|\boldsymbol{\varphi}|}\begin{pmatrix}0\\
0\\
\varphi_{2}\,\sin|\boldsymbol{\varphi}|\\
\varphi_{1}\,\sin|\boldsymbol{\varphi}|\\
0\\
0
\end{pmatrix},\label{eq:location_scp2_limit_2}
\end{align}
 which implies that the image of $\vec{\bf x}$ in this limit is the disk
$\mathds{D}_{R}=\{x,y\in\realv,\,x^{2}+y^{2}\leq R^{2}=c_{N}\frac{N}{2}\}$
which is of course simply the ordinary squashed sphere.
The dispersion~\eqref{disp_scp2} then reduces to 
\begin{align}
\lim_{\varphi_{3},\varphi_{4}\to0}\disp(\boldsymbol{\varphi})= & \frac{3}{8\,(3+N)}\left(7+\cos(2|\boldsymbol{\varphi}|)\right).
\end{align}
 The minima are  given by $|\boldsymbol{\varphi}|=\sqrt{\varphi_{1}^{2}+\varphi_{2}^{2}}=\pi/2$,
which corresponds to the boundary of the disk $\partial\mathds{D}_{R}$.
In \figref{ss2} a picture of the 2-dimensional squashed fuzzy sphere is given indicating the dispersion.

These calculations  will be complemented below 
by a numerical  algorithm to determine optimally localized states, which 
allow to capture the semi-classical limit without relying on any theoretical expectation. 
This algorithm will not only support the above interpretation for squashed $\compv P_{N}^{2}$, but is
applicable to arbitrary matrix configurations which admit an interpretation in terms of a 
semi-classical geometry.

\section{Generalized coherent states \label{sec:Definition-of-Coherent}}

The above definition of Perelomov coherent states is
applicable only to very special matrix backgrounds related to Lie groups. 
Our aim now is to extend the idea of these states to arbitrary matrix configurations,
in order  to extract some approximate classical geometry  
from fuzzy spaces described by such matrices.

To this end, we will introduce the concepts of {\em optimal localized states}
and {\em quasi-coherent states}, which are 
applicable to generic matrix configurations. However, this alone does not quite suffice
to extract the classical geometry, since it applies to any 
point in the embedding space (or target space). Indeed it is clear that not every matrix background
will have a  geometric interpretation. 
In section~\ref{sec:Numerically-finding-coherent} we will then use these quasi-coherent states
to single out backgrounds which do define a reasonable approximate geometry, and to 
to ``measure'' and characterize this  geometry.
This will be possible provided the background admits a certain {\em hierarchy},
which allows to focus on a suitable subset of quasi-minimal states.

\subsection{Optimal localized states and quasi-coherent states}

We recall  the coherent states on the fuzzy sphere $S^2_N$: their expectation values 
$\vec{\bf x}(\Psi_{N})$ 
are points on a sphere with a certain radius $r_N$, and their dispersion is minimal 
among all states, in particular among all states with the given expectation value. 
However given some generic matrix background, we do not know a priori any preferred location;
this is what we want to extract. Moreover as seen in the example of squashed $\C P^2_N$, 
we should expect that the optimal dispersion of the ''coherent states'' depends on the 
location, as the scale of noncommutativity $|\Theta^{ab}(x)|$ may depend on $x$; hence
looking for states with ``globally'' minimal dispersion  is too restrictive.
However, what we can do is  first fix some point $\vec{x} \in \realv^d$, focus on 
those states whose expectation value  $\vec{\bf x}(\Psi_{N})$ is closest to $\vec{x}$, 
and choose  among those the ones with minimal dispersion.  
In a next step, the semi-classical location of the
matrix background can then be identified  as those $\vec{x}\in \realv^d$ where that dispersion is
``small''; the latter will be made more precise below, using the concept of a ``hierarchy''.

This leads to the following definition of an ``optimal localized
state'':
\begin{defn}
[Optimal Localized State]\label{def_optimal_localized_state-1}Let
$\vec{x}$ be a point in  target space $\realv^{d}$. A state $\ket{\Psi}\in\hilbert_{N}$
is called an \emph{optimal localized state at $\vec{x}\in\realv^{d}$},
if the following properties hold:
\begin{enumerate}
\item The\label{enu:exp_opt} expectation values $\vec{\bf x}(\Psi)$ are optimal,
i.e.\
\begin{equation}
\left|\vec{x}-\vec{\bf x}(\Psi)\right|=\min_{\ket{\psi}\in\hilbert_{N}}\left|\vec{x}-\vec{\bf x}(\psi)\right|.
\label{dist-funct-min}
\end{equation}

\item Let $\mathscr{L}_p\subset\hilbert_{N}$ be the set of states obeying
property (\ref{enu:exp_opt}). We demand that the dispersion is minimal
with respect to all states in $\mathscr{L}_p$, i.e.\
\begin{equation}
\disp(\Psi)=\min_{\ket{\lambda}\in\mathscr{L}_p}\disp(\lambda).
\end{equation}

\end{enumerate}
\end{defn}
Since the state space 
\begin{align}
 \hilbert_{N}/U(1) \cong \C P^{M} 
\end{align}
is compact, the space of optimal localized states is non-empty\footnote{The naive approach would be to 
just demand that $\vec{\bf x}(\Psi)=\vec{x}$. However, this cannot always be satisfied
as the example in \secref{Fuzzy-Sphere-revisited} will show.}
for each $\vec{x}\in\realv^d$.
This definition captures the notion of a state which can be thought
of as ``the best'' approximation of a classical point $\vec{x}\in\realv^{d}$. 
However, this definition is rather cumbersome both numerically and analytically.
The following alternative
definition captures both conditions 1. and 2.,  but turns out to be much more useful:
\begin{defn}
[Quasi-coherent states]\label{def_optimal_localized_state-2}Let
$\vec{x}$ be a point in  target space $\realv^{d}$. A state $\ket{\Psi}\in\hilbert_{N}$
is called a \emph{quasi-coherent state at $\vec{ x}\in\realv^{d}$},
if the following property holds:
\begin{equation}
 E(\vec {x}) := \left|\vec x-\vec{\bf x}(\Psi)\right|^2 +  \disp(\Psi)  \qquad \mbox{is minimal for given} \ \vec x .
 \label{dist-funct-comb}
\end{equation}
We will denote $E(\vec x)$ as {\em displacement energy}.
\end{defn}
This will be reformulated 
in terms of an eigenvalue problem in section~\ref{sec:point-probes}.

\vspace{0.2cm}

To get some insight,
consider again the example of the fuzzy sphere $S^2_N$. Clearly the Perelomov coherent states are both 
optimally localized and also quasi-coherent states
for $\vec x = \vec{\bf x}(\psi)$ with $|\vec x| = r_N$. 
If we choose $|\vec x| > r_N$, the corresponding optimally localized state is still  a Perelomov state
localized at the nearest point $x'\in S^2_{r_N}$. In contrast for $|\vec x| < r_N$, the 
states with minimal $\left|\vec{x}-\vec{\bf x}(\Psi)\right|$ are not expected to be the 
Perelomov  states in general, 
and therefore the optimally localized states will have higher dispersion.
Therefore we can choose among all optimally localized states those which have smallest or 
nearly-smallest dispersion $\disp$. Their expectation values $\vec{\bf x}(\psi)$ 
will then reproduce the effective location of $S^2$. 
This can be determined by scanning the state space $\hilbert_{N}/U(1) \cong \C P^M$, as described below.

For the quasi-coherent states, the story is quite different. 
We will show in section~\ref{sec:Fuzzy-Sphere-revisited} that
the quasi-coherent states for $S^2_N$ are always Perelomov states as long as $\vec x \neq 0$, 
located at the point on $S^2$ which is closest to $\vec x$.
In this case, the function $E(\vec x)$ serves as a measure for the deviation of the quasi-coherent state from $\vec x$, and 
$S^2_{r_N}$ is recovered as the minimum locus of $E(\vec x)$. This can be determined by scanning the target space
$\realv^d$. Clearly this approach is much more efficient, and will be described in detail 
in section~\ref{sec:Numerically-finding-coherent}.

\paragraph{Scanning state space.}

Let us briefly describe a scanning procedure for optimal localized states.
For sufficiently generic matrix backgrounds, we must choose some cutoff $\disp_0$
which selects the near-minimal dispersions $\disp \leq \disp_0$. 
Then the (compact) phase space $\hilbert_{N}/U(1) \cong \C P^M$ can be scanned,
selecting those states with dispersion $\leq \disp_0$. This is facilitated by 
equipping  $\C P^M$ with the Fubini-Study metric. 
The selected states $\psi$ with small dispersion will be approximations to 
the optimally localized states at $\vec{\bf x}(\psi)$, and these 
expectation values $\vec{\bf x}(\psi)$ 
 define the effective approximate geometry.

Clearly the cutoff $\disp_0$  must be chosen by hand, and there is no 
global choice of cutoff $\disp_0$ which works for all cases.
This is analogous to the process of ``focusing'' a camera, when trying to take a picture of some object. 
If the cutoff $\disp_0$ is chosen too small, too few points may be selected; this is easily seen 
for the fuzzy ellipsoid, where only the two extremal points might survive. 
On the other hand if $\disp_0$ is chosen too large,
we will obtain a very blurred picture. 

We implemented this idea numerically, and it turns out
to work for simple spaces where a good ``hierarchy'' exists even for low
dimensional matrices.  However for more complicated spaces such as squashed
$\compv P^{2}$, the state space becomes very large, requiring  an unreasonable
computational effort. 
For this reason, we will present a more efficient procedure in the next section, based on 
\emph{point probes} and quasi-coherent states. 
This is inspired by the ideas of intersecting fuzzy branes and
point probes in \cite{Chatzistavrakidis:2011gs,Berenstein:2012ts}.

\section{Point probes and quasi-coherent states\label{sec:Intersecting-Point-Probe}}
\label{sec:point-probes}

In the present section we develop  an efficient approach to 
optimally localized states, based on the idea of  \emph{point probes.} 
This can be viewed as a special case of intersecting noncommutative branes discussed in \cite{Chatzistavrakidis:2011gs},
and it was first applied in  Berenstein~\cite{Berenstein:2012ts}
in the present context. Similar ideas are also used by Ishiki  in~\cite{Ishiki:2015saa}
in the large $N$ limit.
The idea is to measure the energy of strings connecting the probe with the fuzzy
brane, which is minimal at the location $\vec x$ of the brane.
This energy is defined via certain matrix Laplace of Dirac operators.
Quasi-coherent states can then be obtained as {\em ground states} of these operators at $\vec x$,
recovering precisely our previous definition in section~\ref{sec:Definition-of-Coherent}.

\paragraph{Zero-dimensional (point) brane.}

First let us consider the special case of a fuzzy brane defined by
$d$ real $1\times1$ matrices $x^{a}\in Mat_{1}(\compv)\iso\compv,\,a=1,\dots,d$
interpreted as  embedding functions of a point in $\realv^d$
\begin{equation}
\vec{x}=(x^{1},\dots,x^{d})\in\realv^{d}
\end{equation}
This can also be seen via coherent sates: the elements in this algebra have expectation value
$(x^{1},\dots,x^{d})$ and zero dispersion.
Hence these numbers $x^{a}$ describe a zero-dimensional
point-brane embedded in $\realv^{d}$.

\paragraph{Stack of branes.}

We now return to our original matrix configuration~$X^{a}$ which
characterizes some fuzzy brane embedded in $\realv^{d}$ and generates
a matrix algebra~$\mathscr{A} = Mat_{m}(\compv)$. Adding a second
fuzzy brane defined by a set of matrices~$Y^{a}$ generating the
algebra $\mathscr{B}= Mat_{l}(\compv)$, the two branes in  target space $\realv^{d}$  
are described by the direct sum
\begin{equation}
\mbox{\ensuremath{\mathfrak{X}}}^{a}:=X^{a}\oplus Y^{a},\qquad a=1,\dots,d
\end{equation}
acting on $\hilbert_{m}\oplus\hilbert_{l}$ (see~\cite{Chatzistavrakidis:2011gs}), or explicitly in matrix form 
\begin{equation}
\mathfrak{X}^{a}=\begin{pmatrix}X^{a} & 0\\
0 & Y^{a}
\end{pmatrix},\qquad a=1,\dots,d .
\end{equation}
They generate the algebra $\mathscr{A}\oplus\mathscr{B}\subset Mat_{m+l}(\compv)$.

This becomes more interesting
if we consider not only the algebra~$Mat_{m}(\compv)\oplus Mat_{l}(\compv)$
(i.e.\ matrices in block-diagonal form) but the whole matrix algebra~$Mat_{m+l}(\compv)$,
including the off-diagonal blocks which are elements of $\hilbert_{m}\otimes\hilbert_{l}^{*}$
respectively $\hilbert_{l}\otimes\hilbert_{m}^{*}$. Since these blocks
connect the two branes (which means that the branes ``interact'' in some
way), they can be interpreted as oriented strings connecting
the two branes described by the matrices $X^{a}$ and $Y^{a}$.

\paragraph{Point probe.} 

Now we combine the ideas of the point brane and brane interactions
in the following way: Given a fuzzy brane embedded in some target
space $\realv^{d}$ one can place a point brane as a probe at a definite
location in this space. Then we are able to measure the energies of
the strings connecting the brane and the probe. By varying the position
of the probe the energies of the connecting strings will change. In
particular, if the brane is not ``too fuzzy''\footnote{Not ``too fuzzy'' means that there exists a clear hierarchy of energies
of the strings. We again refer to \secref{Numerically-finding-coherent}
for a more precise treatment.}, there should be a region in space where the energies are relatively
low compared to other regions which then can be regarded as an approximation
of the semi-classical limit.
This background consisting of a brane described by $X^a$ and the point brane at $x^a$ is defined by
\begin{equation}
\mathfrak{X}^{a}=\begin{pmatrix}X^{a} & 0\\
0 & x^{a}
\end{pmatrix},\qquad a=1,\dots,d
\end{equation}
with $d$ real numbers $x^{a}$.

\subsection{Laplace operator}

The (matrix) Laplace operator $\boxempty_{\mathfrak{X}}:Mat_{m+1}(\compv)\to Mat_{m+1}(\compv)$
on the above background $\mathfrak{X}^{a}$ is given by 
\begin{equation}
\boxempty_{\mathfrak{X}}=\sum_{a=1}^{d}\com{\mathfrak{X}^{a}}{\com{\mathfrak{X}^{a}}.}.
\end{equation}
It acts on the Hilbert space $End(\hilbert) = End(\hilbert_{m} \oplus \compv) = End(\hilbert_{m}) \oplus \hilbert_{m} \oplus \hilbert_{m}^T \oplus \compv$.
It is natural to interpret the two off-diagonal blocks $\hilbert_{m} \oplus \hilbert_{m}^T$ as (oriented) strings stretching between the 
brane described by $X^a$ and the point brane at $x^a$.
We are only interested here in the energies of this string sector, represented by 
vectors $\Phi\in Mat_{m+1}(\C)$
of the form 
\begin{equation}
\mbox{\ensuremath{\Phi}}=\begin{pmatrix}\begin{array}{ccc}
0 & \cdots & 0\\
\vdots & \ddots & \vdots\\
0 & \cdots & 0
\end{array} & \ket{\phi}\\
\bra{\phi} & 0
\end{pmatrix} \quad \in  \hilbert_{m} \oplus \hilbert_{m}^T
\label{eq:string_elems}
\end{equation}
with $\ket{\phi}\in\hilbert_{m}$. 
Denoting the restriction of the full matrix Laplacian 
to $\hilbert_{m} \oplus \hilbert_{m}^T $ by $\laplacianxx$, this  yields 
\begin{eqnarray}
\boxempty_{\mathfrak{X}}\Phi & = & \sum_{a}\com{\mathfrak{X}^{a}}{\com{\mathfrak{X}^{a}}{\Phi}}=\sum_{a}\left(\mathfrak{X}^{a}\mathfrak{X}^{a}\Phi+\Phi\mathfrak{X}^{a}\mathfrak{X}^{a}-2\mathfrak{X}^{a}\Phi\mathfrak{X}^{a}\right)\nonumber \\
 & = & \begin{pmatrix}\begin{array}{ccc}
0 & \cdots & 0\\
\vdots & \ddots & \vdots\\
0 & \cdots & 0
\end{array} & \sum\limits _{a}(X^{a}-x^{a})^{2}\ket{\phi}\\
\bra{\phi}\sum\limits _{a}(X^{a}-x^{a})^{2} & 0
\end{pmatrix}
\end{eqnarray}
 and thus the Laplace operator~$\laplacianxx$ can be written in terms of 
\begin{equation}
\laplacianxx=\sum\limits _{a=1}^{d}\left(X^{a}- x^{a}\right)^{2} \ 
\end{equation}
acting on $\ket{\phi}$ and $\bra{\phi}$, respectively.

Let us consider the corresponding quadratic form $\frac{1}{2}\mathrm{tr}(\Phi^{\dagger}\boxempty_{\mathfrak{X}}\Phi)$.
It can be written as
\begin{align}
\frac{1}{2}\mathrm{tr}(\Phi^{\dagger}\boxempty_{\mathfrak{X}}\Phi)
=\bra{\phi}\laplacianxx\ket{\phi} & =\sum_{a}\left(\Delta_{\phi}X^{a}\right)^{2}+\sum_{a}\left(\bra{\phi}X^{a}\ket{\phi}-x^{a}\right)^{2}\nonumber \\
 & =\disp(\phi)+|\vec{\bf x}(\phi)-\vec{x}|^{2} \nonumber\\
 &=: E(\vec x)
 \label{eq:lap_quadr_form}
\end{align}
assuming that $\phi$ is normalized. 
We will denote this as displacement energy, which coincides with $E(\vec x)$ in 
the Definition \ref{def_optimal_localized_state-2} 
of quasi-coherent states. 
In particular, the minimum of $E(\vec x)$ is precisely the smallest eigenvalue of $\laplacianxx$, 
and the corresponding quasi-coherent state 
is given by the corresponding eigenvector. 
We can therefore reformulate the definition \eq{def_optimal_localized_state-2} of quasi-coherent 
states as follows:
\begin{defn}
[Quasi-coherent states II]\label{def_optimal_localized_state-laplace}Let
$\vec{x}$ be a point in  target space $\realv^{d}$. 
Then the {\em quasi-coherent state(s)} at $\vec x$ are defined to be the ground state(s) $\Psi $
of $\laplacianxx$, and their eigenvalue 
\begin{align}
 \laplacianxx \Psi = E(\vec {x}) \Psi 
\end{align}
is the  {\em displacement energy}.
\end{defn}
This provides a very efficient and powerful 
way to obtain  quasi-coherent states by solving the eigenvalue problem\footnote{A similar definition of 
coherent states was given in  \cite{Ishiki:2015saa}, however assuming a semi-classical 
limit $N\to\infty$ of a sequence of matrix configurations. In our definition, $E(\vec x)$ contains non-trivial 
information about the dispersion and energy for finite $N$.}.
It also elucidates the relation with the standard definition in quantum mechanics,  interpreting
$\laplacianxx$ as deformed quantum-mechanical harmonic oscillator centered at $\vec x$.

Having solved the problem of finding quasi-coherent states for given $\vec{x}\in \realv^{d}$,
we still have to scan the target space in order to identify the regions with small $E(\vec x)$; 
this will be discussed in detail below. 
In any case, we have reduced the task 
to a $d$-dimensional problem, independent of the size of the matrices. 
This is very important, since $N$ should be sufficiently large 
to obtain a clear hierarchy of energies as discussed below. For example, squashed $\compv P^{2}$ 
corresponding to the $SU(3)$-representations $(0,20)$ has a 
$231$-  dimensional state space, while the
dimension of the target space in this case is $d=6$.

It is remarkable that these stringy ideas greatly simplify the problem of
measuring quantum geometries.

\subsection{Dirac operator}

Similar ideas also work for the Dirac operator instead of the Laplace operator. 
The matrix  Dirac operator 
$\slashed{D}_\mathfrak{X}\in End(\compv^{k}\otimes Mat_{m+1}(\compv))$,
$k=2^{\lfloor\frac{d}{2}\rfloor}$ is defined as
\begin{equation}
\slashed{D}_\mathfrak{X}=\sum_{a=1}^{d}\gamma^{a}\otimes\com{\mathfrak{X}^{a}}.
\end{equation}
with $\left\{ \gamma^{a},\,a=1,\dots,d\right\} $ forming a representation
of the Clifford algebra associated to $\realv^d$.
Again we are only interested in the off-diagonal entries of states
$\Psi\in\compv^{k}\otimes Mat_{m+1}(\compv)$ in the presence of a point brane
\begin{equation}
\Psi=\begin{pmatrix}\begin{array}{ccc}
0 & \cdots & 0\\
\vdots & \ddots & \vdots\\
0 & \cdots & 0
\end{array} & \ket{\psi}\\
\bra{\psi} & 0
\end{pmatrix}
\end{equation}
where $\ket{\psi}\in\compv^{k}\otimes\hilbert_{m}$ is a spinor-valued state. Then
the action of the Dirac operator $\slashed{D}_\mathfrak{X}$ on $\Psi$ is given by
\begin{eqnarray}
\slashed{D}_\mathfrak{X}\Psi & = & \sum_{a}\gamma^{a}\com{\mathfrak{X}^{a}}{\Psi}=\sum_{a}\gamma^{a}(\mathfrak{X}^{a}\Psi-\Psi\mathfrak{X}^{a})\nonumber \\
 & = & \begin{pmatrix}\begin{array}{ccc}
0 & \cdots & 0\\
\vdots & \ddots & \vdots\\
0 & \cdots & 0
\end{array} & \sum\limits _{a}\gamma^{a}(X^{a} - x^{a})\ket{\psi}\\
\bra{\psi}\sum\limits _{a}\gamma^{a}(X^{a} - x^{a}) & 0
\end{pmatrix} .
\end{eqnarray}
Restricted to the off-diagonal $\compv^{k}\otimes\hilbert_{m}$, it reduces to 
\begin{equation}
\diracxx=\sum_{a=1}^{d}\gamma^{a}(X^{a}-x^{a}) 
\end{equation}
which is a hermitian operator, with square
\begin{equation}
\diracxx^{2}=\laplacianxx +\Sigma^{ab}\com{X^{a}}{X^{b}}
\end{equation}
where $\Sigma^{ab}:=\frac{1}{4}\com{\gamma^{a}}{\gamma^{b}}$. 
The corresponding quadratic form reads
\begin{align}
\frac 12 \mathrm{tr}(\Psi^{\dagger}\diracxxx{\mathfrak{X}}^{2}\Psi) 
& =\bra{\psi}\diracxx^{2}\ket{\psi} =\bra{\psi}\laplacianxx\ket{\psi}
+\sum_{a,b=1}^{d}\bra{\psi}\Sigma^{ab}\com{X^{a}}{X^{b}} \ket{\psi} \nonumber \\
 & =|\vec{\bf x}(\psi)-\vec{x}|^{2}+\disp(\psi)+S(\psi)  ,
\end{align}
where we define
\begin{align}
 S(\psi) &:= \sum_{a,b=1}^{d}\bra{\psi}\Sigma^{ab}\com{X^{a}}{X^{b}} \ket{\psi}   \nonumber\\
 \vec{\bf x}(\psi) &:=  \bra{\psi}X^{a}\ket{\psi}
\end{align}
and similarly  $\disp(\psi)$ for spinor-valued states.
Since $\diracxx^2$ is positive, we conclude that $\disp(\psi) + S(\psi) \geq 0$ (because it is independent of $\vec x$!), and therefore
\begin{align}
	|\vec{\bf x}(\psi)-\vec{x}|^{2} \leq \bra{\psi} \diracxx^2 \ket{\psi}.
\end{align}
Furthermore, the ground state $\ket{\psi_0}$ of $\diracxx^2$ of course satisfies
$\bra{\psi_0} \diracxx^2 \ket{\psi_0} \leq \bra{\psi} \diracxx^2 \ket{\psi}$
for every normalized state $\ket{\psi} \in \C^k\otimes \hilbert_m$. 
Choosing $\ket{\psi} = \ket{s} \otimes \ket{\phi}$ for some $\ket{s} \in \C^k$ and for $\ket{\phi}$ being the ground state of the 
Laplace operator $\laplacianxx$, we have
\begin{align}
	\bra{\psi_0} \diracxx^2 \ket{\psi_0} \leq E_L(\vec{x}) + S(s, \phi),
\end{align}
where $E_L(\vec x)$ denotes the Laplace displacement energy \eq{eq:lap_quadr_form}. 
Note that replacing $\ket{s} \to C \ket{s^*}$ with its charge conjugate yields a sign flip of $S(s,\phi)$. 
Thus, we can choose $\ket{s}$ such that $S(s,\phi)$ is negative. We therefore get the following estimate for the groundstate $\ket{\psi_0}$ respectively its eigenvalue:
\begin{align}
	|\vec{\bf x}(\psi_0)-\vec{x}|^{2} \leq \bra{\psi_0} \diracxx^2 \ket{\psi_0} \leq E_L(\vec{x}).
	\label{eq:quadr_form_dirac-2}
\end{align}
If the ground state happens to be\footnote{This is the case for simple spaces such as fuzzy $S^2$ or the Moyal-Weyl quantum plane.} 
a product state $\ket{\psi_0} = \ket{s_0}\otimes \ket{\phi_0}$, the same arguments provide the estimate
\begin{align}
|\vec{\bf x}(\psi_0)-\vec{x}|^{2} \leq \bra{s_0,\phi_0} \diracxx^2 \ket{s_0, \phi_0} \leq | \vec{\bf x}(\psi_0)-\vec{x}|^{2} + \disp(\psi_0).
\end{align}
This leads to the following definition:
\begin{defn}
[Quasi-coherent spinor states]\label{def_optimal_localized_state-dirac}Let
$\vec{x}$ be a point in  target space $\realv^{d}$. 
Then the quasi-coherent spinor state(s) at $\vec x$ are defined to be the ground state(s) 
of $\diracxx^{2}$, and their eigenvalue 
\begin{align}
\diracxx^{2}\Psi = E(\vec {x}) \Psi 
\end{align}
is the  {\em (spinor) displacement energy}.
\end{defn}
The function $E(\vec x)$ satisfies the estimate \eq{eq:quadr_form_dirac-2}
\begin{align}
 |\vec{\bf x}(\psi)-\vec{x}|^{2} \leq E(\vec x) \leq E_L(\vec{x}).\end{align}
It turns out that this $E(\vec x)$ is very powerful to determine the location of the 
noncommutative brane: in all cases under consideration, $\diracxx$ 
appears to have exact (!)  zero modes on the branes\footnote{For $d=3$ the existence of zero 
modes\footnote{} follows from an index theorem as 
 shown by Berenstein  \cite{Berenstein:2012ts}, and we 
 will argue in section~\ref{sec:exact-zero-modes} that such  the 
 zero modes arise quite genrically. Analogous zero modes arise for intersecting noncommutative 
 higher-dimensional branes,  as shown in  \cite{Chatzistavrakidis:2011gs}.}.
 On the other hand, $E(\vec x)$ does not provide immediate information on the 
 dispersion $\disp(\Psi)$, hence on the quality of 
 the semi-classical approximation. In the same vein, the ground state(s) $\Psi$
 may or may not be product states in $\compv^{k}\otimes\hilbert_{m}$.
 This information could be extracted by keeping track of additional information 
 (e.g. the behavior under charge conjugation, possible degeneracies, the dispersion etc.), 
 or simply by taking into account also the Laplace operator. 
 Therefore in the present paper, we will consider both approaches using $\diracxx$ 
 and $\Box$, and apply the same algorithm in section \ref{sec:Numerically-finding-coherent} 
to extract the location of the brane from quasi-minima of the functions $E(\vec x)$.
This approach applies independently of possible degeneracies of the ground state(s).

\section{Measuring the quantum geometry\label{sec:Numerically-finding-coherent}}

Having the quasi-coherent states from the lowest eigenvectors
of  $\laplacianxx$ or $\diracxx^2$ at our disposal, 
we now address the problem of scanning the target space 
$\realv^{d}$ to  determine a subset $\manifold_E \subset \realv^d$ with quasi-minimal displacement energy
$E$, and corresponding \emph{quasi-minimal states} $\mathscr{S}_E\subset \compv P^{M}$.
Their expectation values produce a manifold 
\begin{align}
 \manifold :=\vec{\bf x}(\mathscr{S}_E) \ \approx \manifold_E,
\end{align}
which represents the semi-classical limit of the matrix geometry. 

\subsection{Quasi-minimal energy regions and hierarchy\label{sub:num_proc}}

Assume that we have found (numerically or analytically) the  smallest eigenvalue 
$E(\vec x)$ of $\laplacianxx$ for each $x\in \realv^{d}$,
or the smallest eigenvalue of $\diracxx^2$. 
This defines the ``displacement energy``
function\footnote{Such a function was also considered in \cite{Ishiki:2015saa} in the limit $N\to\infty$.}
\begin{align}
E:\,\realv^{d} & \to\realv\label{eq:func}\\
\vec{x} & \mapsto E(\vec x) = \min\left({\rm spec}\,\laplacianxx\right)\nonumber 
\end{align}
for the Laplacian, or similarly
\begin{align}
E:\,\realv^{d} & \to\realv\label{eq:func_dirac}\\
\vec{x} & \mapsto \min\left({\rm spec}\,\diracxx^2\right) \nonumber 
\end{align}
for the Dirac operator. 
Let us focus on the Laplacian for simplicity, and  
assume that the multiplicity of its lowest eigenspace is one\footnote{If the multiplicity is $k>1$, this indicates that 
the brane is really a stack of $k$ coincident branes. 
An example of this is the fuzzy 4-sphere $S^4_N$ 
\cite{Castelino:1997rv,Ramgoolam:2001zx,Steinacker:2015dra}.}.
Then the function $E(\vec x)=  \left|\vec x-\vec{\bf x}(\Psi)\right|^2 +  \disp(\Psi)$
 \eq{eq:lap_quadr_form} can be interpreted as
zero point energy of a string stretching from $\vec x$ to the fuzzy brane, 
and thereby encodes its location.
$E(\vec x)$ is differentiable everywhere except on
points where the two smallest eigenvalues cross each other;
however,  we will completely ignore this issue,
since for our numerical purpose  we will be working with finite differential quotients anyway.

Now assume that the matrix background describes some quantized manifold $\manifold\subset \realv^{d}$ of dimension $k < d$.
The difficulty in determining $\manifold$ is that $E(\vec x)$ is in general not constant on $\manifold$,
hence it is not sufficient to look for minimal surfaces.
However, in view of the explicit form \eq{eq:lap_quadr_form}
 we expect that $E(\vec x)$ grows like the square distance 
in the directions transversal to  $\manifold$, while it should change only slowly  in the directions along $\manifold$.
This means that the Hessian 
\begin{align}
 H_{\mu\nu} = \nabla_\mu \nabla_\nu E
\end{align}
at $x\in \manifold$ should have $k$ small eigenvalues which characterize the embedding  of 
$\manifold\subset \realv^d$, 
and $d-k$ large eigenvalues  of order one. 
This essential observation will be exploited in the procedure described below.

\paragraph{Quasi-minima of the $E$-function and its  hierarchy.}

We now describe a scanning procedure which allows to select a set of quasi-minima  $\manifold_E$ of $E$,
while verifying a manifold-like structure.
Again, this is not a universal procedure, and it may require some ``focusing'' by hand.
The quasi-classical nature of $\manifold_E$ is ensured by noting that 
for each $x\in \manifold_E$ there is a quasi-coherent state
$\psi$, which is the corresponding lowest
eigenstate of $\laplacianxx$ (or $\diracxx^2$).
Its expectation value differs from $\vec{x}$ by at most 
\begin{align}
 |\vec{\bf x}(\psi) - \vec x| \leq E(x) 
\end{align}
which is small by construction. This subset of quasi-coherent states 
 $\psi$ will be denoted by  $\mathscr{S}_E$.
 We could hence consider either $\manifold_E$  or $\manifold :=\vec{\bf x}(\mathscr{S}_E)$ as 
the semi-classical geometry, as long as $E$ is small.
However, it turns out that $\manifold =\vec{\bf x}(\mathscr{S}_E)$ works much
better, because it is remarkably in-sensitive to small perturbations of $\manifold_E$.
This will be understood  in the 
example of the fuzzy sphere in section~\ref{sec:Fuzzy-Sphere-revisited},
where we will show that the states  $\mathscr{S}_E$ are always Perelomov states
on the sphere even if $\manifold_E$ is slightly off.

Now start with a global minimum $x_0$ of $E$. The change of the function $E$ in some direction $\varepsilon$
up to second order in $|\varepsilon|$ is given by
\begin{equation}
|E(x_{0})-E(x_{0}+\varepsilon)|=\frac{1}{2}|\varepsilon^{T}H_{x_{0}}\varepsilon|+\mathcal{O}(|\varepsilon|^{3})
\end{equation}
where $H_{x_{0}} = \nabla\nabla E$ is the Hesse matrix  at  $x_{0}$. 
Assuming that the brane has a slowly varying noncommutative structure 
$|\nabla\nabla\Theta^{\mu\nu}|\ll 1$ \eq{Poisson-tensor},
one should observe a clear \textbf{hierarchy of ``small'' and ``large''
eigenvalues of $H$}. Moreover, the eigenvectors corresponding to the small
eigenvalues should constitute a basis of the tangent space~$T_{x_{0}}\manifold_E$ at $x_{0}$,
while the eigenvectors corresponding to the large eigenvalues (of order one) 
 should constitute a basis of the normal space to $\manifold_E$ at $x_{0}$.
Hence $x_{0}+\varepsilon$ is approximately
an element of $\manifold$ for $\varepsilon$ being an eigenvector
corresponding to a ``small'' eigenvalue and $|\varepsilon|$ 
sufficiently small. In particular,  the dimension of the manifold $\manifold$ is obtained as
the number $k$ of small eigenvalues of $H$.
This characterization of $\manifold$ is clear-cut as long as there is a clear hierarchy separating the 
 small and the   large eigenvalues of $H$.

If $x_{0}$ is not a local minimum of $E$, then the change of $E(x)$
in a direction $\varepsilon$ is given by 
\begin{equation}
E(x_{0}+\varepsilon)-E(x_{0})= \varepsilon\cdot\nabla E_{x_{0}}+\frac{1}{2}\varepsilon^{T}H_{x_{0}}\varepsilon+\mathcal{O}(|\varepsilon|^{3})
\end{equation}
with $\nabla E_{x_{0}}$
denoting the gradient of $E$ at point $x_{0}$. 
We can separate $\nabla E_{x_{0}} = (\nabla E_{x_{0}})^\parallel + (\nabla E_{x_{0}})^\perp$ 
into the tangential and transversal components, as defined by the Hessian.
On the quasi-classical manifold $\manifold$, 
we expect that $\nabla E$ points in the tangential directions (i.e.\ in the span of the 
low eigenvectors of $H$), so that $(\nabla E)^\perp \approx 0$.
If this is no longer the case, this indicates that we are moving away from $\manifold$.


\medskip{}

This leads to the following strategy:

\medskip{}
\emph{}%
\fbox{\begin{minipage}[t]{0.9\columnwidth}%
\emph{Given some quasi-minimum $x_{0}\in \manifold_E\subset\realv^{d}$ of $E$,  
the Hesse matrix $H=\nabla\nabla E$ at $x_{0}$ should exhibit a clear hierarchy
of small and large eigenvalues.}

\begin{enumerate}
\item \emph{We select all small eigenvalues $\lambda_{i} \ll 1$
of  $H$.}

\item \emph{New quasi-minimal points can be obtained as
$x_i=x_{0}+\varepsilon_{i}$ where $\varepsilon_{i}$ is an eigenvector
corresponding to the eigenvalue $\lambda_{i}$ with $\left|\varepsilon_{i}\right|$
being sufficiently small.}

 \item \emph{The points $x_i$ are replaced by the expectation values
$\vec{\bf x}(\Psi_{x_i})$ with $\Psi_{x_i}$ being the quasi-coherent state at $x_i$.
}

 \item \emph{This procedure should be iterated, verifying 
the hierarchy of $H$ at each step.}%
\end{enumerate}

\emph{The collection of these points constitutes our semi-classical approximation
$\manifold :=\vec{\bf x}(\mathscr{S}_E)$ with $\mathscr{S}_E$ being the set of
quasi-coherent states corresponding to $\manifold_E$.}
%
%
\end{minipage}}

\medskip{}

This strategy allows to identify a semi-classical approximation 
for a large class of examples. 
The procedure is justified as long as there is a clear hierarchy in the eigenvalues of $H$.
It needs to be specified by some cutoffs for this hierarchy, and possibly for $E$,  
$(\nabla E)^\perp$ and/or $|\nabla E|$. 
Since the hierarchy of eigenvalues depends on the example under consideration,
there is no universal prescription; the appropriate parameters 
must be adjusted for each example individually. 
This can be compared to taking a picture with a camera, 
where the photographer needs to adjust the focus to get a sharp image. 

Clearly the same procedure applies for the (squared) Dirac operator instead of the Laplace operator.
It is an interesting question whether the states selected by the Dirac
operator yield the same manifold as the Laplacian. We will provide some numerical and analytical examples in
sections  \secref{Fuzzy-Sphere-revisited} and \secref{Squashed--revisited}, and we will see that 
this is the case in simple examples, but not always.

If there is no clear hierarchy of eigenvalues of $H$, the matrices
simply do not contain enough information to extract a meaningful semi-classical manifold.

\subsection{Numerical procedure}

The considerations in the previous section provide the basis for an algorithm
to rasterize the manifold $\manifold \approx \manifold_E$. Locally we can find the tangent
space $T_{x_{0}}\manifold_E$ by using algorithm~\ref{alg:getdirections}, 
written as so-called pseudo-code.

\begin{algorithm}[h]

\begin{algorithmic}[1]

\Function{GetDirections}{function $f$, point $x_0$, dimension $k$}
\State $H$ $\leftarrow$ \Call{HesseMatrix}{$x_0$} 
\Comment Calculate Hesse Matrix $H_{x_{0}}$ at point $x_0$.
\State $\{(\lambda,\varepsilon)\}$ $\leftarrow$ \Call{Diag}{H} 
\Comment Diagonalize $H_{x_{0}}$ with eigenvectors $\varepsilon$ corresponding to eigenvalues 
$\lambda$. \\

\Return \Call{Select}{$\{(\lambda,\varepsilon)\}$,$k$} 
\Comment Return eigenvectors corresponding to the choice of $k$ ``small'' eigenvalues.
\EndFunction{}

\end{algorithmic}

\caption{Select directions corresponding to a ``small'' change in $f$\label{alg:getdirections}}

\end{algorithm}

On each point obtained by $x=x_{0}+\varepsilon$ as explained above, 
we can subsequently apply algorithm~\ref{alg:getdirections}, identify new tangential directions,
and gather new points of $\manifold_E$ respectively $\manifold$. 
These will sample an open neighborhood $U_{|\varepsilon|}(x)\subset \manifold$
for each point $x$, and by repeating this procedure the entire manifold should be covered. 
If this is done blindly, then many areas will be covered more than once. To avoid this redundancy,
one can attempt a \emph{nearest neighborhood
search} \cite{zezula2006SimSea-MetSpaApp} with respect to the Euclidean
metric, to prevent accepting points which are already covered.

One also has to implement some stopping mechanism which prevents
points to be accepted when the above criteria for $\manifold_E$ respective $\manifold$ are no longer satisfied.
A local stop can be imposed if
\begin{itemize}
\item $E$ exceeds a certain value $E_{\mathrm{crit}}$,
\item the norm of the gradient $|\nabla E|$ or of $(\nabla E)^\perp$ 
exceeds a certain value $\left(\nabla E\right)_{\mathrm{crit}}$,
\item the hierarchy of eigenvalues of the Hessian matrix $H_{x}$ no longer holds, 
i.e.\ the highest ``small'' eigenvalue $\lambda$ exceeds
a certain value $\lambda_{\mathrm{crit}}$.
\end{itemize}
A simple unoptimized form of an algorithm which provides a complete
point cloud of~$\manifold$ is presented by algorithm~\ref{alg:Successively-apply-alg.}
where standard programming structures are used.\footnote{See for reference~\cite{watt2004Prolandescon}.}

\begin{algorithm}[h]

\begin{algorithmic}[1]

\Function{Rasterize}{function $f$, startpoint $x_0$, dimension $k$}
\State $pc\leftarrow List$ \Comment List of points which constitute the point cloud.
\State $q\leftarrow Queue$ \Comment FIFO Queue which holds new unchecked points.
\\
\State add point $x_0$ to $pc$
\State add point $x_0$ to $q$
\\
\While{$q$ not empty}

  \\

  \State $x\leftarrow$ take next element from $q$ \Comment Current point to process.
  \State $\mathrm{dirs}\leftarrow$ \Call{GetDirections}{$f$,$x$,$k$}

  \\
  
  \ForAll{directions $\varepsilon$ in $\mathrm{dirs}$}
    
    \State project $x_\mathrm{new}=x+\varepsilon$ to its corresponding expectation
    value $\bf{x}_\mathrm{new}$

    \If{\Call{IsLegal}{$\bf{x}_\mathrm{new}$} is true} \Comment This refers to the considerations above.
    
      \State add $\bf{x}_\mathrm{new}$ to $pc$
      \State add $\bf{x}_\mathrm{new}$ to $q$
    
    \EndIf
  \EndFor

\\
\EndWhile

\\

\\ \Return $pc$

\EndFunction{}

\end{algorithmic}

\caption{Successively  apply alg.~\ref{alg:getdirections} to gather a complete
point cloud of~$\protect\manifold$.\label{alg:Successively-apply-alg.} }
\end{algorithm}

Furthermore, we clearly have to restrict the search
to a compact subset of $\realv^{d}$. For compact manifolds $\manifold$
described by finite-dimensional matrices, 
these bounds can be easily extracted from the spectrum of $X^a$. 
The result is a point cloud which is an approximation of $\manifold=\vec{\bf x}(\mathscr{S}_E)$.
The quality of the approximation depends on three aspects. Obviously
it is determined by the step length $|\varepsilon|$. Furthermore,
the separation of the hierarchy plays an important role, and depends
on the dimension of the given $M\times M$ matrices. Finally, the
approximation is affected by the above mentioned cut-off parameters
$E_{\mathrm{crit}}$, $\left(\nabla E\right)_{\mathrm{crit}}$ or
$\lambda_{\mathrm{crit}}$.

\subsection{Exact zero modes and brane location from $\diracxx$}
\label{sec:exact-zero-modes}

Even though the approach using $\laplacianxx$ is conceptually simpler at first sight, 
it turns out that the Dirac operator $\diracxx$ has typically  better properties. 
In fact for 2-dimensional fuzzy spaces embedded in $\realv^3$, 
the quasi-classical manifold $\manifold$ 
can be obtained from {\em exact} zero modes of $\diracxx$, as shown by Berenstein \cite{Berenstein:2012ts}.
The corresponding exact zero modes of $\diracxx$ deserve the name {\em coherent (spinor) states}.
Because the argument is so beautiful we shall repeat it here.
One defines the index of $\diracxx$ as difference of positive and negative eigenvalues,
\begin{align}
 {\rm ind}(\diracxx) = \frac 12(n_+ - n_-)
\end{align}
Clearly this index is locally constant and can only jump by $\pm 1$ at locations $\vec x$ where 
$\diracxx$ has a zero mode.
It is easy to see that for $\vec x \to \infty$ this index always vanishes, 
${\rm ind}(\slashed{D}_\infty) = 0$,
while for $\vec x=0$ it is one for a fuzzy sphere background\footnote{E.g.\ for $S^2_N$, 
the eigenvalues of 
$\sigma_a J^a$ fall into a positive and a negative multiplet whose dimension differs by one. } 
(and a large class of deformations thereof).
This implies that  ${\rm ind}(\diracxx) =0$ defines a 
surface around the origin, which is the location of  the fuzzy sphere.

This argument is  very compelling, and our numerical
studies (see section~\ref{sec:applic}) indicate that $\diracxx$ has exact zero modes also for 
many higher-dimensional fuzzy spaces
including the fuzzy torus and squashed fuzzy $\compv P^2$,  even if there 
is no well-defined notion of ``interior'' and ``exterior'' space. 
We can provide a  heuristic argument why this is the case. 
Consider the case of a flat $2n$-dimensional quantum plane $\realv^{2n}_\theta$
with commutation relations $[X^a,X^b] = i \Theta^{ab} \matunity$, embedded in target space $\realv^{2n+1}$
at the $x^{2n+1}=0$ hyperplane. It is then easy to see 
(cf.~\cite{Chatzistavrakidis:2011gs,Karczmarek:2015gda}) 
that the minimal eigenvalue $\lambda^{(2n+1)}(\vec x)$ of the
point probe Dirac operator\footnote{With $\gamma_{2n+1}$ given by the chirality 
operator $\gamma_1 ... \gamma_{2n}$ on $\realv^{2n}$.} $\diracxx^{(2n+1)}$ for a test-brane at $x^{2n+1}\neq 0$ is given by 
the transversal
distance, $\lambda^{(2n+1)}(\vec x) = x^{2n+1}$, with sign set by the orientation form
$\omega^{\wedge n}$ on the  quantum plane. This function divides target space $\realv^{2n+1}$
into  ``left'' and ``right'' half-spaces defined by 
$\lambda^{(2n+1)}(\vec x) > 0$ and  $\lambda^{(2n+1)}(\vec x) < 0$, respectively. Moreover, the corresponding 
minimal energy state is a product state $\ket{s,\psi}=\ket s\otimes\ket{\psi}$ of a 
coherent state on $\realv^{2n}_\theta$ and a spinor with definite chirality.

Now consider the 
generic case of a quantized symplectic space $\manifold^{2n} \hookrightarrow \R^d$ 
with $d>2n$, and a point probe brane at $\vec x \in \realv^d$. 
Assume that $\manifold$ is sufficiently flat near $\vec x$, and denote with $\vec x_0$
the closest point on $\manifold$ to $\vec x$. 
Then $\manifold$ can be well approximated by a 
quantum plane $\realv^{2n}_\theta$ through the tangent space $T_{x_0}\manifold$, and 
we can consider the reduced $2n+1$-dimensional Dirac operator $\diracxx^{(2n+1)}$  in the 
reduced target space $T_{x_0}\manifold\oplus \realv(\vec x-\vec x_0)$, with transversal coordinate  $x^{2n+1}$.
According to the above discussion, that 
Dirac operator has a minimal energy state\footnote{The localization properties of these coherent states were studied further in \cite{Karczmarek:2015gda}
for $d=3$.} with eigenvalue
$\lambda^{(2n+1)}(\vec x) \sim x^{2n+1}$.
Now the full Dirac operator $\diracxx$ can be considered as a
small perturbation of $\diracxx^{2n+1}\otimes \Gamma_k$ with $\Gamma_k^2 = \matunity$.
Standard arguments in perturbation theory 
then imply that $\diracxx$  has $k$ zero modes going from positive to negative 
$x^{2n+1}$, provided the perturbation is sufficiently small. 
Here $k\geq 2$ in the case of several transversal dimensions.

This argument already suggests that in the case of several transversal dimensions $d-2n>1$,
one should in general not expect that the zero modes of $\diracxx$ coincide on some lower-dimensional manifold.
In some cases of interest, $\diracxx$ might not even have any exact zero modes.
Our general method as explained above does not require the existence of 
exact zero modes, and it provides more information which allows to measure 
the quality of the semi-classical approximation, its effective dimension 
as well as the dispersion.

The localization properties of these coherent spinor states
on 2-dimensional surfaces were studied  in \cite{deBadyn:2015sca}.

\section{Applications and examples}
\label{sec:applic}

We elaborate the above results and apply the numerical scanning algorithm for several examples, 
starting with the fuzzy sphere.

\subsection{Fuzzy sphere revisited\label{sec:Fuzzy-Sphere-revisited}}

We recall the fuzzy sphere $S^2_N$, which is defined by the three matrices $X^{a},\,a=1,\dots,3$
which satisfy the commutation relations 
\begin{equation}
\com{X^{a}}{X^{b}}=i\,\frac{2}{\sqrt{N^{2}-1}}\,\varepsilon_{abc}X^{c}.\label{eq:fuzzysphere_emb_2}
\end{equation}

\subsubsection{The point probe Laplacian $\protect\laplacianxx$}

For the simple case of the fuzzy sphere we can explicitly evaluate
the  function $E(\vec x)$ ~(\ref{eq:func}) exactly, i.e.\ the minimal eigenvalue
of the point probe Laplacian\footnote{This is also calculated in~\cite{Ishiki:2015saa}.}
\begin{equation}
\laplacianxx=\sum_{a=1}^{3}\left(X^{a}-x^{a}\right)^{2}=\matunity-2\sum_{a=1}^{3}x^{a}X^{a}+\sum_{a=1}^{3}x^{a}x^{a}.
\end{equation}
Since this expression is invariant under $SO(3)$-rotations, it suffices
to consider the operator at the north pole~$\vec{x}=(0,0,x^{3})$
where
\begin{equation}
\laplacianxx =\matunity+|\vec{x}|^{2} - 2|\vec{x}|\,X^{3}\label{eq:rotated_lapl}
\end{equation}
assuming $x^{3}>0$ to be specific.
Obviously, eigenvectors of the
operator $\laplacianxx $ are eigenvectors of $X^{3}$ and
vice versa. Since the eigenvalues of $X^3$ are smaller than one, it follows that
the smallest eigenvalue $E$ of $\laplacianxx$ arises for the highest state vectors $\ket{\frac{N-1}{2},\frac{N-1}{2}}$.
Using $X^{3}\ket{\frac{N-1}{2},\frac{N-1}{2}}=\sqrt{\frac{N-1}{N+1}}\ket{\frac{N-1}{2},\frac{N-1}{2}}$,
we obtain
\begin{equation}
E(\vec{x})=1+|\vec{x}|^{2}-2|\vec{x}|\sqrt{\frac{N-1}{N+1}}
\end{equation}
whose minima are given by $|\vec{x}_{\min}|=\sqrt{\frac{N-1}{N+1}}=1+\mathcal{O}(\frac{1}{N})$.
Hence the quasi-minimal space $\manifold_E$ is sharply defined by 
\begin{equation}
\manifold_E=\{\vec{x}\in\realv^{3}:\,|\vec{x}|=\sqrt{\frac{N-1}{N+1}}=1+\mathcal{O}(\frac{1}{N})\}.
\end{equation}
It is remarkable that the quasi-coherent states $\mathscr{S}$ as defined by the point probe Laplacian 
always coincide with the Perelomov coherent states on $S^2_N$, even if $\vec x$ has the wrong length.
Therefore the minimal energy 
manifold coincides precisely with the expectation values
of the coherent states $\manifold_E = \vec{\bf x}(\mathscr{S})$, and minimizes
the dispersion $\disp(\Psi)=\sum_{a=1}^{3}(\Delta_{\Psi}X^{a})^{2}$.

\paragraph{A test of the numerical procedure.}

Independent of the above exact computations we check the implementation
of the numerical algorithm in this well understood case. As input
we take three $10\times10$ matrices $X^{1},\,X^{2},\,X^{3}$ given by the rescaled
generators of $SU(2)$ with the correct normalization
factor as in ~\eqref{fuzzysphere_emb_2}. They are
explicitly given by 
\begin{align}
X^{1} & =\frac{2}{\sqrt{99}}\times\mathrm{diag}_{2}\left(\frac{3}{2},2,\frac{\sqrt{21}}{2},\sqrt{6},\frac{5}{2},\sqrt{6},\frac{\sqrt{21}}{2},2,\frac{3}{2}\right)+\mathrm{h.c.},\nonumber \\
X^{2} & =-\frac{2\,i}{\sqrt{99}}\times\mathrm{diag}_{2}\left(\frac{3}{2},2,\frac{\sqrt{21}}{2},\sqrt{6},\frac{5}{2},\sqrt{6},\frac{\sqrt{21}}{2},2,\frac{3}{2}\right)+\mathrm{h.c.},\nonumber \\
X^{3} & =\mathrm{\frac{2}{\sqrt{99}}\times diag}\left(\frac{9}{2},\frac{7}{2},\frac{5}{2},\frac{3}{2},\frac{1}{2},-\frac{1}{2},-\frac{3}{2},-\frac{5}{2},-\frac{7}{2},-\frac{9}{2}\right)\label{eq:explicit_su2_matrices}
\end{align}
where $\mathrm{diag_{2}}(\dots)$ represents a matrix with entries
in the second diagonal. 

The numerical procedure then picks a global minimum $\vec{x}_{\min}$ of $E$,
with corresponding displacement energy 
\[
E(\vec{x}_{\min})\approx0.181818
\]
 which is in perfect agreement with the theoretical minimum $1-\frac{N-1}{N+1}=1-\frac{9}{11}\approx0.181818$
and the norm $|\vec{x}_{\min}|=\sqrt{\frac{N-1}{N+1}}\approx0.904534$.
The moduli of the eigenvalues of the Hesse matrix at this point $\vec{x}_{\min}$
are given by $\approx(1.1\times10^{-6},\,2.4\times10^{-6},\,2.)$, which exhibits a clear
hierarchy. Obviously, two directions are classified as ``small'', so that the effective dimension is 
found to be $\dim \manifold_E = 2$.

\begin{figure}
\begin{centering}
\hspace*{\fill}
\subfloat[Full picture of $\protect\manifold=S^2_{10}$.]{\begin{centering}
\includegraphics[width=0.35\textwidth]{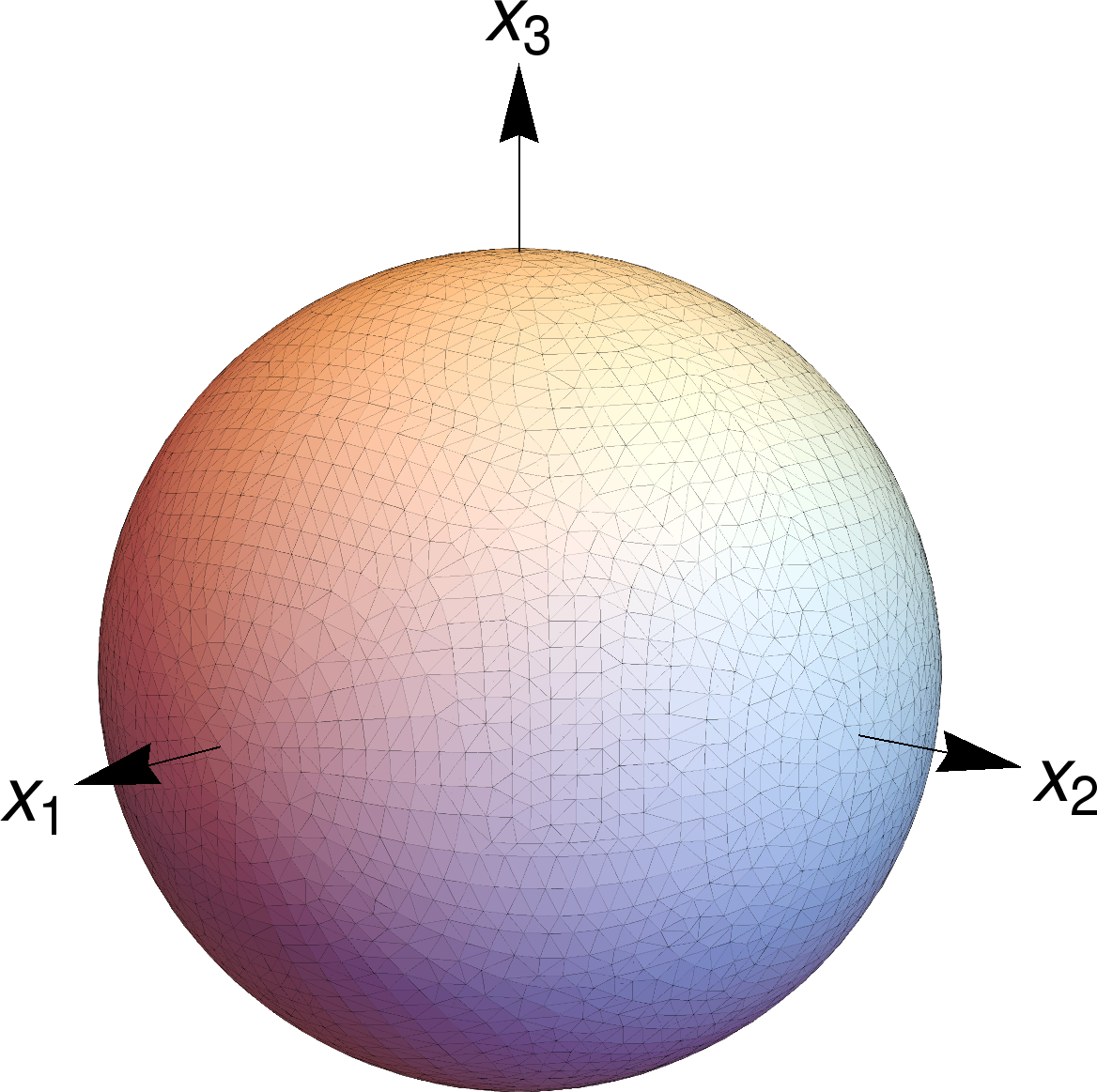}
\par\end{centering}

}\hfill\subfloat[A cut through the $x_{1}-x_{3}$-plane to show that there are no points
inside.]{\begin{centering}
\includegraphics[width=0.25\textwidth]{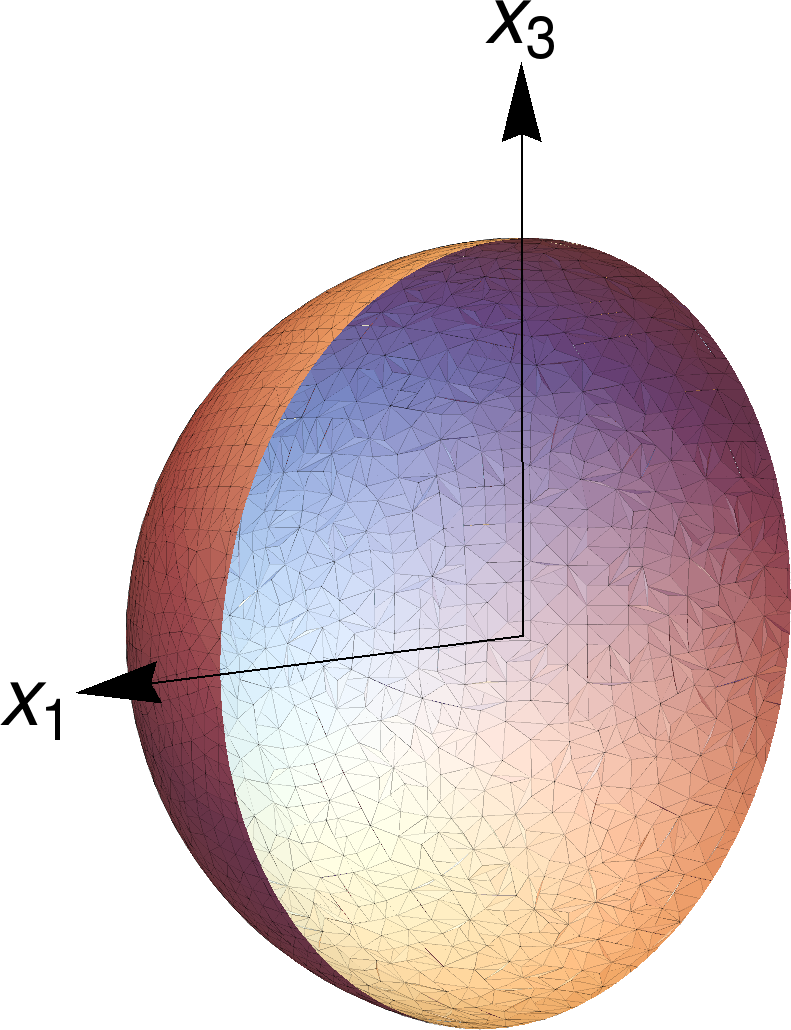}
\par\end{centering}
}\hspace*{\fill}
\par\end{centering}

\caption{Visualization of the semi-classical limit of the fuzzy sphere $S_{N}^{2}$
constructed from $S_{10}^{2}$.\label{fig:Visualization-of-the}}

\end{figure}

After applying algorithm~\ref{alg:Successively-apply-alg.} the result
is a point cloud representing the manifold $\manifold_E$. Under the
assumption that the point cloud constitutes a two-dimensional manifold,
one can build a mesh of polygons connecting these points to create
a visualization of $\manifold_E$. A picture is shown in \figref{Visualization-of-the}.
As expected one recovers the sphere $S^{2}$ with radius $R=\sqrt{\frac{N-1}{N+1}}$
to a good approximation.

\subsubsection{The point probe Dirac operator $\protect\diracxx$}

Similar considerations are possible for the Dirac operator 
\begin{equation}
\diracxx=\sum_{a=1}^{3}\sigma^{a}\left(X^{a}-x^{a}\right)
\end{equation}
 with $\sigma^{a}$ being the Pauli matrices 
\begin{equation}
\sigma^{1}=\begin{pmatrix}0 & 1\\
1 & 0
\end{pmatrix},\qquad\sigma^{2}=\begin{pmatrix}0 & -i\\
i & 0
\end{pmatrix},\qquad\sigma^{3}=\begin{pmatrix}1 & 0\\
0 & 1
\end{pmatrix}.
\end{equation}

Again due to the $SU(2)$ symmetry it is enough to consider points
$\vec{x}=(0,0,x^{3})$  on the positive $x^3$ axis, so that 
\begin{equation}
\diracxx =\sum_{a=1}^{3}\sigma^{a}X^{a}-\sigma^{3}|\vec{x}| \ .
\end{equation}
The eigenvector with minimal (absolute) eigenvalue~$\left(\sqrt{\frac{N-1}{N+1}}-\left|\vec{x}\right|\right)$
is given by $\ket{\uparrow}\otimes\ket{\frac{N-1}{2},\frac{N-1}{2}}$\footnote{The 
vector $\ket{\uparrow}$ is defined in the usual way as eigenvector
of $\sigma^{3}$ with the positive eigenvalue $1$.}. Thus the minimal energy function $E$ 
with respect to the squared Dirac operator $\diracxx^2$ 
is given by
\begin{equation}
E(\vec{x})=\Big(\sqrt{\frac{N-1}{N+1}}-|\vec{x}|\Big)^2.
\end{equation}
Its minima (which at the same time are roots of $E$) are obviously
given by $|\vec{x}_{\min}|=\sqrt{\frac{N-1}{N+1}}$ as
in the Laplacian case, which means that we get the same result for
the semi-classical limit $\manifold_E$ 
\begin{equation}
\manifold_E=\{\vec{x}\in\realv^{3}:\,|\vec{x}|=\sqrt{\frac{N-1}{N+1}}=1+\mathcal{O}(\frac{1}{N})\}.
\end{equation}
Remarkably, $\diracxx$ has {\em exact} zero modes for $\vec{x}\in \manifold_E \cong S^2$,  
which was first observed  in \cite{Berenstein:2012ts} and explained
on topological grounds. 
This interesting phenomenon\footnote{Note that the present Dirac operator does not 
anti-commute with any chirality operator, nevertheless it is the right choice in the 
present context.} generalizes to many higher-dimensional cases, 
including $S^4_N$ \cite{Karczmarek:2015gda,Steinacker:2015dra} and even squashed $\C P^2$ as discussed below.

We have seen that for the fuzzy sphere, the definition of the coherent states 
with respect to the Laplacian~$\laplacianxx$ is essentially
equivalent to the definition with respect to the Dirac operator~$\diracxx$.
Nevertheless, in a numerical context the method using the Dirac operator
has advantages, due to the fact that the minima of $f$ are typically exact roots
for all $N$. This leads to a greatly improved precision and hierarchy for
small $N$, and quasi- coherent states can be clearly identified even for small $N$.
However, we will use the same $10\times10$
matrices in (\ref{eq:explicit_su2_matrices}) to compare the numerical results with the bosonic approach.

\paragraph*{Numerical test.}

For the Dirac case our numerical implementation finds a global minimum
with
\[
E(\vec{x}_{\min}) \approx 0,
\]
within numerical accuracy. The eigenvalue moduli of the numerical Hesse matrix at point $\vec{x}_{\min}$
are given by $\approx (2.3\times 10^{-6}, \,7.3\times 10^{-6}, \,2.0)$.

The visual result is the same as in
the Laplacian case and is not displayed again.

\subsection{Fuzzy torus revisited\label{sec:Fuzzy-Torus-revisited}}

Recall, the fuzzy torus $T_{N}^{2}$ is defined by the quantized embedding
functions $X^{a}\climit x^{a}$ given by four $N\times N$ matrices
\begin{align}
X^{1} & =(U+U^{\dagger})/2\\
X^{2} & =-i\,(U-U^{\dagger})/2\nonumber \\
X^{3} & =(V+V^{\dagger})/2\nonumber \\
X^{4} & =-i\,(V-V^{\dagger})/2\nonumber 
\end{align}
with $U$ and $V$ being the \emph{shift} and \emph{clock} matrix,
\begin{equation}
U=\begin{pmatrix}0 & 1\\
 & 0 & 1\\
 &  & \ddots & \ddots\\
 &  &  & 0 & 1\\
1 &  &  &  & 0
\end{pmatrix},\qquad V=\begin{pmatrix}1\\
 & q\\
 &  & q^{2}\\
 &  &  & \ddots\\
 &  &  &  & q^{N-1}
\end{pmatrix},
\end{equation}
and $q=e^{2\pi i/N}$.

\subsubsection{The point probe Laplacian $\protect\laplacianxx$}

The Laplace operator $\laplacianxx$ is then defined as 
\begin{align}
\laplacianxx & =\sum_{a=1}^{4}\left(X^{a}- x^{a}\right)^{2}=\matunity+|\vec{x}|^{2}-2\sum_{a=1}^{4}X^{a}x^{a}\nonumber \\
 & =\matunity+|\vec{x}|^{2}-U\left(x^{1}-ix^{2}\right)-U^{\dagger}\left(x^{1}+ix^{2}\right)-V\left(x^{3}-ix^{4}\right)-V^{\dagger}\left(x^{3}+i x^{4}\right).
\end{align}

\begin{figure}
\subfloat[The raw point cloud which is generated by the numerical procedure.\label{fig:torus_point_cloud}]{\begin{centering}
\includegraphics[width=0.3\textwidth]{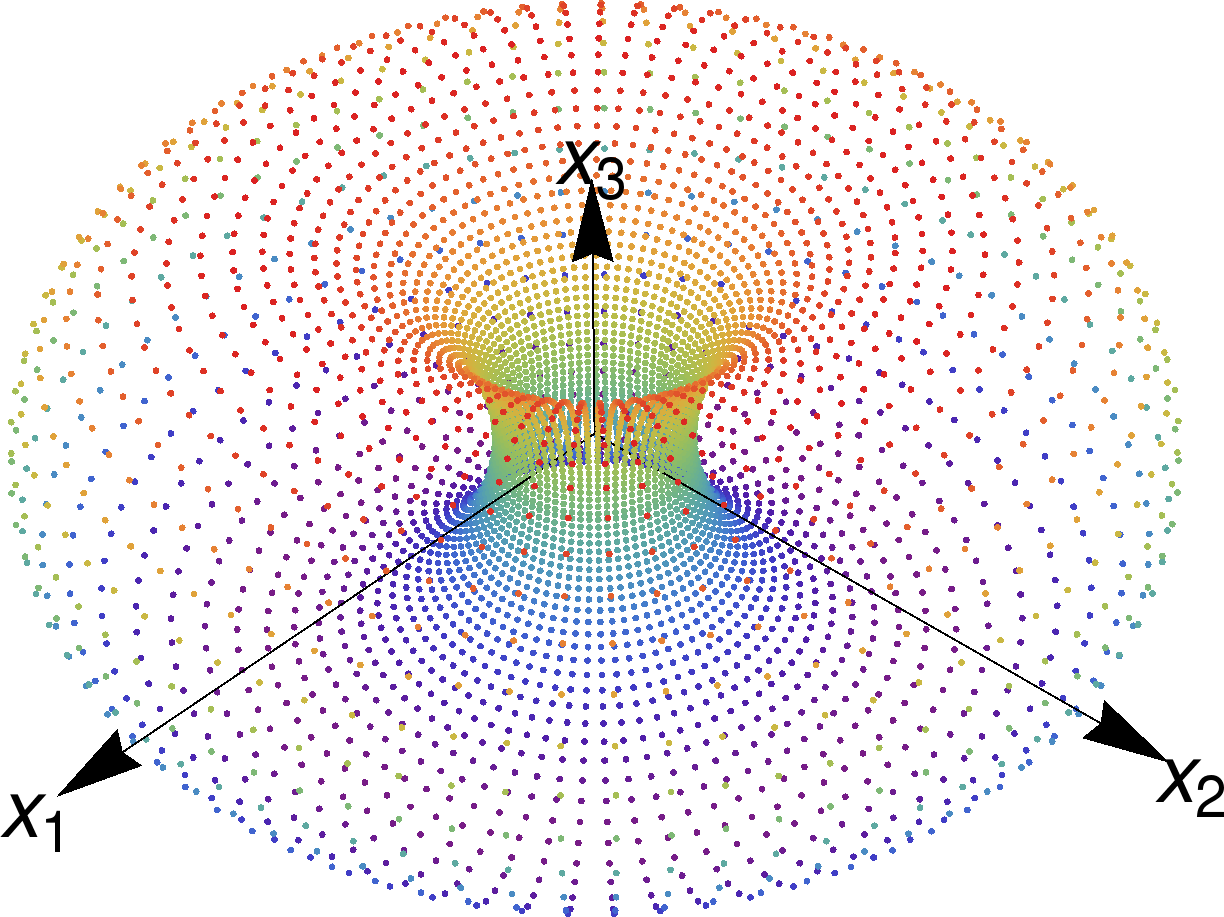}
\par\end{centering}

}\hfill{}\subfloat[A polygon mesh generated from the point cloud in \figref{torus_point_cloud}.]{\begin{centering}
\includegraphics[width=0.3\textwidth]{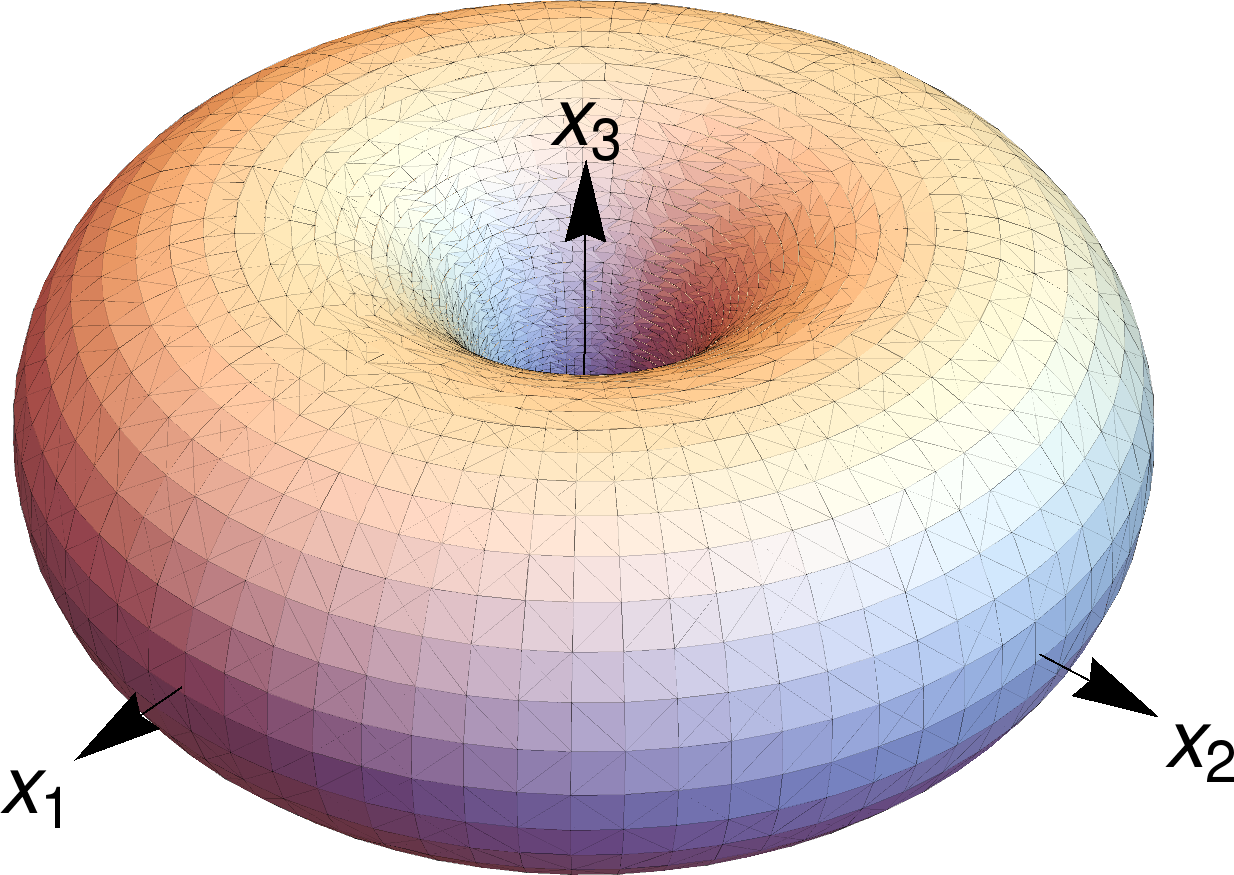}
\par\end{centering}

}\hfill{}\begin{centering}
\subfloat[A cut through the torus to show that it is hollow.]{\begin{centering}
\includegraphics[width=0.25\textwidth]{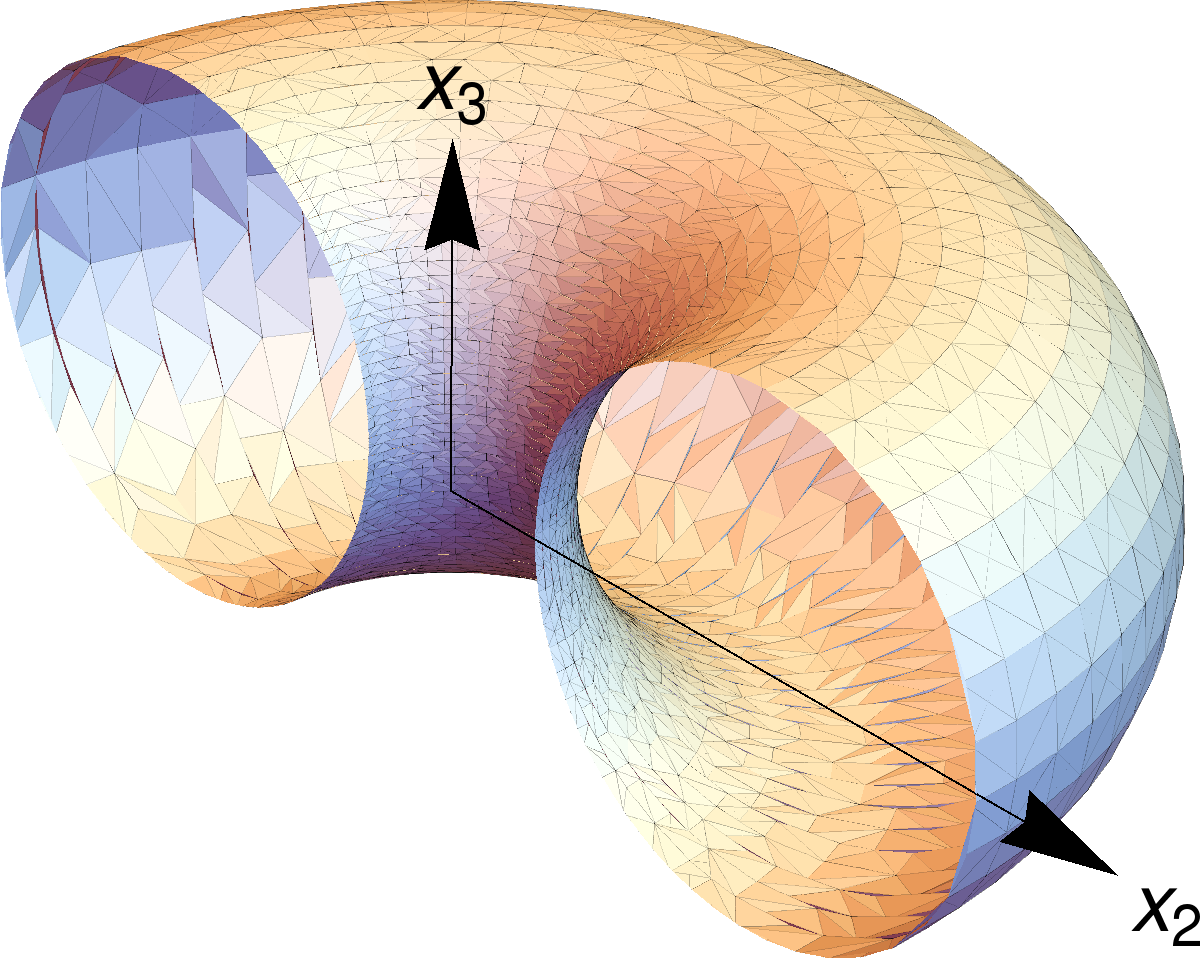}
\par\end{centering}

}
\par\end{centering}

\caption{Visualization of the semi-classical limit of the fuzzy torus $T_{N}^{2}$
constructed from $T_{20}^{2}$.\label{fig:vis_fuzzy_torus}}

\end{figure}

In this case we will skip an exact treatment -- although possible
(cf.~\cite{Ishiki:2015saa}) -- and turn directly  towards
the numerical results.
To visualize the resulting point cloud which represents a two-dimensional
manifold embedded in  $\manifold\subset S^{3}\subset\realv^{4}$,
one can use a generalized stereographic projection $\mathcal{P}:S^{3}\to\bar{\realv}^{3}$
defined by 
\begin{align}\label{torus-projection}
\mathcal{P}:S^{3}\subset\realv^{4} & \to\bar{\realv}^{3}\\
\begin{pmatrix}x_{1}\\
x_{2}\\
x_{3}\\
x_{4}
\end{pmatrix} & \mapsto\frac{1}{1-x_{4}}\begin{pmatrix}x_{1}\\
x_{2}\\
x_{3}
\end{pmatrix}\nonumber.
\end{align}
The result $\mathcal{P}(\manifold)$ is a two-dimensional manifold
embedded in $\realv^{3}$ which is shown in \figref{vis_fuzzy_torus}.

\subsubsection{The point probe Dirac operator $\protect\diracxx$}

The Dirac operator $\diracxx$ is given by
\begin{equation}
\diracxx=\sum_{a=1}^{4}\gamma^{a}\left(X^{a}-x^{a}\right)
\end{equation}
where the following matrices can be used as representation for the
Clifford algebra $C\ell_{4}(\realv)$:\begingroup \allowdisplaybreaks
\begin{alignat}{2}
\gamma^{1} & =\left(\begin{array}{cccc}
0 & 0 & 1 & 0\\
0 & 0 & 0 & 1\\
1 & 0 & 0 & 0\\
0 & 1 & 0 & 0
\end{array}\right), & \gamma^{2}=\left(\begin{array}{cccc}
0 & 0 & -i & 0\\
0 & 0 & 0 & -i\\
i & 0 & 0 & 0\\
0 & i & 0 & 0
\end{array}\right),\nonumber \\
\gamma^{3} & =\left(\begin{array}{cccc}
0 & 1 & 0 & 0\\
1 & 0 & 0 & 0\\
0 & 0 & 0 & -1\\
0 & 0 & -1 & 0
\end{array}\right),\qquad & \gamma^{4}=\left(\begin{array}{cccc}
0 & -i & 0 & 0\\
i & 0 & 0 & 0\\
0 & 0 & 0 & i\\
0 & 0 & -i & 0
\end{array}\right).
\end{alignat}
\endgroup
The numerical procedure yields the same point cloud as in~\figref{torus_point_cloud}.
Remarkably, even for low dimensional matrices the hierarchy of the Hessian is clearly
visible. For $5\times5$ matrices the Hesse matrix eigenvalues at the determined minimum are
$(1.6\times 10^{-5},\,3.5\times 10^{-4},\,1.26,\, 1.86)$.
The numerically obtained points fulfill  
\[
E(\vec{x})<8.6\times10^{-14}
\]
 which suggests that the Dirac operator $\diracxx$ has exact zero
modes at $T^{2}\subset\realv^{4}$.

\subsection{Squashed $\protect\compv P_{N}^{2}$ revisited\label{sec:Squashed--revisited}}

Now let us turn to squashed fuzzy $\compv P^{2}$ given by six matrices
$X^{a}\climit x^{a},\,a\in\mathcal{I}=\{1,2,4,5,6,7\}$ 
as explained in section \ref{sec:saqushed-cp2}.

\subsubsection{The point probe Laplacian $\protect\laplacianxx$}

The point probe Laplacian $\laplacianxx$ is as always given by 
\begin{equation}
\laplacianxx=\sum_{a\in\mathcal{I}}\left(X^{a}-x^{a}\right)^{2}.\label{eq:scp2_laplc}
\end{equation}
To visualize the numerical result, one can consider the intersection 
with a 3-dimensional subspace of the target space $\realv^{6}$. An
interesting choice is to consider the limit $x^{2}=x^{5}=x^{7}\to0$\footnote{Again recall that the indices take values in $\mathcal{I}=\{1,2,4,5,6,7\}$
for reasons discussed above.} which can be achieved by simply setting $x^{2}=x^{5}=x^{7}=0$
in~\eqref{scp2_laplc}. This corresponds to the limit considered in~\secref{Coherent-States-on_scp2}.
Taking $N=30$, i.e.\ the representation~$(0,30)$, one finds a global
minimum at 
\[
\vec{x}_{\min}\approx\left(0.546391,\,-0.546396,\,0.546392\right)
\]
 with corresponding energy  
\[
E(\vec{x}_{\min})\approx0.052891.
\]
The expectation values~$\vec{\bf x}(\phi)$ of the coherent state~$\phi$
corresponding to this minimum agree with $\vec{x}_{\min}$
for at least $6$ decimal digits. Hence $E(\vec{x}_{\min})\approx\disp(\phi)$
to a very good approximation.\footnote{See the general considerations in \secref{point-probes}, especially~\eqref{lap_quadr_form}.}

In~\secref{Coherent-States-on_scp2} the
Perelomov states were studied and led to a minimal energy value
\[
E_{\min}^{\mathrm{th}}=\frac{2}{3+N}\stackrel{N=30}{\approx}0.0606061
\]
 which implies a relative deviation of $\approx15\%$ from the numerical
value. The predicted norm of the expectation values $|\vec{\bf x}(\phi)|$
is 
\[
|\vec{\bf x}(\phi)|=\sqrt{3}c_{N}N\stackrel{N=30}{\approx}0.953463
\]
 which deviates by $0.7\%$ from the numerically obtained norm.

This numerical evidence suggests that the coherent states
for squashed~$\compv P^{2}_n$ obtained from $\laplacianxx$ are not exactly the Perelomov
coherent states but even better ones, if one takes the dispersion~$\disp(\phi)$
as a measure for quality. Nevertheless, the deviations of the expectation
values are small, therefore the Perelomov states should suffice as
a useful approximation. This can also be seen in~\figref{scp2_numerically}
which approximately looks like~\figref{scp2_3section} in~\secref{Coherent-States-on_scp2}.

\subsubsection{The point probe Dirac operator $\protect\diracxx$
and exact zero modes.}

It remains to examine the Dirac operator $\diracxx$ which in this
case is defined as 
\begin{equation}
\diracxx=\sum_{a\in\mathcal{I}}\gamma^{a}\left(X^{a}-x^{a}\right)
\end{equation}
with the six gamma matrices $\gamma^{a}$ given by\begingroup \allowdisplaybreaks
\begin{align}
\gamma^{1} & =\left(\begin{array}{cccccccc}
0 & 0 & 0 & 0 & 1 & 0 & 0 & 0\\
0 & 0 & 0 & 0 & 0 & 1 & 0 & 0\\
0 & 0 & 0 & 0 & 0 & 0 & 1 & 0\\
0 & 0 & 0 & 0 & 0 & 0 & 0 & 1\\
1 & 0 & 0 & 0 & 0 & 0 & 0 & 0\\
0 & 1 & 0 & 0 & 0 & 0 & 0 & 0\\
0 & 0 & 1 & 0 & 0 & 0 & 0 & 0\\
0 & 0 & 0 & 1 & 0 & 0 & 0 & 0
\end{array}\right), & \gamma^{2} & =\left(\begin{array}{cccccccc}
0 & 0 & 0 & 0 & -i & 0 & 0 & 0\\
0 & 0 & 0 & 0 & 0 & -i & 0 & 0\\
0 & 0 & 0 & 0 & 0 & 0 & -i & 0\\
0 & 0 & 0 & 0 & 0 & 0 & 0 & -i\\
i & 0 & 0 & 0 & 0 & 0 & 0 & 0\\
0 & i & 0 & 0 & 0 & 0 & 0 & 0\\
0 & 0 & i & 0 & 0 & 0 & 0 & 0\\
0 & 0 & 0 & i & 0 & 0 & 0 & 0
\end{array}\right),\nonumber \\
\gamma^{4} & =\left(\begin{array}{cccccccc}
0 & 0 & 1 & 0 & 0 & 0 & 0 & 0\\
0 & 0 & 0 & 1 & 0 & 0 & 0 & 0\\
1 & 0 & 0 & 0 & 0 & 0 & 0 & 0\\
0 & 1 & 0 & 0 & 0 & 0 & 0 & 0\\
0 & 0 & 0 & 0 & 0 & 0 & -1 & 0\\
0 & 0 & 0 & 0 & 0 & 0 & 0 & -1\\
0 & 0 & 0 & 0 & -1 & 0 & 0 & 0\\
0 & 0 & 0 & 0 & 0 & -1 & 0 & 0
\end{array}\right), & \gamma^{5} & =\left(\begin{array}{cccccccc}
0 & 0 & -i & 0 & 0 & 0 & 0 & 0\\
0 & 0 & 0 & -i & 0 & 0 & 0 & 0\\
i & 0 & 0 & 0 & 0 & 0 & 0 & 0\\
0 & i & 0 & 0 & 0 & 0 & 0 & 0\\
0 & 0 & 0 & 0 & 0 & 0 & i & 0\\
0 & 0 & 0 & 0 & 0 & 0 & 0 & i\\
0 & 0 & 0 & 0 & -i & 0 & 0 & 0\\
0 & 0 & 0 & 0 & 0 & -i & 0 & 0
\end{array}\right),\nonumber \\
\gamma^{6} & =\left(\begin{array}{cccccccc}
0 & 1 & 0 & 0 & 0 & 0 & 0 & 0\\
1 & 0 & 0 & 0 & 0 & 0 & 0 & 0\\
0 & 0 & 0 & -1 & 0 & 0 & 0 & 0\\
0 & 0 & -1 & 0 & 0 & 0 & 0 & 0\\
0 & 0 & 0 & 0 & 0 & -1 & 0 & 0\\
0 & 0 & 0 & 0 & -1 & 0 & 0 & 0\\
0 & 0 & 0 & 0 & 0 & 0 & 0 & 1\\
0 & 0 & 0 & 0 & 0 & 0 & 1 & 0
\end{array}\right), & \gamma^{7} & =\left(\begin{array}{cccccccc}
0 & -i & 0 & 0 & 0 & 0 & 0 & 0\\
i & 0 & 0 & 0 & 0 & 0 & 0 & 0\\
0 & 0 & 0 & i & 0 & 0 & 0 & 0\\
0 & 0 & -i & 0 & 0 & 0 & 0 & 0\\
0 & 0 & 0 & 0 & 0 & i & 0 & 0\\
0 & 0 & 0 & 0 & -i & 0 & 0 & 0\\
0 & 0 & 0 & 0 & 0 & 0 & 0 & -i\\
0 & 0 & 0 & 0 & 0 & 0 & i & 0
\end{array}\right).
\end{align}
\endgroup
In this case we carry out the computations taking $N=3$, which turns out to be 
sufficient to obtain a clear hierarchy. 
Searching for a global minimum yields 
\[
\vec{x}_{\min}\approx\left(-0.0608956,\,-0.582712,\,-0.291495\right)
\]
with displacement energy 
\[
E(\vec{x}_{\min})\approx0.
\]

\begin{figure}
\begin{centering}
	\hspace*{\fill}
	\subfloat[Laplace coherent states of $\Pi(\protect\compv P^2_{30})$.\label{fig:scp2_numerically}] {
		\centering{}
		\includegraphics[width=0.35\textwidth]{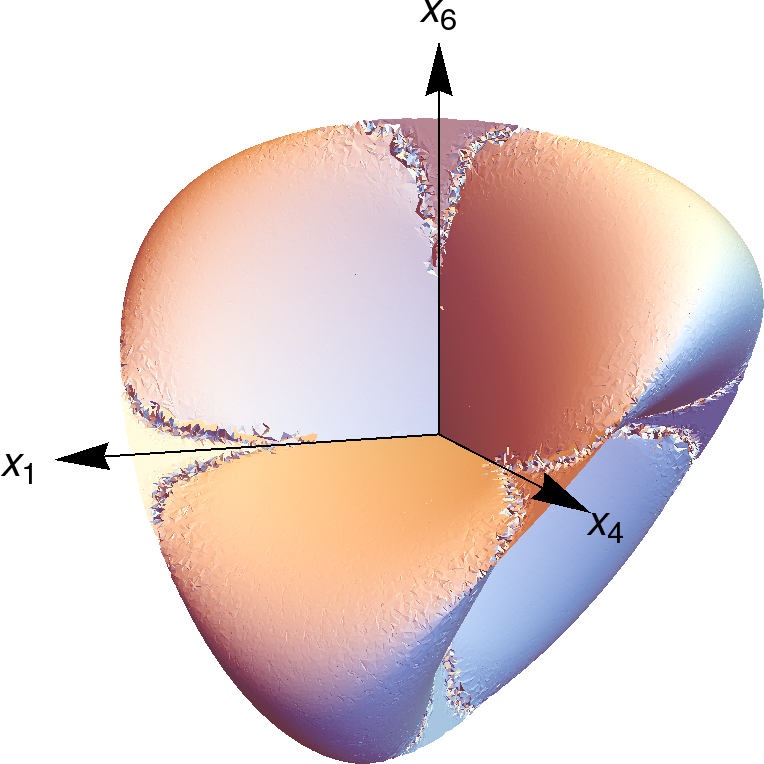}
	}
	\hfill{}
	\subfloat[Dirac coherent states of $\Pi(\protect\compv P_{3}^{2})$.\label{fig:vis_dirac_scp2}]{
		\centering{}
		\includegraphics[width=0.35\textwidth]{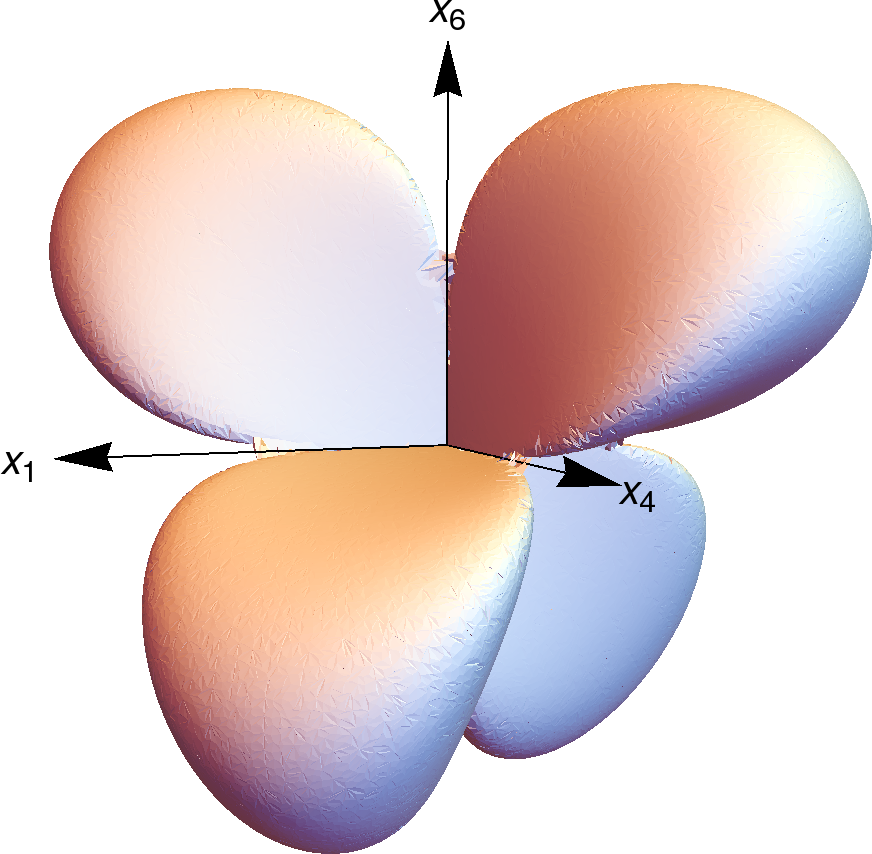}
	}
	\hfill{}
	\subfloat[A cut through one of the vaults of~\ref{fig:vis_dirac_scp2} to show that it is hollow.]{
		\centering{}
		\includegraphics[width=0.2\textwidth]{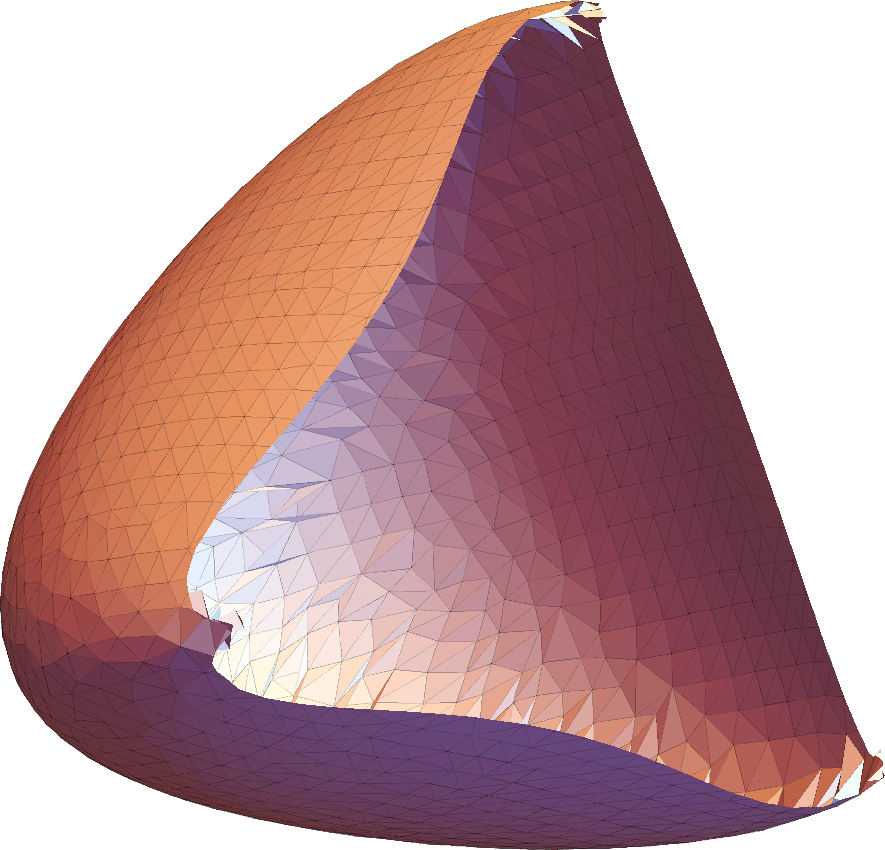}
	}
	\hspace*{\fill}
\par\end{centering}

\caption{Visualization of numerically obtained coherent states of squashed $\protect\compv P_{N}^{2}$.\label{fig:dirac_coh_states}}
\end{figure}

A visualization of the computation result for Dirac coherent states
can be seen in~\figref{dirac_coh_states}.

This is the first example where the Dirac coherent states do not agree
with the Laplace coherent states, although some similarity can be
recognized. Remarkably, the calculated points satisfy
\[
E(\vec{x})<1.14\times10^{-13}
\]
which again suggests that these states are {\em exact} zero modes of
the Dirac operator $\diracxx$. Indeed the general argument in section \ref{sec:exact-zero-modes}
strongly suggests that as there are two transversal dimensions,
there should be two exact zero modes of $\diracxx$ at the 
effective location of the semi-classical 4-dimensional manifold $\manifold$. 
Moreover since the target space 
is even-dimensional, the Dirac operator anti-commutes with the 6-dimensional 
chirality operator, so that its zero modes must always come in pairs of 
two\footnote{This is so because we are working with finite-dimensional operators
whose index in even dimensions  vanishes, cf.~\cite{Steinacker:2013eya}.}. Therefore there should be 
a 4-dimensional variety of exact zero modes. This is indeed seen numerically. 
To demonstrate this, let us consider a smooth curve $\gamma:\realv\to\realv^{3}_{135} \subset \R^6$
through the $135$-plane,
and the corresponding smooth curve of Dirac operators
\begin{equation}
t\mapsto \slashed{D}_{\gamma(t)} .
\end{equation}
We are able to numerically generate a smooth functions
$t\mapsto\lambda(t)$, where $\lambda(t)$ are eigenvalues of $\slashed{D}_{\gamma(t)}$,
to arbitrary high resolution. 
Moreover we can follow these smooth functions as they pass through zero, where they
cross with their chiral counterparts.
This confirms the existence of zero modes on a one parameter curve~$\gamma$.
Let us  choose the curve~$\gamma$ as follows 
\begin{equation}
\gamma(t)=\frac{1}{\sqrt{3}}\begin{pmatrix}1\\
1\\
-1
\end{pmatrix}t+\begin{pmatrix}0\\
\kappa\\
0
\end{pmatrix}e^{-(t-1/10)^{2}}
\end{equation}
setting~$\kappa=10^{-3}$. 
The (small) second term is added is to resolve additional degeneracies, which would occur 
e.g.\ on $\bar{\gamma}(t)=(1,1,-1)\,t$.

\begin{figure}
\begin{centering}
\subfloat[A plot of the $12$ lowest eigenvalues. The highest and the third
highest eigenvalues possess a multiplicity of $2$ 
(which can not be seen in this picture).\label{fig:ev_track_12}] {
\begin{centering}
  \includegraphics[width=0.45\textwidth]{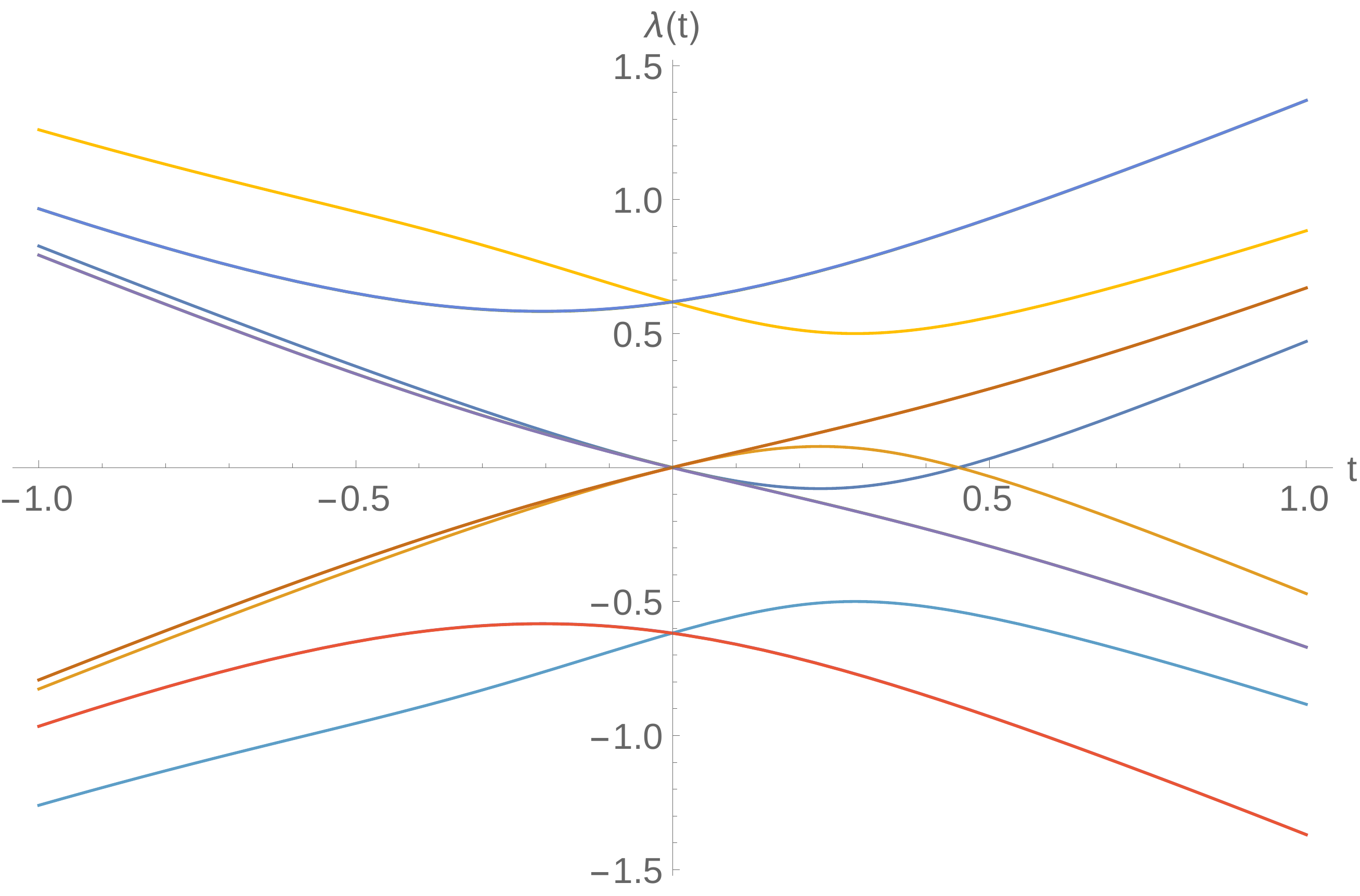}
\par\end{centering}}\hfill\subfloat[A plot of the two lowest eigenvalues. They clearly exhibit two roots.\label{fig:ev_track_2}]{
\begin{centering}
\includegraphics[width=0.45\textwidth]{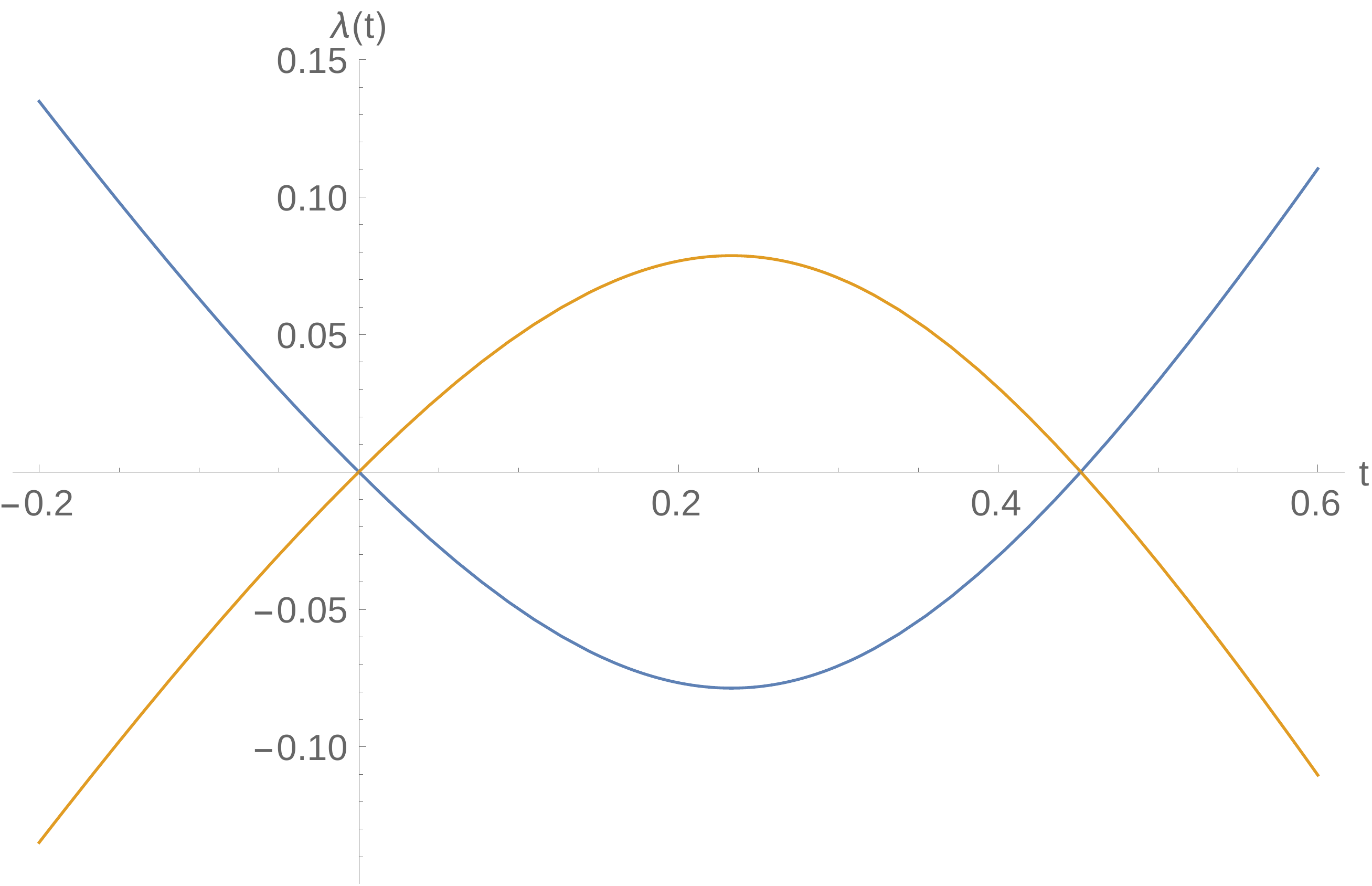}
\par\end{centering}}

\caption{Visualization of tracked eigenvalues of~$\diracxxx{\gamma(t)}$,
indicated by differing colors, for $\gamma(\protect\realv)$ approximately being
a straight line through the origin and the point~$(1,1,-1)$.}
\end{centering}
\end{figure}

To get an idea of the behavior of the eigenvalues of $\slashed{D}_{\gamma(t)}$
we track the $12$ lowest eigenvalues on the set~$\gamma([-1,1])$.
They are visualized in \figref{ev_track_12}, and a more detailed plot of lowest
two eigenvalues is given in~\figref{ev_track_2}.
One clearly observes the two sign changes: one at the origin and one
near~$\gamma(t\approx0.45)$. Having a zero mode at~$\gamma(t=0)$
is not surprising. However, the second roots near ~$\gamma(t\approx0.45)$ are not evident
a priori, and could only be asserted by numerical means.
This nicely confirms the discussion in section \ref{sec:exact-zero-modes}, which strongly suggests the 
existence of exact zero modes on a semi-classical 4-dimensional variety.

\section{Conclusion}

In this paper, we introduced generalized coherent states for generic quantum 
or fuzzy geometries 
defined by a set of matrices $X^a, \ a=1,...,d$, and used them to extract 
the effective geometry of such a configuration.
Adapting and generalizing ideas in 
\cite{Chatzistavrakidis:2011gs,Berenstein:2012ts,Ishiki:2015saa}, 
we propose to define quasi-coherent states
as ground states of a matrix Laplacian $\laplacianxx$ or a matrix Dirac operator $\diracxx$ in the presence of 
a point brane in target space.
The eigenvalues of the lowest off-diagonal modes 
are interpreted as displacement energy,
corresponding to the energy of strings stretching between the test brane and the background brane.
This string-inspired idea is independent of more traditional 
notions in noncommutative geometry such as spectral geometry or differential calculi,
and turns out to be very powerful.
These quasi-coherent states 
can be obtained numerically by a simple scanning procedure in target space, 
possibly selected by some cutoff criteria. 
Since they have good localization properties, 
the expectation values of the $X^a$ with these states provides a natural way to measure 
the location in target space, leading to an 
approximate location of some  (possibly degenerate) variety embedded in target space.
We provide analytical and numerical tests and illustrations of these ideas in various 
examples of fuzzy geometries.

Although similar ideas how to measure  matrix geometries
were put forward previously  \cite{Ishiki:2015saa}, we emphasize that our approach allows to address the 
case of a single, given matrix background, without relying on the 
existence of some limit $N\to\infty$. We discuss in particular  
ways to measure the quality of the geometric approximation, notably in terms of the hierarchy 
of eigenvalues of a  Hessian.

In this paper, only the basic principles of an algorithm to measure such quantum geometries
are presented. However, we also provide a full implementation 
as a \emph{Wolfram Mathematica} package {\text BProbe} \cite{lukas_schneiderbauer_2016_45045}, which is open to further developments.
Both approaches using the Laplacian $\laplacianxx$ and the Dirac operator $\diracxx$ are implemented.
This nicely reproduces the 
expected semi-classical geometry of  standard examples 
such as the fuzzy sphere and fuzzy tori, 
including the non-trivial case of squashed fuzzy $\compv P^2_N$.
We hope that this should provide a useful tool to assist further studies in this field.

While our numerical procedure to collect coherent states 
is mainly used here to visualize the corresponding classical
manifolds, they have much broader applications. 
The quasi-coherent states can be used to compute expectation values of various observables.
For example, the observables $[X^a,X^b]$ are expected to approximate a 
semi-classical Poisson structure, and more complicated observables 
such as $[X^a,X^b][X_b,X^c]$ are related to the effective ``open string'' metric as discussed 
in \cite{Steinacker:2010rh}. Similarly, the full differential geometry of 
embedded manifolds corresponding to the closed string metric
could in principle be extracted \cite{Ishiki:2015saa}.

One particularly interesting application of our method would be to measure the 
geometry of the matrix configurations obtained by numerical simulations of the IIB 
matrix model \cite{Kim:2012mw,Ito:2015mxa,Kim:2011cr}. These results suggest that 3+1-dimensional 
cosmological backgrounds are dynamically generated. The present methods should allow a much 
more detailed analysis of these geometries, and we hope to be  able to carry this 
out in the future\footnote{We thank J. Nishimura and A. Tsuchiya for support in this context.}.

\paragraph{Acknowledgements.}

This work was supported in part by the Austrian Science Fund (FWF) grants P24713 and 
 P28590 and by the Action MP1405 QSPACE from the European Cooperation in Science and Technology (COST).
H.S. would like to thank D. Berenstein, D. O'Connor, H. Grosse, J. Karczmarek, J. Madore
and A. Tsuchiya
for related discussions and correspondence, 
and J. Zahn for related collaboration.

\appendix

\section{Conventions for $SU(3)$}
\label{sec:appendix_su3_def}

A standard orthonormal basis of $\mathfrak{su}(3)$ is 
given by the Gell-Mann matrices:
\begin{eqnarray*}
\lambda_{1}=\begin{pmatrix}0 & 1 & 0\\
1 & 0 & 0\\
0 & 0 & 0
\end{pmatrix},\quad & \lambda_{2}=\begin{pmatrix}0 & -i & 0\\
i & 0 & 0\\
0 & 0 & 0
\end{pmatrix},\quad & \lambda_{3}=\begin{pmatrix}1 & 0 & 0\\
0 & -1 & 0\\
0 & 0 & 0
\end{pmatrix},\\
\lambda_{4}=\begin{pmatrix}0 & 0 & 1\\
0 & 0 & 0\\
1 & 0 & 0
\end{pmatrix},\quad & \lambda_{5}=\begin{pmatrix}0 & 0 & -i\\
0 & 0 & 0\\
i & 0 & 0
\end{pmatrix},\quad & \lambda_{6}=\begin{pmatrix}0 & 0 & 0\\
0 & 0 & 1\\
0 & 1 & 0
\end{pmatrix},\\
\lambda_{7}=\begin{pmatrix}0 & 0 & 0\\
0 & 0 & -i\\
0 & i & 0
\end{pmatrix},\quad & \lambda_{8}=\frac{1}{\sqrt{3}}\begin{pmatrix}1 & 0 & 0\\
0 & 1 & 0\\
0 & 0 & -2
\end{pmatrix}
\end{eqnarray*}
We use the rescaled basis   $t_{a}:=\lambda_{a}/2$, which 
 satisfy the commutation relations 
\begin{equation}
\com{t_{a}}{t_{b}}=i\,c_{abc}\,t_{c}
\end{equation}
where $c_{abc}$ are the so called \emph{antisymmetric structure constants}
of $\mathfrak{su}(3)$ given by 
\begin{eqnarray}
c_{123} & = & 1\label{eq:astructure_constants}\\
c_{147}=c_{165}=c_{246}=c_{257}=c_{345}=c_{376} & = & 1/2\nonumber \\
c_{458}=c_{678} & = & \sqrt{3}/2,\nonumber 
\end{eqnarray}
 while all the others vanish. They  determine the structure
of the Lie algebra respectively Lie group, and obey
the relations 
\[
\sum_{a,b=1}^{8}c_{abi}c_{abj}=3\,\delta_{ij} .
\]
Then the totally symmetric tensor
$d_{abc}$ of $\mathfrak{su}(3)$ can be defined by the relation 
\begin{equation}
\com{t_{a}}{t_{b}}_{+}=\frac{1}{3}\delta_{ab}+d_{abc}\,t_{c}
\end{equation}
where $\com ab_{+}$ is the anti-commutator.
They are given by
\begin{eqnarray}
d_{118}=d_{228}=d_{338}=-d_{888} & = & 1/\sqrt{3}\\
d_{448}=d_{558}=d_{668}=d_{778} & = & -1/(2\sqrt{3})\nonumber \\
d_{146}=d_{157}=-d_{247}=d_{256}=d_{344}=d_{355}=-d_{366}=-d_{377} & = & 1/2.\nonumber 
\end{eqnarray}
The root generators (or ladder
operators) 
\begin{align}
t_{1}^{\pm} & :=t_{4}\pm it_{5},\\
t_{2}^{\pm} & :=t_{6}\pm it_{7},\nonumber \\
t_{3}^{\pm} & :=t_{1}\pm it_{2}=\pm\com{t_{1}^{\pm}}{t_{2}^{\mp}}.\nonumber 
\end{align}
together with the Cartan generators $t_{3}$ and $t_{8}$ form
a Cartan-Weyl basis of $\mathfrak{su}(3)$ .

\section{Calculations for Coherent States of Squashed $\protect\compv P^{2}$\label{chap:scp2_calc}}

\subsection{Expectation values}

Let the rotation $U(\boldsymbol{\varphi})$ with $\boldsymbol{\varphi}=(\varphi_{1},\varphi_{2},\varphi_{3},\varphi_{4})$
be defined as 
\begin{equation}
U(\boldsymbol{\varphi})=e^{i\varphi_{1}T^{4}+i\varphi_{2}T^{5}+i\varphi_{3}T^{6}+i\varphi_{4}T^{7}}
\end{equation}
 with $T^{a}=\pi_{(0,N)}(t^{a})$. Additionally, let us define the
rotated vector $\ket{\boldsymbol{\varphi}}$ as
\begin{equation}
\ket{\boldsymbol{\varphi}}:=U(\boldsymbol{\varphi})\ket{\Psi_{0}},
\end{equation}
 where $\ket{\Psi_{0}}$ is the highest weight vector of a $(0,N)$
representation. We want to calculate the quantity 
\begin{equation}
\vec{\bf x}(\boldsymbol{\varphi})_{a}=\bra{\boldsymbol{\varphi}}X^{a}\ket{\boldsymbol{\varphi}}.\label{eq:calc_quant}
\end{equation}
To this end consider the adjoint action of $SU(3)$ on $Mat_{M}(\compv)$
given by $U^{-1}AU$ for some $A\in Mat_{M}(\compv)$ and $U=\Pi_{(0,N)}(g)$
belonging to the $(0,N)$ representation. Since $Mat_{M}(\compv)\repiso\bigoplus_{p=0}^{N}\hilbert_{(p,p)}$
where $T^{a}\in\hilbert_{(1,1)}$, the $SU(3)$ action leaves $\hilbert_{(1,1)}$
invariant and we can write 
\begin{equation}
Ad(T^{a})=U^{-1}T^{a}U=\sum_{b=1}^{8}R_{ab}T^{b}\label{eq:transformation}
\end{equation}
 for an orthogonal $8\times8$ matrix $R$. Since the representations
$Ad(T^{a})$ and $Ad(t^{a})$ are equivalent there exists an isomorphism
$f:\hilbert_{(1,1)}\to\mathfrak{su}(3)$ such that $f(Ad(T^{a}))=Ad(f(T^{a}))$.
Applying $f$ to~\eqref{transformation} we get 
\begin{equation}
Ad(t^{a})=\sum_{b=1}^{8}R_{ab}t^{b}.
\end{equation}
Using the natural scalar product on $\mathfrak{su}(3)$ given by $(A,B):=2\,\mathrm{tr}(A\cdot B)$
chosen such that the set $\left\{ t^{a},\,a=1,\dots,8\right\} $ forms
an orthonormal basis we can explicitly calculate the matrix coefficients
of $R$ via 
\begin{equation}
R_{ab}=(t^{a},\,Ad(t^{b}))=2\,\mathrm{tr}(t^{a}U^{-1}t^{b}U)
\end{equation}
which for $U=U(\boldsymbol{\varphi})$ can be carried out by computer
algebra systems.
Expression (\ref{eq:calc_quant}) can now be written as 
\[
\vec{\bf x}(\boldsymbol{\varphi})_{a}=\sum_{b=1}^{8}R_{ab}\bra{\Psi_{0}}X^{b}\ket{\Psi_{0}}=c_{N}\sum_{b=1}^{8}R_{ab}\bra{\Psi_{0}}T^{b}\ket{\Psi_{0}}
\]
 and since $\bra{\Psi_{0}}T^{8}\ket{\Psi_{0}}=\frac{N}{\sqrt{3}}$
is the only non-zero component we get 
\[
\vec{\bf x}(\boldsymbol{\varphi})_{a}=c_{N}R_{a8}\frac{N}{\sqrt{3}}
\]
and after plugging in the coefficients $R_{a8}$ we recover~\eqref{location_scp2}:
\begin{align}
\vec{\bf x}(\boldsymbol{\varphi}) & =c_{N}\frac{N}{2}\frac{1}{|\boldsymbol{\varphi}|}\begin{pmatrix}\frac{(\varphi_{1}\varphi_{3}+\varphi_{2}\varphi_{4})}{|\boldsymbol{\varphi}|}(\cos|\boldsymbol{\varphi}|-1)\\
2\frac{(\varphi_{1}\varphi_{4}-\varphi_{2}\varphi_{3})}{|\boldsymbol{\varphi}|}\sin^{2}\frac{|\boldsymbol{\varphi}|}{2}\\
\varphi_{2}\,\sin|\boldsymbol{\varphi}|\\
-\varphi_{1}\,\sin|\boldsymbol{\varphi}|\\
\varphi_{4}\,\sin|\boldsymbol{\varphi}|\\
-\varphi_{3}\,\sin|\boldsymbol{\varphi}|
\end{pmatrix}.\label{eq:location_scp2_copy}
\end{align}

\subsection{Dispersion}

Next we want to evaluate the dispersion~(\ref{eq:disp_squashed_gen})
which reads 
\begin{equation}
\disp(\boldsymbol{\varphi})=1-\sum_{i=3,8}\bra{\boldsymbol{\varphi}}(X^{i})^{2}
\ket{\boldsymbol{\varphi}}-|\vec{x}(\boldsymbol{\varphi})|^{2}.\label{eq:disp_squashed-calc}
\end{equation}
Having calculated the third term already we are left with the second
term $\sum_{i=3,8}\bra{\boldsymbol{\varphi}}(X^{i})^{2}\ket{\boldsymbol{\varphi}}$
which can be written as 
\begin{equation}
\sum_{i=3,8}\bra{\boldsymbol{\varphi}}(X^{i})^{2}\ket{\boldsymbol{\varphi}}=\sum_{i=3.8}\sum_{a,b=1}^{8}R_{ia}R_{ib}\bra{\Psi_{0}}X^{a}X^{b}\ket{\Psi_{0}}.\label{eq:disp_calc_step}
\end{equation}
The expression $M_{ab}:=\bra{\Psi_{0}}X^{a}X^{b}\ket{\Psi_{0}}=c_{N}^{2}\bra{\Psi_{0}}T^{a}T^{b}\ket{\Psi_{0}}$
can be calculated explicitly and yields 
\begin{equation}
M=c_{N}^{2}\frac{N}{4}\begin{pmatrix}0 & 0 & 0 & 0 & 0 & 0 & 0 & 0\\
0 & 0 & 0 & 0 & 0 & 0 & 0 & 0\\
0 & 0 & 0 & 0 & 0 & 0 & 0 & 0\\
0 & 0 & 0 & 1 & i & 0 & 0 & 0\\
0 & 0 & 0 & -i & 1 & 0 & 0 & 0\\
0 & 0 & 0 & 0 & 0 & 1 & i & 0\\
0 & 0 & 0 & 0 & 0 & -i & 1 & 0\\
0 & 0 & 0 & 0 & 0 & 0 & 0 & \frac{4}{3}N
\end{pmatrix}.
\end{equation}

With this we can compute (\ref{eq:disp_calc_step}) and get a long
expression for the second term  
\begin{align}
\sum_{i=3,8}\bra{\boldsymbol{\varphi}}(X^{i})^{2}\ket{\boldsymbol{\varphi}} & =c_{N}^{2}\frac{N}{48}\frac{1}{|\boldsymbol{\varphi}|^{4}}e^{-2i|\boldsymbol{\varphi}|}\times\nonumber \\
\times & \left(12e^{i|\boldsymbol{\varphi}|}(N-1)(\varphi_{1}^{2}+\varphi_{2}^{2})(\varphi_{3}^{2}+\varphi_{4}^{2})+12e^{3i|\boldsymbol{\varphi}|}(N-1)(\varphi_{1}^{2}+\varphi_{2}^{2})(\varphi_{3}^{2}+\varphi_{4}^{2})\right.\nonumber \\
+ & 3(N-1)\left(\varphi_{1}^{4}+\varphi_{2}^{4}+\varphi_{2}^{2}(\varphi_{3}^{2}+\varphi_{4}^{2})+(\varphi_{3}^{2}+\varphi_{4}^{2})^{2}+\varphi_{1}^{2}(2\varphi_{2}^{2}+\varphi_{3}^{2}+\varphi_{4}^{2})\right)\nonumber \\
+ & 3e^{4i|\boldsymbol{\varphi}|}(N-1)\left(\varphi_{1}^{4}+\varphi_{2}^{4}+\varphi_{2}^{2}(\varphi_{3}^{2}+\varphi_{4}^{2})+(\varphi_{3}^{2}+\varphi_{4}^{2})^{2}+\varphi_{\text{1}}^{2}(2\varphi_{2}^{2}+\varphi_{3}^{2}+\varphi_{4}^{2})\right)\nonumber \\
+ & 2e^{2i|\boldsymbol{\varphi}|}\left((3+5n)\varphi_{1}^{4}+(3+5n)\varphi_{2}^{4}+(15+N)\varphi_{2}^{2}(\varphi_{3}^{2}+\varphi_{4}^{2})+(3+5m)(\varphi_{3}^{2}+\varphi_{4}^{2})^{2}\right.\nonumber \\
 & \left.\left.+\varphi_{1}^{2}\left(2(3+m)\varphi_{2}^{2}+(15+m)(\varphi_{3}^{2}+\varphi_{4}^{2})\right)\right)\vphantom{12e^{i|\boldsymbol{\varphi}|}}\right).\label{eq:disp_calc_quadrterm}
\end{align}
Plugging (\ref{eq:location_scp2_copy}) and (\ref{eq:disp_calc_quadrterm})
into~\eqref{disp_squashed-calc} and simplifying thankfully yields
a more compact relation for the dispersion 
\begin{align}
\disp(\boldsymbol{\varphi})= & \frac{3}{8\,(3+N)}\frac{1}{|\boldsymbol{\varphi}|^{4}}\left\{ 4(\varphi_{1}^{2}+\varphi_{2}^{2})(\varphi_{3}^{2}+\varphi_{4}^{2})\cos|\boldsymbol{\varphi}|\right.\nonumber \\
+ & \left.\left(\varphi_{1}^{4}+\varphi_{2}^{4}+\varphi_{1}^{2}(2\,\varphi_{2}^{2}+\varphi_{3}^{2}+\varphi_{4}^{2})+\varphi_{2}^{2}(\varphi_{3}^{2}+\varphi_{4}^{2})+(\varphi_{3}^{2}+\varphi_{4}^{2})^{2}\right)\cos2|\boldsymbol{\varphi}|\right.\nonumber \\
+ & \left.\left(7\,(\varphi_{1}^{4}+\varphi_{2}^{4})+7\,(\varphi_{3}^{2}+\varphi_{4}^{2})^{2}+11\,\varphi_{2}^{2}(\varphi_{3}^{2}+\varphi_{4}^{2})\right)\right\} \label{eq:disp_scp2-copy}
\end{align}
which concludes the calculation.

\bibliography{papers}
\bibliographystyle{diss}

\end{document}